\documentclass[]{JFM-FLM_Au}

\lefttitle{P. Hegde, W. Herreman and G. M. Horstmann}
\righttitle{Journal of Fluid Mechanics}

\title{Stability theory for metal pad roll with consideration of viscous and magnetic damping}

\author{Pranav Hegde\aff{1},
	Wietze Herreman\aff{2}
	\and Gerrit Maik Horstmann\aff{1}}

\affiliation{
	\aff{1}Helmholtz-Zentrum Dresden-Rossendorf, Bautzner Landstr. 400, 01328 Dresden, Germany
	\aff{2}LIMSI, CNRS, Université Paris-Sud, Université Paris-Saclay, Orsay, F-91405, France
}

\corresau{Gerrit Maik Horstmann, \email{g.horstmann@hzdr.de}}

\newcommand{\h}[1]{\textbf{\textit{#1}}}
\newcommand{\ii}[1]{\text{#1}}

\newcommand{\nn}{\nonumber}
\newcommand{\overbar}[1]{\mkern 1.5mu\overline{\mkern-2mu#1\mkern-2mu}\mkern 1.5mu}
\newcommand{\RN}[1]{%
	\textup{\uppercase\expandafter{\romannumeral#1}}%
}
\newcommand{\Rn}[1]{%
	\textup{\expandafter{\romannumeral#1}}%
}
\usepackage{siunitx}
\usepackage{bm} 
\usepackage{amsmath} 
\usepackage{amssymb} 
\usepackage{physics} 
\DeclareDocumentCommand\dpdv{}{\displaystyle\partialderivative}
\usepackage{multirow}
\usepackage{cancel}
\usepackage{scalerel} 

\usepackage{etoolbox}
\makeatletter
\patchcmd{\NAT@citex}
{\@citea\NAT@hyper@{%
				\NAT@nmfmt{\NAT@nm}%
				\hyper@natlinkbreak{\NAT@aysep\NAT@spacechar}{\@citeb\@extra@b@citeb}%
				\NAT@date}}
{\@citea\NAT@nmfmt{\NAT@nm}%
		\NAT@aysep\NAT@spacechar\NAT@hyper@{\NAT@date}}{}{}

\patchcmd{\NAT@citex}
{\@citea\NAT@hyper@{%
				\NAT@nmfmt{\NAT@nm}%
				\hyper@natlinkbreak{\NAT@spacechar\NAT@@open\if*#1*\else#1\NAT@spacechar\fi}%
				{\@citeb\@extra@b@citeb}%
				\NAT@date}}
{\@citea\NAT@nmfmt{\NAT@nm}%
		\NAT@spacechar\NAT@@open\if*#1*\else#1\NAT@spacechar\fi\NAT@hyper@{\NAT@date}}
{}{}
\makeatother

\begin{document}
\maketitle

\begin{abstract}
We present a new theoretical model for the analytical prediction of the metal pad roll instability in idealised aluminium reduction cells consisting of two stably stratified liquid layers, which carry a vertical electric current and are exposed to a likewise vertical magnetic field. In contrast to the numerous previous models, we strive with this contribution for the most complete, and at the same time, a very approachable description of the linear metal pad roll instability. The model avoids the shallow-water approximation and incorporates some capillary effects, allowing it to be further applied to small-scale liquid metal batteries. As a major novelty the model also properly accounts for the stabilising effects of viscous and magnetic damping. To this end, we have derived explicit analytical formulae for the associated decay rates valid for arbitrary interfacial wave modes, which can also find application in other fields such as two-liquid sloshing. Our principal result is a set of explicit analytical solutions for stability thresholds and growth rates that are straightforward to implement and are intended to serve as theoretical benchmarks for multiphase magnetohydrodynamic solvers and model experiments. All of the results presented are carefully validated against five different and independent studies, with the particularity that we make use of quantitative stability threshold measurements for the first time. 
\end{abstract}

\begin{keywords}
Authors should not enter keywords on the manuscript, as these must be chosen by the author during the online submission process and will then be added during the typesetting process (see \href{https://www.cambridge.org/core/journals/journal-of-fluid-mechanics/information/list-of-keywords}{Keyword PDF} for the full list).  Other classifications will be added at the same time.
\end{keywords}


\section{Introduction}
\label{sec:intro}

\renewcommand{\figureautorefname}{figure}

Over the past 50 years, the modelling of the metal pad roll (MPR) instability has emerged as one of the flagship applications in the discipline of magnetohydrodynamics and can be found in textbooks today \citep{Davidson2001,gerbeau2006a}. The high level of interest is attributable in part to the physically intriguing and insightful self-excitation mechanism underlying the MPR instability, but more importantly to the high practical significance for aluminium production. Aluminium reduction cells (ARCs), which are employed in the highly energy-intensive Hall-Héroult process \citep{Evans2007} for the electrochemical reduction of alumina (${\rm Al}_2{\rm O}_3$) to aluminium, are known to be vulnerable to magnetohydrodynamic interfacial wave instabilities. In terms of hydrodynamics, ARCs can be regarded as stably stratified systems of two liquid layers carrying a strong vertical electric current, see \autoref{fig:cell_schematic}. The alumina is dissolved in molten cryolite (${\rm Na}_3{\rm Al}{\rm F}_6$) floating on top of the reduced pure aluminium, which deposits on the bottom owing to its higher density and forms the lower liquid layer. Between both layers, a free interface is formed, which can be destabilised and set in motion by Lorentz forces. Destabilising Lorentz forces can arise from several interactions between the cell current and induced or external magnetic fields. Many studies have shown that, of all possible situations, the most critical instability is the one resulting between the interaction of horizontal closing currents in the aluminium layer and the vertical component of an external magnetic field \citep[chapter~6.1]{gerbeau2006a}. External magnetic fields are induced in the power supply lines and cannot be fully circumvented during operation. In this paper, we use the term MPR instability to refer only to this particular scenario in line with the majority of the literature. The MPR instability causes a rotating sloshing motion in the aluminium (metal pad), which results from the superposition of at least two standing gravity waves coupled by the Lorentz force, see \S \ref{sec:lsa}.

Sloshing must be strictly controlled throughout the electrolysis as the cells can be short-circuited if the liquid aluminium makes contact with the upper carbon anodes. For ARCs to remain stable, the cryolite layer must not be thinner than about $3.5$ \textendash \ $4.5\,{\rm cm}$. This is, however, very unfavourable as the cryolite constitutes the anode-to-cathode distance, where, by far, the highest Ohmic losses occur. About $40\,\%$ of a cell’s electrical energy reduces no aluminium, but is instead converted into waste heat. Keeping in mind that aluminium production consumes around $8\,\%$ of the electricity used by the global industrial sector \citep{Kermeli2015}, the importance of understanding magnetohydrodynamic wave instabilities is self-evident.

The first successful attempts at explaining the MPR instability mechanism date back to the studies of \citet{Urata1985} and \citet{Sele1977}. On the basis of some heuristic arguments, the latter author succeeded in deducing an essential stability parameter
\begin{align}
	\beta_{\rm S} = \frac{I_0 B_z}{(\rho_{\rm a} - \rho_{\rm c})gh_{\rm a}h_{\rm c}},
\end{align}
today denoted as the Sele parameter, that was later proven to be indeed appropriate. In this definition, $I_0$ and $B_z$ denote the total cell current and the vertical magnetic field, $g$ is the standard acceleration due to gravity and $\rho_{\rm a}, \rho_{\rm c}$ and $h_{\rm a},h_{\rm c}$ refer to the densities and heights of the aluminium and cryolite layers, respectively. ARCs are rendered MPR unstable once the Sele parameter exceeds a critical value  $\beta_{\rm S} > \beta_{\rm crit}$, which largely depends on the lateral cell geometry, but also on viscous and magnetic damping, making it difficult to determine exactly. The scaling behaviour $\beta_{\rm S} \sim 1/h_{\rm c}$ explains why the cryolite layer thickness cannot be reduced indefinitely,
although it must be taken into account that viscous damping can likewise increase considerably with decreasing layer thicknesses, which becomes particularly important in the design of downscaled model experiments.

Further progress in gaining a deeper understanding of the MPR instability was made in the 1990s, most notably by the studies of \cite{Sneyd1994}, \cite{Bojarevics1994} and \cite{Davidson1998}. These authors applied perturbation techniques to study the stability of the linearised problem in the framework of the shallow-water approximation. It 
was revealed that the Lorentz force can couple different standing gravity wave modes that form a set of orthogonal functions. The MPR instability is triggered whenever the natural frequencies $\omega$ and $\omega'$ of at least two transverse gravity wave modes are sufficiently close to one another. The natural frequencies of the gravity wave modes are modified by the MHD coupling in a way that $\omega$ and $\omega'$ are gradually shifted towards each other with increasing $I_0 B_z$ until they coincide at $\beta_{\rm S} = \beta_{\rm crit}$. If $I_0 B_z$ is then increased even further ($\beta_{\rm S} > \beta_{\rm crit}$), the wave frequencies escape into the complex plane and give rise to exponential growth of the rotating metal pad wave. The stability threshold $\beta_{\rm crit}$ scales with the difference between the squared eigenfrequencies $\beta_{\rm crit} \sim |\omega^2 - \omega'^{2}|$. Cells in which pairs of gravity waves with identical
natural frequencies $\omega = \omega'$ exist (e.g., this is the case for square and cylindrical cell geometries) are therefore said to be inherently unstable;
here the stability threshold only depends on the integral damping in the system. An alternative view on the origin of the MPR instability was given by \citet{Lukyanov2001}, who argued that in closed domains the instability takes place as a result of multiple self-amplified wave reflections at the sidewalls. The MHD coupling mechanism is thereby inherently intertwined with the lateral walls, coining the term \textit{wall modes} \citep{Molokov2011}.

In the decade that followed, much attention was devoted to the development of nonlinear numerical models \citep{Zikanov2000,Zikanov2004,Bojarevics2006,gerbeau2006a} that can describe wave dynamics well beyond the stability threshold, e.g., for estimating saturation amplitudes. 
This has allowed ARCs to reach a certain level of maturity today, and a great deal of effort has gone into the optimisation of magnetic field and electric current distributions, for which mathematical modelling is still a primary design tool \citep{bojarevics2015}. 

Interest in the MPR instability, and in particular in its fundamental aspects, has resurfaced since 2015 in the context of liquid metal batteries (LMBs)\textemdash a promising technology for large-scale storage of renewable energies\textemdash in which several flow instabilities can appear that can have a crucial impact on the cell's efficiency and operational safety \citep{Kelley2018,Duczek2024}. LMBs are physically similar to ARCs, except that they consist of three liquid layers instead of two (metal\textendash salt\textendash alloy), and contain two free interfaces, which can be coupled both mechanically and electrically. Many studies have therefore addressed the coupled MPR wave physics and shown that two fundamental coupling states exist (fast symmetric in-phase and slow anti-symmetric out-of-phase wave motion), and that their destabilisation essentially depends on the densities of the selected active materials, or more precisely on the ratio of the density differences at the two interfaces \citep{Horstmann2018,Zikanov2018,Tucs2018,Molokov2018,Xiang2019,Herreman2023}. 

The great resurgence of interest in MPR instability has also brought substantial progress in the understanding of conventional two-layer systems, with two studies being particularly noteworthy. \citet{Herreman2019} calculated explicit solutions for the linear MPR growth rates in cylindrical cells using a perturbation theory approach, and for the first time calculated and included accurate approximations for viscous and magnetic damping instead of embedding purely phenomenological friction parameters. The presented linear damping rates are not directly relevant for industrial ARCs, in which complex turbulent flows predominate, but they are an important ingredient for benchmarking direct numerical solvers and decisively define stability onsets in downscaled MPR model experiments \citep{Nore2021,Eltishchev2022,Hegde2025}, and also in small and mid-size LMBs. The second study is authored by \citet{Politis2021} and, following many previous studies, addresses again the linear MPR stability in rectangular geometries representing idealised ARCs. Remarkably, the authors revealed that the previous understanding of the MPR instability was still lacking after more than 40 years of research. They showed that the degeneracy of the natural frequencies of gravity wave pairs $\omega = \omega'$ is only a necessary, but not a sufficient condition for instability. Instead, in the absence of damping, rectangular cells are inherently unstable ($\beta_{\rm crit} = 0$) if and only if the lateral aspect ratio fulfils the condition $L_x /L_y = \sqrt{\bm{m}/\bm{n}}$, where $\bm{m}$ and $\bm{n}$ are any two odd natural numbers. This implies an interesting fractal distribution of critical aspect ratios, which form an absolutely discontinuous dense set of unstable points. Even though the fractality is gradually smoothed out by damping, the superordinate structure remains preserved and describes the actual MPR physics. Both studies can be considered as state-of-the-art today; \cite{Herreman2019} on cylindrical and \cite{Politis2021} on rectangular domains, both providing correct and the most complete descriptions of the MPR instability for idealised two-layer cells.

Despite these major advances, the long-standing problem remains in the literature of not having been able to experimentally validate the MPR models to date. Although rotating waves have often been observed indirectly in reduction cells, e.g., by measuring fluctuations of the cell current \citep{bojarevics2015}, interface motions cannot be
measured directly in the high-temperature environment of ARCs. For this reason, several attempts have been made to develop downscaled model experiments that can synthesise or imitate MPR waves decoupled from the electrolytic process under ambient conditions. Even this has turned out to be rather difficult and, despite courageous attempts, there have been unsuccessful endeavours \citep{Grants2021}. The pioneering MPR experiment was devised by \citet{Pedchenko2009,Pedcenko2017}, who replaced the cryolite layer by a grid of vertical thin rods imitating its electrodynamic role, and the aluminium layer by the room-temperature liquid eutectic alloy GaInSn. The experiment succeeded in demonstrating the MPR instability, but also suffered some limitations. Most restrictive was the fact that the rods were intrusive and impaired the fluid mechanics, while the instability was artificially intensified when the GaInSn detached from the grid under wave motion. For these reasons the experiment cannot be employed as a benchmark for MPR theories and numerical solvers. This is exactly what our recent MPR experiment is designed for, which works with only one liquid layer of GaInSn and synthesises the MPR instability through a feedback loop mechanism \citep{Hegde2025}. Yet, the feedback mechanism still needs to be carefully calibrated to the true MPR physics in two-layer cells, which is one of the goals behind our theoretical work presented here. At this point, we would like to take the opportunity to emphasise that there is another, very sophisticated MPR experiment, which has, most unfortunately, only been published in a Russian-language journal and has therefore been overlooked by almost the entire MPR community. Nevertheless, we are pleased to pay tribute to the authors{'}\,work, albeit belatedly. \cite{Borisov2010} stratified two immiscible transparent solutions of nitric acid ($\rm{HNO}_{\rm3(aq)}$ at the bottom and $\rm{HNO}_{3}$ + ${\rm C}_5\rm{H}_{12}\rm{O}$ at the top) in rectangular cells with different aspect ratios. As both liquids have very similar densities, the setup facilitates destabilisation of the interface with comparatively small currents and weak magnetic fields. It is the only study that has ever reported experimental stability onsets including the associated most unstable wave pairs as a function of the lateral aspect ratio. For the sake of completeness, we conclude our experimental survey by mentioning \cite{Eltishchev2022,Eltishchev2024}, who destabilised rotating surface waves in a cylindrical cell through a MHD forcing mechanism that is quite similar to that of the MPR instability.

Spurred on by these recent experimental advances and the rediscovery of Borisov's experiment, we present a refined theoretical MPR model in this paper, in a final attempt to harmonise the previous analytical approaches and to provide a complete description of the \textit{linear} MPR instability for rectangular geometries. Despite the high degree of maturity demonstrated by the recent theoretical studies, there are still a number of shortcomings that have rendered experimental benchmarking unfeasible. First, all rectangular MPR models are formulated in the framework of the shallow water approximation. This is accurate for ARCs, but the MPR model experiments almost all operate considerably outside the shallow water regime, as do LMBs. Second, previous analytical models do not take dissipation into account or simply include phenomenological friction parameters \citep{Politis2021}. This is again sufficient for ARCs, since the turbulent dissipation can only be estimated or at best measured anyway. However, stability onsets of downscaled MPR experiments and LMBs are decisively affected by linear dissipation effects, primarily by viscous and magnetic damping rates. 

We seize on this in the present study and derive explicit analytical expressions for both viscous and magnetic damping rates. Our viscous damping solution is an extension of the  free-surface damping problem first treated by \cite{Keulegan1959} and is widely applicable beyond the MPR instability, e.g. in two-liquid sloshing \citep{LaRocca2002}. Magnetic damping rates due to a vertical magnetic field were first calculated by \cite{Sreenivasan2005}, but only for the particular unidirectional wave mode $(m=1,n=0)$, and only in the shallow water limit. We calculate magnetic damping rates for arbitrary wave modes and also present an exact solution for the wave-induced electric current distribution. To formulate the actual linear stability problem, we apply a perturbation ansatz similar to the one \cite{Herreman2019} have used to describe the MPR instability in the cylindrical domains and express the solvability condition for a pair of two arbitrary gravity-capillary wave modes, resulting in a $2\times2$ interaction matrix. This approach will allow us, for the first time, to derive explicit analytical solutions for growth rates and stability thresholds while considering dissipation effects. As our stability criteria do not require any numerics, our solution is much more approachable than preceding studies and can be easily implemented with little prior knowledge. We will show that in the limit of shallow layers, constant damping, strongly contrasting electrical conductivities between the two layers, and negligible interfacial tension, our analytical predictions are fully identical to the numerical calculations by \cite{Politis2021}, and we further carefully validate our outcomes against experimental and theoretical results gathered five different and independent studies, including the hitherto overlooked experiment by \cite{Borisov2010}.

The manuscript is organised as follows: In \S \,\ref{sec:lsa} we establish the framework of our problem and the mathematical formulation on the basis of perturbation methods. Initially, we present and discuss the stability solution as a function of a phenomenological damping parameter in \S\,\ref{sec:growth_rate_gen}. For precise estimation of the damping parameter in small-scale cells, we calculate analytical decay rates comprising viscous and magnetic damping separately . This was achieved by deriving  analytical solutions for the Stokes boundary layers (\S\,\ref{sec:visc_damp}) and wave motion-induced current distributions (\S\,\ref{sec:mag_damp}). In \S\,\ref{sec:results}, 
we individually compare the derived stability onsets and damping rates with experimental and numerical results by 
\cite{LaRocca2002} (\S\,\ref{sec:visc_LaRocca}), \cite{Sreenivasan2005} (\S\,\ref{sec:Sreen}), \cite{Politis2021} (\S\,\ref{sec:fractal}) and \cite{Borisov2010} (\S\,\ref{sec:Borisov}).   
In \S\,\ref{sec:discuss} we collate our results and discuss open questions.

\section{Linear stability analysis}
\label{sec:lsa}
\subsection{The theoretical model for a reduction cell}
\begin{figure}
  \centerline{\includegraphics{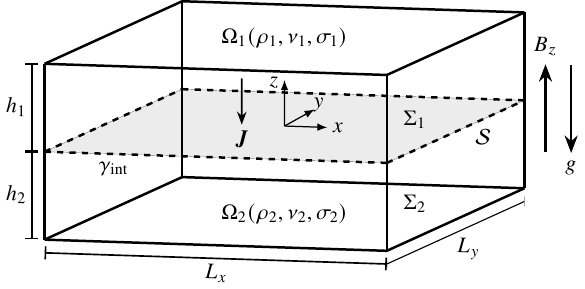}}
  \caption{Schematic of the cell with origin $O(x,y,z)$ consisting of two fluid mediums $\Omega_1$ (upper), $\Omega_2$(lower) bounded by a rigid boundary (container walls) $\Sigma_1$ and $\Sigma_2$ respectively and their respective densities $\rho_1,\rho_2$, viscosities $\nu_1,\nu_2$ and electrical conductivity $\sigma_1,\sigma_2$, with layer of thickness $h_1$ and $h_2$. Gravity $g$ acts in the negative $z$ direction, stratifying the two media with an interface $\mathcal{S}$ and interfacial tension $\gamma_{\ii{int}}$.}
\label{fig:cell_schematic}
\end{figure}

Our mathematical framework is shown in \autoref{fig:cell_schematic}. We consider a system of two immiscible liquid layers ($\Omega_1$ and $\Omega_2$) filling a rectangular domain ($x$, $y$, $z$) of dimensions $L_x$, $L_y$ and $H=h_1+h_2$. For both layers, the material properties of density $\rho_i$, kinematic viscosity $\nu_i$ and electrical conductivity $\sigma_i$ are taken into account. The subscripts $i=1,2$ refer to the upper and lower layer respectively. The origin of the coordinate system $\mathcal{O}$ is placed at the centre of the unperturbed interface $\mathcal{S}$ located at $z=\eta(x,y,t)$, and the rigid wall boundaries of each liquid layer are denoted as $\Sigma_i$. We include interfacial tension $\gamma_{\ii{int}}$ to our model, which acts in addition to gravity $g$ as a restoring force, but neglect capillary effects at the side walls, i.e., we assume that the contact line slides freely along the side walls while maintaining a static contact angle of $90^{\circ}$. The entire cell is exposed to an external homogeneous magnetic field oriented vertically $\h B = B_z \h e_z$ and gravity acts in the negative $z$ direction $\h g = -g \h e_z$. At equilibrium, both layers carry a homogeneous vertical electrical current $\h J=J_0 \h e_z$ with $J_0 = -I_0/(L_xL_y)$, where $J_0$ denotes the electrical current density and $I_0$ the total cell current. The equilibrium electrical potential appears in the form of $\Phi_i = \Phi_0 - J_0z/\sigma_i$, where $\Phi_0$ is an arbitrary constant.

\subsection{Governing equations and boundary conditions}
Considering low magnetic field strength $\bm B$ in both layers, the Lundquist number for a given layer reads
\begin{equation}
	Lu = \sigma B \sqrt{\frac{L_x L_y\mu_0}{\rho}} < 1.
\end{equation} 
The term $B$ here is the magnitude of the max magnetic field strength and $\mu_0$ is the permeability of free space. At low Lundquist numbers, the time derivative $\partial\h B/\partial t$ can be neglected. This assumption reduces the magnetic problem to a static case in both the liquid layers. Therefore we can impose a quasi-magnetostatic approximation on the system where the electric field is a dependent variable of the electric potential. The perturbed quantities $(\h u_i, p_i, \h j_i, \varphi_i, \h b_i)$, for the velocity, pressure, current density, electrical potential and induced magnetic field, are included in the following governing equations
\begin{subequations}
	\begin{align}
		\rho_i \partial_t{\h u}_i + {\nabla}p_i 
		&= \h J \times \h b_i + \bm{j}_i \times B_z\h e_z + \rho_i\nu_i\nabla^2\h u_i, \label{eq:gov_momentum}
		\\
		\bm{\nabla} \cdot \h{u}_i &= 0, \label{eq:gov_cont}
		\\
		\bm\nabla \cdot \h j_i &= 0,  \label{eq:charge_cont}
		\\
		\h j_i &= -\sigma_i\bm\nabla\varphi_i 
		+ \sigma_i (\bm u_i \times \bm B_z\h e_z),
		\label{eq:gen_ohms_law}
		\\
		\bm \nabla \times \h b_i &= \mu_0\h J, \qquad \bm \nabla \cdot \h b_i = 0.
		\label{eq:gauss+amp_max_law}
	\end{align} \label{eq:all_gov}
\end{subequations}
The equation \eqref{eq:gov_momentum} is the linearised conservation of momentum relation; \eqref{eq:gov_cont} for the continuity of velocity and \eqref{eq:charge_cont} for the compensation current. Equation (\ref{eq:gen_ohms_law}) refers to quasi-magnetostatic approximation of Ohm's law \citep{Davidson2001} (the magnetostatic approximation ignores the effect of induction, $\sigma_i\,\h u_i \times B_z\h e_z \approx 0$). Though we state the system of equations consisting of the induced magnetic field perturbation $\h b_i$ in equation \eqref{eq:gauss+amp_max_law}, it is of no importance in this study. Previous studies based on shallow water theories have largely ignored the contribution of the induced magnetic field on the perturbed Lorentz force \citep{Bojarevics1994, Davidson1998}. In a cylindrical system, it was shown that $\h J \times \h b_i$ does not contribute to the amplitude of the wave but rather to the frequency shift \citep{Herreman2019}. This is because the Lorentz force due to the horizontal component of the induced magnetic field $\h J \times \h b_i$ is out of phase with $\h J \times B_z\h e_z$ (the dominant contribution) and therefore, cannot inject power into the wave.

For the formulation of the boundary conditions at the tank walls, we account for the fact that the upper carbon anode in ARCs is a much better electrical conductor than the cryolite, and the bottom steel cathode often a much weaker conductor than the aluminium. For this idealised case, in which excess currents close entirely in the aluminium layer, we obtain the following linearised boundary conditions:
\begin{subequations}
	\begin{align}
		\bm{j}_1\cdot \h e_x\bigl|_{x=-L_x/2} &= \bm{j}_2\cdot \h e_x\bigl|_{x=L_x/2} = 0,  \label{eq:bc1}
		\\
		\h j_1\cdot \h e_y\bigl|_{y=-L_y/2} &= \h j_2\cdot \h e_y\bigl|_{y=L_y/2} = 0,  \label{eq:bc2}
		\\
		\varphi_1 &= 0 \,\bigl|_{z=h_1}, \label{eq:bc3}
		\\
		\h j_2\cdot \h e_z &= 0\,\bigl|_{z=-h_2} . \label{eq:bc4}
	\end{align} 
\end{subequations}
Equations \eqref{eq:bc1} and \eqref{eq:bc2} ensure that no current can escape through the insulating side walls. The perfectly conducting anode is modelled as an iso-potential surface \eqref{eq:bc3} and the weakly conducting cathode is treated as the perfect insulator \eqref{eq:bc4} for the excess current component $\h j$. In \S\,\ref{sec:Borisov}, we further consider the case where the cathode is taken as a perfect conductor as condition \eqref{eq:bc4} is not valid for all reduction cells and MPR model experiments. 
As kinematic boundary conditions we simply apply  impermeability at all walls
\begin{equation}
	\h u_i \cdot \h n_i|_{\Sigma_i} = 0,
	\label{eq:no-outflow_bc}
\end{equation}
for the inviscid formulation \S\,\ref{sec:Ansatz} to which we must add
\begin{equation}
	\h u_i \times \h n_i|_{\Sigma_i} = 0.
\end{equation}
to account for viscous effects we address in \S\,\ref{sec:visc_damp}. At the interface $\mathcal{S}$, we have to take into account the following linearised  hydrodynamic and electrical boundary conditions:
\begin{subequations}
	\begin{align}
		\partial_t \eta &= u_{1,z} \,\bigl|_{z = 0} = u_{2,z} \bigl|_{z = 0},  \label{eq:int_bc1}
		\\
		-\gamma_{\ii{int}} \nabla^2\eta + (\rho_2-\rho_1)g\eta &= p_2 \,\bigl|_{z=0} - p_1 \,\bigl|_{z=0},  \label{eq:int_bc2}
		\\
		\h	j_1 \cdot \h e_z &= \h j_2\cdot \h e_z \,\bigl|_{z = 0}, \label{eq:int_bc3}
		\\
		J_0(\sigma_1 - \sigma_2)\eta &= \varphi_2 \,\bigl|_{z=0} -\varphi_1 \,\bigl|_{z=0}. \label{eq:int_bc4}
	\end{align}
\end{subequations}
Equations \eqref{eq:int_bc1} and \eqref{eq:int_bc2} represent the classic kinematic and dynamic boundary conditions for gravity-capillary waves. In \S\,\ref{sec:visc_damp}, we state the corrections to \eqref{eq:int_bc2} incorporating the continuity of the tangential velocities and viscous stresses in order to solve the Stokes velocities (for viscous damping). The electrical boundary conditions are derived from the basic principle that both the normal current density $\h n \cdot (\h J+\h j_i)$ and the total electric potential $(\Phi_i+\varphi_i)$ must remain continuous. This results in the linearised interfacial conditions \eqref{eq:int_bc3} and \eqref{eq:int_bc4} for the perturbed quantities \citep{Munger2006}. This jump condition is essential  for the MPR instability as it describes how the current density redistributes in both layers as a response to arbitrary (small) interface elevations $\eta$. Because the horizontal component of the redistributed current is responsible for the Lorentz force that reinforces interface displacements $\eta$, the self-destabilisation is mainly anchored in the interfacial boundary condition. 

\subsection{Conditions to apply perturbation method}
In order to conduct linear stability analysis using perturbation methods, the Lorentz force and the viscous force terms can only be small perturbative quantities in the momentum balance equation. In terms of orders of magnitude (denoted using square brackets), this implies the limitations
\begin{equation}
	[\bm j_i \times B_z\bm e_z], [\rho_i\nu_i\nabla^2\bm u_i] \ll 
	[\rho_i\partial_t \bm u_i], [\bm\nabla p_i]
	\label{eq:inequal_cond}
\end{equation}
for the parameter space wherein we operate. The physical quantities can be assigned to the characteristic scales of our problem. The term $[\bm u_i] = U$ refers to the velocity scale, $[\bm j_i]=J$ to the characteristic current density and for the magnetic field we assign the critical vertical component $[B_z\bm e _z] = B$. Considering absolute values $U,J,B,\omega > 0$, we can write the momentum balance terms as
\begin{equation}
	[\rho_i\partial_t \bm u_i] \sim \rho_i \omega U, \qquad
	[p_i] \sim \rho_i\omega L U.
\end{equation}
The magnitude of the viscous term is of the order
\begin{equation}
	[\rho_i\nu_i\nabla^2\bm u_i] \sim \rho_i\nu_i U/L^2.
\end{equation}
The characteristic length is defined as $L=\sqrt{L_xL_y}$. From the quasi-static approximation of Ohm's law \eqref{eq:gen_ohms_law}, the current excess is a function of the perturbed electric potential and the induced current and depends on the deformations of the interface, estimated by the jump condition \eqref{eq:int_bc4}. The orders of magnitude of the current excess is the  given by,
\begin{equation}
	[\bm j_i] = [-\sigma_i\bm \nabla\varphi_i] + [\sigma_i\bm u_i \times B_z\bm e_z].
\end{equation}
The magnitude of the first term is
\begin{equation}
	[-\sigma_i\bm \nabla\varphi_i] \sim J U/{(\omega L)}
\end{equation}
and the magnitude of the induction term scales as
\begin{equation}
	[\sigma_i\bm u_i \times B_z\bm e_z] \sim \sigma_iUB.
\end{equation}
Re-exploring the initial constraint \eqref{eq:inequal_cond} we had put in place finally yields 
\begin{equation}
	[\rho_i\nu_i\nabla^2\bm u_i] \ll [\rho_i\partial_t \bm u_i]
	\quad \Leftrightarrow \quad
	Re_i = \frac{\omega L^2}{\nu_i} \gg 1,
\end{equation}
and
\begin{equation}
	[\bm j_i \times B_z\bm e_z] \ll [\rho_i\partial_t \bm u_i]
	\quad \Leftrightarrow \quad
	\left\{
	\begin{array}{l}
		\dfrac{JB}{\rho_i\omega^2 R} \ll 1
		\\[10pt]
		\dfrac{\sigma_i B^2}{\rho_i\omega} \ll 1
	\end{array}
	\right. .
\end{equation}
These inequalities allow us to safely apply our perturbation techniques described below. However, this does not necessarily entail that our model will fail if these conditions are (slightly) violated. At this point, it cannot be verified yet how necessary these limitations will be. 

\subsection{Two-wave perturbation approach}
\label{sec:Ansatz}
Our perturbative approach is based on the (inviscid) potential flow solutions of free gravity-capillary interfacial waves that form a set of orthogonal functions in the absence MHD effects ($\bm{B},\bm{J}=0$) and without viscosity ($\nu = 0$). Viscosity only affects the flow in the close vicinity of the Stokes boundary layers forming at the container walls, and below and above the interface. In \S\,\ref{sec:visc_damp}, we calculate Stokes boundary layers corrections to the potential flow solutions to predict viscous damping rates associated with interfacial wave motions. In our formulation, however, only perturbations of the inviscid bulk flow are considered and all dissipative effects are integrated into a global damping factor $\lambda$. A second key assumption of our approach is that we consider the superposition of two independent wave modes $\eta + \eta'$ as the base state, which are given as follows:
\begin{align}
	&\eta(x,y,t) = A_{mn} \; \cos\left[\dfrac{m\pi}{L_x}\left( x + \dfrac{L_x}{2} \right)\right]
	\cos\left[\dfrac{n\pi}{L_y}\left( y + \dfrac{L_y}{2} \right)\right]e^{(i\omega - \lambda)t}
	\nonumber \\
	& \textcolor{white}{AAAAAAAAAAAAAAAAAA}  + \nonumber \\
	&\eta'(x,y,t) = A_{m'n'} \; \cos\left[\dfrac{m'\pi}{L_x}\left( x + \dfrac{L_x}{2} \right)\right]
	\cos\left[\dfrac{n'\pi}{L_y}\left( y + \dfrac{L_y}{2} \right)\right]e^{(i\omega' - \lambda')t} 
	\label{eq:wave ansatz}.
\end{align}
Here, $A_{mn}$ are mode-dependent wave amplitudes, $(m,n) \in \mathbb{N}^{0}\times\mathbb{N}^{0}$ specify the lateral wave numbers defining the number of crests and troughs in the $x$- and $y$-direction, $\omega \in \mathbb{R}$ is the natural frequency of the wave and $\lambda \in \mathbb{R}^+$ denotes the exponential decay rate. In contrast to most previous studies, which examine the interaction of infinitely many wave modes and truncate the system \textit{a posteriori}, we calculate the two-wave interaction directly allowing us to derive exact solutions for the growth rates and stability onsets. The disadvantage of this method is that the MPR instability can sometimes arise from three (or more)-wave interactions. Accordingly, these cases are not represented in our model. As a final presumption, we suppose that both waves are magnetohydrodynamically coupled within their base state and therefore both oscillate with the same mean frequency $\bar{\omega} = (\omega + \omega')/2$ and also attenuate simultaneously with the mean damping rate $\bar{\lambda} = (\lambda + \lambda')/2$. In light of these assumptions, we propose the following two-wave ansatz:
\begin{equation}
	\fontsize{10.5pt}{10.5pt}
	[\textbf{\textit{u}}_i, p_i, \eta, \textbf{\textit{j}}_i ]= 
	\biggl( 
	C 
	\left[ 
	\widehat{\h{u}}_{i}, \widehat{p}_{i}, \widehat{\eta}, \widehat{\h{j}}_{i} \right] 
	+ 
	C' 
	\left[ 
	\widehat{\h{u}}'_{i}, \widehat{p}'_{i}, \widehat{\eta}', \widehat{\h{j}}'_{i} \right] 
	+\left[
	\widetilde{\h{u}}_i, \widetilde{p}_{i}, \widetilde{\eta}, \widetilde{\h{j}}_{i} \right] 
	\biggr) 
	\;
	\ii e^{ \bigl(\alpha -\overbar\lambda 
		\,+\, \ii{i}\overbar{\omega}
		\bigr)t} .
	\label{eq:perturb_method_ansatz}
\end{equation}
\renewcommand{\subsectionautorefname}{Appendix}
In this notation, the tilded variables are small perturbations with respect to the hatted variables representing the basic two-wave state. The eigenvalues $C$ and $C'$ are to be determined and $\alpha$ is the potentially destabilising contribution to the growth rate that we seek to derive. The wave pair is MPR unstable if and only if $\alpha$ is both a real number and exceeding the mean damping rate $\alpha > \bar{\lambda}$. The total growth rate $\alpha_{\rm MPR}$ is then given by the difference $\alpha_{\rm MPR} = \alpha - \bar{\lambda}$ and stability onsets are defined by the condition $\alpha = \bar{\lambda}$. Considering the wave pair to be in a coupled motion $\omega = \omega' = \bar{\omega}$ is justified by the fact that the wave frequencies are shifted by Lorentz forces ($\bar{\omega} = \omega - \delta \omega$ and $\bar{\omega} = \omega' + \delta \omega$ with $\delta\omega = (\omega - \omega')/2$) in such a way that they coincide exactly at the point of marginal stability \citep{Davidson1998}. This point defines the base state, but it does not mean that our approach does not also allow both waves to oscillate freely and independently. In the absence of any Lorentz forces, precisely the classic gravity-capillary wave modes given in equation (\ref{eq:wave ansatz}) result as two independent solutions of our system, as we prove it in \autoref{appA_1}. This rigorously justifies our approach, and our theory is capable of describing the complete transition from purely hydrodynamic waves passing through magnetohydrodynamically modified waves regimes up to the (linear) self-amplifying MPR wave.

To formulate the base state, we assign the hatted quantities with the classical gravity-capillary wave solutions that can be obtained by expressing the bulk velocities as gradients of scalar potentials $\widehat{\h u}_i = \bm\nabla \widehat\phi_i$ and solving the laplace equation $\bm\nabla^2\widehat\phi_i = 0$ together with the no outflow boundary condition \eqref{eq:no-outflow_bc} and the interface conditions \eqref{eq:int_bc1} and \eqref{eq:int_bc2}. The linear solutions can be represented as follows:
\begin{equation}
	\begin{bmatrix}
		\widehat{\h u}_i \\[5pt]
		\widehat{p}_i \\[5pt]
		\widehat{\eta}
	\end{bmatrix}
	=
	\begin{bmatrix}
		\bm{\nabla}\widehat{\phi}_i \\[5pt]
		-\rho_i(\ii i\omega - \lambda)\widehat{\phi}_i\\[5pt]
		\partial_z \widehat{\phi}_i/(\ii i\omega - \lambda)
	\end{bmatrix},
	\label{eq:pot_wave_soln}
\end{equation}
with the leading-order flow potentials given as
\begin{equation}
	\begin{bmatrix}
		\widehat{\phi}_{1} \\[12pt]
		\widehat{\phi}_{2} 
	\end{bmatrix} = 
	\left[
	\begin{array}{cc}
		\dfrac{\omega}{k} \dfrac{\cosh[k(z-h_1)]}{\sinh[k h_1]} & 
		\multirow{3}{*}{$
			A_{mn} \; \cos\left[\dfrac{m\pi}{L_x}\left( x + \dfrac{L_x}{2} \right)\right]
			\cos\left[\dfrac{n\pi}{L_y}\left( y + \dfrac{L_y}{2} \right)\right]
			$} \\[10pt]
		-\dfrac{\omega}{k} \dfrac{\cosh[k(z+h_2)]}{\sinh[k h_2]} &
	\end{array} 
	\right] 
	\label{eq:vel_potential}
\end{equation}
along with the wave function reading
\begin{equation}
	\widehat\eta = A_{mn} \; \cos\left[\dfrac{m\pi}{L_x}\left( x + \dfrac{L_x}{2} \right)\right]
	\cos\left[\dfrac{n\pi}{L_y}\left( y + \dfrac{L_y}{2} \right)\right] .
	\label{eq:eta_eq}
\end{equation} 
The vertical wave number $k$ is a function of the lateral wave numbers $(m,n)$
\begin{equation}
	k =  \pi\sqrt{\frac{m^2}{L_x^2} + \frac{n^2}{L_y^2} }\,
\end{equation}
and the natural frequencies of the gravity-capillary waves are obtained as
\begin{equation}
	\omega = \pm \sqrt{
		\dfrac{(\rho_2-\rho_1)gk + \gamma_{\scaleto{\ii{int}}{5pt}}{k}^3}
		{\rho_1 \coth[k\: h_1]
			+ \rho_2 \coth[k\: h_2]}
	} \,.
\end{equation}
The sign associated with $\omega$ plays an important role in determining the nature of instability. Wave motions in alignment with the direction of the Lorentz forces tend to amplify, whereas those in the opposite direction tend to dampen. Waves with contrary natural frequencies form standing waves. 

It remains to calculate the leading-order current perturbations through Ohm's law $\widehat{\h j}_i = \sigma_i\bm\nabla \widehat\varphi_i$. The electric potentials must also fulfill the Laplace equation $\bm{\nabla}^2\widehat{\varphi} = 0$ with boundary conditions given in equations (\ref{eq:bc1} -- \ref{eq:bc4}), \eqref{eq:int_bc3} and \eqref{eq:int_bc4}. As the electrical boundary conditions at the insulating side walls are identical to the hydrodynamic outflow condition \eqref{eq:no-outflow_bc}, the solutions of $\widehat\varphi_i$ result in a very similar form:
\begin{equation}
	\begin{bmatrix}
		\widehat{\varphi}_1 \\[12pt]
		\widehat{\varphi}_2 
	\end{bmatrix} = 
	\left[
	\begin{array}{cc}
		-\ii i \dfrac{\sinh[k (z-h_1)]}{\sigma_1 \cosh[kh_1]} &
		\multirow{3}{*}{$
			J_0\:A_{mn} \Lambda \left(
			\cos\left[\dfrac{m\pi}{L_x}\left( x + \dfrac{L_x}{2} \right)\right]
			\cos\left[\dfrac{n\pi}{L_y}\left( y + \dfrac{L_y}{2} \right)\right]
			\right)
			$} \\[10pt]
		-\ii i \dfrac{\cosh[k (z+h_2)]}{\sigma_2 \sinh[kh_2]} & 
	\end{array} 
	\right] ,
	\label{eq:elec_potential}
\end{equation}
where $\Lambda$ can be referred to as the conductivity-jump parameter reading
\begin{equation}
	\Lambda = \frac{(\sigma_1^{-1} - \sigma_2^{-1})}{\sigma_1^{-1} \:\tanh[kh_1] + \sigma_2^{-1} \:\coth[kh_2]} .
\end{equation}

\subsection{Calculation of the growth rate}
\label{sec:growth_rate_gen}
All the preparations have now been made to calculate the growth rate $\alpha_{\ii {MPR}}$. By inserting the ansatz \eqref{eq:perturb_method_ansatz} in the governing equations \eqref{eq:gov_momentum} and \eqref{eq:gov_cont}, and considering only the leading order terms, we can obtain the next order in the perturbation problem through the equations:
\begin{subequations}
	\begin{equation}
		\begin{bmatrix}
			C([\ii{i}\overbar\omega - \overbar\lambda] + \alpha)\rho \widehat{\h{u}} + C\bm{\nabla}\widehat{p} \\
			+ \\
			C'([\ii{i}\overbar\omega - \overbar\lambda] + \alpha)\rho \widehat{\h{u}}' + C'\bm{\nabla}\widehat{p}'
		\end{bmatrix}
		+
		\begin{bmatrix}
			(\ii{i}\overbar\omega - \overbar\lambda) \rho\widetilde{\h{u}} 
			+ \bm{\nabla}\widetilde{p} \\
			+ \\
			(\ii{i}\overbar\omega - \overbar\lambda) \rho\widetilde{\h{u}}' 
			+ \bm{\nabla}\widetilde{p}'
		\end{bmatrix}
		=
		\begin{bmatrix}
			C \left( \; \widehat{\h{j}} \times B_z\bm{e}_z\right) \\
			+ \\
			C' \left( \; \widehat{\h{j}}' \times B_z\bm{e}_z\right)
		\end{bmatrix} ,
		\label{eq:momentum_fredholm}
	\end{equation}
	\begin{equation}
		\bm\nabla \cdot \widetilde{\h u}_i = 0 .
		\label{eq:cont_fredholm}
	\end{equation}
\end{subequations}
\renewcommand{\sectionautorefname}{Appendix}
To find an equation for $\alpha$, we have to state a suitable solvability condition. For that we apply the Fredholm alternative in an analogue manner to \citet{Herreman2019}. All technical calculations are provided in \autoref{appA} in detail. The solvability condition results in an eigenvalue problem in form of a $2\times 2$ interaction matrix that can be expressed in the following form:
\begin{equation}
	\begin{bmatrix}
		\dfrac{I_{\eta\eta}}{2K_{\eta\eta}} + \ii{i}\delta\omega - \delta\lambda -\alpha & \dfrac{I_{\eta\eta'}}{2K_{\eta\eta}} \\
		\\
		\dfrac{I_{\eta'\!\eta}}{2K_{\eta'\!\eta'}} & \dfrac{I_{\eta'\!\eta'}}{2K_{\eta'\!\eta'}} - \ii{i}\delta\omega + \delta\lambda -\alpha
	\end{bmatrix}
	\begin{bmatrix}
		C \\
		\\
		C'
	\end{bmatrix}
	= 0 ,
	\label{eq:alpha_matrix_initial}
\end{equation}
with,
\begin{subequations}
	\begin{align}
		I_{\eta\eta} &= \sum_{i=1,2}\int_{V_i} \widehat{\h u}_i^{*} \cdot 
		\Bigl( \; \widehat{\h{j}}_i \times B_z\bm{e}_z \Bigr) 
		= \sum_{i=1,2}\int_{V_i} \bm\nabla\widehat{\phi}_i^{*} \cdot 
		\Bigl( \; \bm\nabla\widehat{\varphi}_i \times B_z\bm{e}_z \Bigr) 
		= 0, \label{eq:A_11} \\
		I_{\eta\eta'} &= \sum_{i=1,2}\int_{V_i} \widehat{\h u}_i^{*} \cdot 
		\Bigl( \; \widehat{\h{j}}'_i \times B_z\bm{e}_z \Bigr) 
		= \sum_{i=1,2}\int_{V_i} \bm\nabla\widehat{\phi}_i^{*} \cdot 
		\Bigl( \; \bm\nabla\widehat{\varphi}'_i \times B_z\bm{e}_z \Bigr) ,
		\label{eq:A_12}\\
		I_{\eta'\!\eta} &= \sum_{i=1,2}\int_{V_i} \widehat{\h u}'^{*}_i \cdot 
		\Bigl( \; \widehat{\h{j}}_i \times B_z\bm{e}_z \Bigr) 
		= \sum_{i=1,2}\int_{V_i} \bm\nabla\widehat{\phi}'^{*}_i \cdot 
		\Bigl( \; \bm\nabla\widehat{\varphi_i} \times B_z\bm{e}_z \Bigr) ,
		\label{eq:A_21}\\
		I_{\eta'\!\eta'} &= \sum_{i=1,2}\int_{V_i} \widehat{\h u}'^{*}_i \cdot 
		\Bigl( \; \widehat{\h{j}}'_i \times B_z\bm{e}_z \Bigr)
		= \sum_{i=1,2}\int_{V_i} \bm\nabla\widehat{\phi}'^{*}_i \cdot 
		\Bigl( \; \widehat{\h{j}}' \times B_z\bm{e}_z \Bigr)
		= 0 . \label{eq:A_22}
	\end{align}
\end{subequations}
The terms given above are associated with the power that can be injected by Lorentz forces. Equations \eqref{eq:A_11} and \eqref{eq:A_22} describe the interaction of the wave modes $\eta$ and $\eta'$ with their self-induced Lorentz forces. However, these integrals disappear, which proves that standing waves cannot be destabilised. The terms $K_{\eta\eta}$ and $K_{\eta'\!\eta'}$ are associated with the potential (or equivalently kinetic) energy of the respective wave modes and are given by
\begin{subequations}
	\begin{align}
		\nn
		K_{\eta\eta} &= \int_{\mathcal{S}} \Bigl( -\gamma_{\ii{int}} |\bm{\nabla}\widehat{\eta}|^2 
		+ (\rho_2 - \rho_1)g |\widehat{\eta}|^2 \Bigr)\Bigl|_{z=0} \ii{d}S  ,
		\\
		& = \xi_{m\times n} \: L_xL_y \: \Bigl[\gamma_{\ii{int}}k^2 + (\rho_2 - \rho_1)g \Bigr] ,
		\label{eq:K_11}
		\\
		K_{\eta'\!\eta'} &=  \xi_{m'\times n'} \: L_xL_y \: \Bigl[\gamma_{\ii{int}}{k'}^2 + (\rho_2 - \rho_1)g \Bigr] .
		\label{eq:K_22}
	\end{align}
\end{subequations}
For the sake of simplicity, we introduced an auxiliary function $\xi_r$ that is defined as follows
\[
\xi_r = \biggl\{
\begin{array}{c}
	1/2 , \quad r = 0  \\
	1/4 , \quad r \neq 0
\end{array} .
\]
Finally, the secular equation of the matrix \eqref{eq:alpha_matrix_initial} yields the solutions for $\alpha$
\begin{equation}
	\alpha = \pm \; \sqrt{\dfrac{I_{\eta\eta'}I_{\eta'\!\eta}}{4K_{\eta\eta}K_{\eta'\!\eta'}} + \bigl[ \ii i{\delta \omega} -\delta\lambda \bigr]^2 }.
\end{equation}
By subtracting the damping rate $\overbar\lambda$ and solving the integrals \eqref{eq:A_12} and \eqref{eq:A_21}, we obtain the MPR growth rate, the central result of our study, that can be represented as follows
\begin{equation}
	\alpha_{\ii {MPR}} = \alpha - \overbar\lambda
	= \pm \; 
	\sqrt{ J^2{B_z}^2{\Theta}^2
		\; (\overbar\omega^2 - \delta\omega^2)
		\;\frac{\mathcal{T}\mathcal{T}'}{\mathcal{U}\mathcal{U}'}
		+ \bigl[ \ii i{\delta \omega} -\delta\lambda \bigr]^2
	} \;-\; \overbar\lambda ,
	\label{eq:main_gr}
\end{equation}
where 
\begin{subequations}
	\begin{gather}
		\Theta = \pm \; \sqrt{\delta_{m \times m'}\;\delta_{n \times n'}} \; 
		\frac{((-1)^{m+m'}-1)((-1)^{n+n'}-1)(m^2{n'}^2 - {m'}^2n^2)}
		{(m^2 - {m'}^2)(n^2 - {n'}^2) },
		\label{eq:form_function}
		\\
		\delta_r = \left\{ \begin{array}{cc}
			1, & r=0\\
			2, & r>0
		\end{array} \right. \!\!.
		\label{eq:delta_func}
		\intertext{$\Theta$ here defines the wave mode selection function, which is nonzero only if $m+m'$ and $n+n'$ are both odd, and}
		\mathcal{U} = L_xL_y\Bigl(\gamma_{\ii{int}} \: k^2 + (\rho_2 - \rho_1)g \Bigr), \qquad
		\mathcal{U}' = L_xL_y\Bigl(\gamma_{\ii{int}}\: {k'}^2 + (\rho_2 - \rho_1)g \Bigr) ,
		\intertext{are representations of the potential energy of the standing waves and}
		\mathcal{T} = \frac{\Lambda}{k(k^2 - k'^2)}
		\Biggl[ 
		\frac{k}{\tanh[k'h_2]} - \frac{k'}{\tanh[kh_2]}
		+ \frac{k'\:\sech[k'h_1]}{\sinh[kh_1]}
		-\frac{k'}{\tanh[kh_1]}
		+ k\tanh[k'h_1] \Biggr] ,
		\label{eq:T_11}
		\\
		\mathcal{T}' = \frac{\Lambda'}{k'({k'}^2-k^2)}
		\Biggl[ 
		\frac{k'}{\tanh[kh_2]} - \frac{k}{\tanh[k'h_2]}
		+ \frac{k\: \sech[kh_1]}{\sinh[k'h_1]}
		-\frac{k}{\tanh[k'h_1]}
		+ k'\tanh[kh_1] \Biggr] , 
		\label{eq:T_22}
		\intertext{root from the power injected by the Lorentz force and represent the non-shallow part of the formula. At the limit $k' \rightarrow k$ these terms simplify to}
		\mathcal{T} = \mathcal{T}' = \frac{\Lambda}{4k^2\:\sinh^2[kh_2]}
		\biggl[ \frac{\sinh[k(h_1+2h_2)]}{\cosh[kh_1]} +2kh_2 - \tanh[kh_1] \biggr] .
	\end{gather}
\end{subequations}
The perturbation analysis results in two distinct solutions for the growth rate $\alpha_{\rm MPR}$. The two solutions correspond to MPR waves that rotate either clockwise or anti-clockwise (when looking at the cell from above), or equivalently, to standing wave pairs ($\eta + \eta'$) with a positive or negative phase shift. One of the two wave solutions is always suppressed by the Lorentz force (when the induced Lorentz force $\h j\times B_z\h e_z$ anti-aligns with the rotating wave motion $\h u$) and the other one is amplified in the unstable situations (when $\h j\times B_z\h e_z$ aligns with $\h u$). In our configuration of \autoref{fig:cell_schematic} anti-clockwise rotating waves are destabilised. 

From the growth rate \eqref{eq:main_gr} one can readily derive the stability onset that is defined by the equivalence of the real part of the square root in \eqref{eq:main_gr} with $\overbar\lambda$. In order to find a preferably simple expression, we introduce a modified Sele parameter $\beta_{\gamma_{\ii{int}}}$, similar to \citet{Horstmann2018}, which takes into account interfacial tension in addition to gravity. In this way, the stability criterion can be represented compactly in a dimensionless form as follows
\begin{equation}
	\beta_{\gamma_{\ii{int}}} > \beta_{\ii{crit}} =  \frac{L_x^2 L_y^2}{h_1h_2 \Theta}
	\sqrt{ \frac{\overbar\lambda^2 - \bigl[ \ii i{\delta \omega} -\delta\lambda \bigr]^2} {(\overbar\omega^2 - {\delta\omega}^2) \mathcal{T}\mathcal{T}'} },
	\quad \left\{
	\beta_{\gamma_{\ii{int}}} = \frac{I_0B_z}{\left[ (\rho_2-\rho_1)g + \gamma_{\ii{int}}\bar{k}^2 \right] h_1h_2} \right. .
	\label{eq:beta_crit-onset}
\end{equation}
In order to analyse the stability of large-scale aluminium reduction cells, the criterion can be further simplified. We reformulate \eqref{eq:beta_crit-onset} in the shallow-water limit and neglect interfacial tension. Then the interfacial waves become non-dispersive ($\omega = ck$, $\omega' = ck'$), with
\begin{equation}
	c = \sqrt{ \frac{(\rho_2 - \rho_1)g}{\rho_1/h_1 + \rho_2/h_2 } } .
	\label{eq:wave-speed}
\end{equation}
The product $\mathcal{T}\mathcal{T'}$ simplifies as
\begin{equation}
	\mathcal{T}\mathcal{T'} = \frac{1}{k^4k'^4h_1^2h_2^2},
\end{equation}
and therewith equation \eqref{eq:beta_crit-onset} reduces to
\begin{equation}
	\beta_{\ii{crit}} = \frac{L_x^2 L_y^2}{c \,\Theta}
	\sqrt{k^3k'^3 \, \left( \overbar\lambda^2 - \bigl[ \ii i{\delta \omega} -\delta\lambda \bigr]^2 \right) } .
	\label{eq:beta_crit_shallow}
\end{equation}
This stability threshold criterion is equivalent to the numerical state-of-the-art solution of linear MPR instability presented in \citet{Politis2021}, see \S\,\ref{sec:fractal}. As a final simplification, we shall take a look at the artificial case of square cells, which is physically relevant for some MPR model experiments. For a square cell we have $L_x=L_y$ and $\lambda=\lambda'=\overbar\lambda$ and the parameters further reduce to $\Theta = 4$, $k=\pi/L$, $\delta\omega=0$, $\delta\lambda=0$, yielding
\begin{equation}
	\beta_{\ii{crit}} = \frac{\pi^3L}{4\, c}\,\lambda .
	\label{eq:beta_crit_square-shallow}
\end{equation} 
As expected, the critical Sele parameter in this case is not mainly dictated by the difference in natural frequencies anymore but solely by the damping parameter $\lambda$, as is true for all infinitely many inherently unstable aspect ratios, refer to \S\,\ref{sec:fractal}, equation (\ref{eq:fractal}).

\subsection{The viscous damping rate}
\label{sec:visc_damp}

Up to this point, we have been treating $\lambda$ as a purely phenomenological damping parameter. To evaluate the stability of industrial ARCs, in which turbulent flows usually prevail, $\lambda$ must be estimated or determined experimentally. In small-scale models and liquid metal batteries, where flows remain largely laminar (at least in the boundaries or around the point of marginal instability), viscous and magnetic damping rates can be calculated analytically. This is a key ingredient for our model to serve as a benchmark. We start with the viscous damping rate. In the leading order, magnetohydrodynamics effects can be neglected (no consideration of Hartmann layers) and it remains to calculate the total dissipation produced in the Stokes boundary layers, which are created solely by wave motions. We follow the approach by \citet{Keulegan1959}, who has calculated damping rates of free-surface gravity waves in rectangular containers. We extend his formulation to two liquid layers, which is not straightforward as viscous boundary layers are formed above and below the interface in two-liquid systems that do not develop on free surfaces and must be calculated separately. In absence of MHD effects, the linear governing equations \eqref{eq:gov_momentum} and \eqref{eq:gov_cont} reduce to
\begin{equation}
	\left.
	\begin{array}{c}
		\rho_i \partial_t \h{u}_i + \bm{\nabla}p_i = \rho_i\nu_i\nabla^2\h u_i 
		\\[5pt]
		\bm{\nabla} \cdot \h{u}_i = 0 
	\end{array}
	\right\} .
	\label{eq:visc_gov1}
\end{equation}
For the upper and lower layer ($i=1,2$), the no slip condition at the side wall (represented as rigid boundary $\Sigma_i$) is  
\begin{subequations}
	\begin{equation}
		\h u_i \big|_{\Sigma_i} = 0,
	\end{equation}
	and the tangential velocities (denoted by $\perp$) must be equal at the interface:
	\begin{equation}
		\h u_{1, \perp}\big|_\mathcal{S} = \h u_{2, \perp}\big|_\mathcal{S}.
		\label{eq:visc_bc1}
	\end{equation}
	The dynamic boundary condition \eqref{eq:int_bc2} under the consideration of viscous effects now changes to
	\begin{equation}
		p_2|_\mathcal{S} - p_1|_\mathcal{S} = 
		(\rho_2 -\rho_1)g\eta -\gamma_{\ii{int}}\nabla^2\eta 
		+ 2( \rho_2\nu_2\partial_z u_{2,z}  - \rho_1\nu_1\partial_z u_{2,z}), 
		\label{eq:visc_bc2}
	\end{equation}
	while the continuity of the tangential stresses are
	\begin{equation}
		\rho_1\nu_1 \Bigl( \partial_z u_{1,x} + \partial_x u_{1,z} \Bigr)\Bigl|_\mathcal{S}
		=
		\rho_2\nu_2 \Bigl( \partial_z u_{2,x} + \partial_x u_{2,z} \Bigr)\Bigl|_\mathcal{S},
		\label{eq:visc_bc3}
	\end{equation}
	\begin{equation}
		\rho_1\nu_1 \Bigl( \partial_z u_{1,y} + \partial_y u_{1,z} \Bigr)\Bigl|_\mathcal{S}
		=
		\rho_2\nu_2 \Bigl( \partial_z u_{2,y} + \partial_y u_{2,z} \Bigr)\Bigl|_\mathcal{S} .
		\label{eq:visc_bc4}
	\end{equation}
\end{subequations}
In a stratified two-layer system, say an interfacial gravity wave mode of $(m,n)$ of amplitude $A_{mn}$ oscillating at angular frequency $\omega$ undergoes viscous decay over time at a rate of $\lambda_{\ii v}$. The fluid velocity taking into account the Stokes boundary layer corrections can be formulated as
\begin{equation}
	\h u_i = C \Bigl[  \widehat{\h u}_i + \widehat{\bar{\h u}}_i
	\Bigr] e^{(\ii{i}\omega - \lambda_{\ii v}) t} .
	\label{eq:visc_ansatz}
\end{equation}
where the $\widehat{\h u}_i$ are the potential flow velocities and the barred term $\widehat{\bar{\h u}}_i$ are the viscous corrections to be calculated. We introduce wall-normal coordinates $\zeta$ that tend to decay the viscous boundary layer corrections to zero as they traverse inwards in the bulk. We define $\zeta$ as follows:
\begin{equation}
	\zeta = \begin{cases}
		z - h_1, \qquad &\text{near } z=h_1 \\
		\dfrac{L_x}{2} - x, \qquad &\text{near } x = L_x/2 \\[7pt]
		\dfrac{L_y}{2} - y, \qquad &\text{near } y = L_y/2 \\
		z, \qquad &\text{near } z=0
	\end{cases}, 
	\qquad
	\zeta = \begin{cases}
		z + h_2, \qquad &\text{near } z=-h_2 \\
		x + \dfrac{L_x}{2}, \qquad &\text{near } x = -L_x/2 \\[7pt]
		y + \dfrac{L_y}{2}, \qquad &\text{near } y = -L_y/2 \\
		-z, \qquad &\text{near } z=0
	\end{cases} .
\end{equation}
The effect of the diffusive boundary layer tends to diminish in the bulk region. Inserting the ansatz \eqref{eq:visc_ansatz} into equation \eqref{eq:visc_gov1}, we obtain $\widehat{\bar{\h u}}_i  \cdot \h n_i = 0$ and $\widehat{\overbar{p}}_i= 0$; the wall-normal boundary layer correction for velocity and leading-order boundary layer correction for pressure, both equal to zero. We compute the Stokes boundary layer at the domain walls (with no-slip condition) as
\begin{equation}
	\ii{i}\omega 
	\widehat{\overbar{\h u}}_{i,\perp}
	\approx 
	\nu_i \partial^2_{\zeta\zeta}
	\widehat{\overbar{\h u}}_{i,\perp} .
\end{equation}
The mode-dependent velocity at the solid walls are
\begin{equation}
	\widehat{\bar{\h u}}_{i,\perp}
	=
	- \widehat{\h u}_{i,\perp}|_{\Sigma_i}
	\ii{e}^{-\Gamma \sqrt{|\omega|/\nu_i}\zeta -\lambda_{\ii v} t},
	\qquad
	\Bigl\{ \Gamma = \bigl(1 + \ii{i}\omega\bigr)/\sqrt{2} .
	\label{eq:u_wall}
\end{equation}
From equation \eqref{eq:u_wall} we can obtain the characteristic length of the viscous boundary layer of each medium as $\sqrt{\nu_i/|\omega|}$. At the interface, from (\ref{eq:visc_bc3}) and (\ref{eq:visc_bc4}) we deduce
\begin{equation}
	\widehat{\h u}_{1,\perp} +  \widehat{\bar{\h u}}_{1,\perp}|_\mathcal{S} 
	\approx
	\widehat{\h u}_{2,\perp} +  \widehat{\bar{\h u}}_{2,\perp}|_\mathcal{S} 
	\qquad
	\begin{cases}
		\rho_1\nu_1\partial_z\widehat{\bar{u}}_{1,x}|_\mathcal{S} 
		\approx
		\rho_2\nu_2\partial_z\widehat{\bar{u}}_{2,x}|_\mathcal{S} \\[5pt]
		\rho_1\nu_1\partial_z\widehat{\bar{u}}_{1,y}|_\mathcal{S} 
		\approx
		\rho_2\nu_2\partial_z\widehat{\bar{u}}_{2,y}|_\mathcal{S}
	\end{cases}\hspace{-3mm},
\end{equation}
\begin{align}
	\left. \widehat{\bar{\h u}}_{1,\perp}\right\vert_{m,n}
	&=
	\frac{\Lambda_{\ii v}}{\rho_1\sqrt{\nu_1}}
	\: \frac{\omega}{k}
	\begin{bmatrix}
		\frac{m\pi}{L_x}
		\;\sin\left[\frac{m\pi}{L_x}\left( x + \frac{L_x}{2} \right)\right]
		\cos\left[\frac{n\pi}{L_y}\left( y + \frac{L_y}{2} \right)\right]
		\h e_x 
		\\+\\
		\frac{n\pi}{L_y}
		\;\cos\left[\frac{m\pi}{L_x}\left( x + \frac{L_x}{2} \right)\right]
		\sin\left[\frac{n\pi}{L_y}\left( y + \frac{L_y}{2} \right)\right]
		\h e_y
	\end{bmatrix} \ii{e}^{-\Gamma \sqrt{\frac{|\omega|}{\nu_1}z} -\lambda_{\ii v} t},
	\label{eq:u_int1}
	\\[5pt]
	\left. \widehat{\bar{\h u}}_{2,\perp}\right\vert_{m,n}
	&=
	\frac{-\Lambda_{\ii v}}{\rho_2\sqrt{\nu_2}}
	\: \frac{\omega}{k}
	\begin{bmatrix}
		\frac{m\pi}{L_x}
		\;\sin\left[\frac{m\pi}{L_x}\left( x + \frac{L_x}{2} \right)\right]
		\cos\left[\frac{n\pi}{L_y}\left( y + \frac{L_y}{2} \right)\right]
		\h e_x 
		\\+\\
		\frac{n\pi}{L_y}
		\;\cos\left[\frac{m\pi}{L_x}\left( x + \frac{L_x}{2} \right)\right]
		\sin\left[\frac{n\pi}{L_y}\left( y + \frac{L_y}{2} \right)\right]
		\h e_y
	\end{bmatrix} \ii{e}^{\Gamma \sqrt{\frac{|\omega|}{\nu_2}z} -\lambda_{\ii v} t},
	\label{eq:u_int2}
\end{align}
\begin{flalign*}
	\ii{with,} \qquad
	\Lambda_{\ii v} = \frac{(\coth[kh_1] + \coth[kh_2]) }
	{1/\bigl(\rho_1\sqrt{\nu_1}\bigr) + 1/\bigl(\rho_2\sqrt{\nu_2}\bigr)} . &&
\end{flalign*}
The energy balance \eqref{eq:visc_energy_bal} for viscous dissipation can be equated to the rate of change of kinetic and potential energy. It is given as
\begin{align}
	\nonumber
	\dv{}{t}
	\underbrace{	
		\Biggl( \sum_{i=1,2} \frac{1}{2} \int_{V_i} \rho_i |\h u_i|^2 \ii{d}V ]\Biggr)}_{E_{\ii{kin}}}
	+
	\dv{}{t}
	\underbrace{
		\Biggl( \int_S \biggl[ \frac{1}{2} (\rho_2 -\rho_1)g\eta^2 + \gamma_{\ii{int}}\sqrt{1+|\nabla\eta|^2}
		\biggr]
		\ii{d}S}_{E_{\ii{pot}}}
	\\
	= -\underbrace{
		\sum_{i=1,2} 2\rho_i\nu_i \int_{V_i} \bm{\varepsilon}_i \bm{:} \bm{\varepsilon}_i \; \ii{d}V
	}_{\mathcal{D}} .
	\label{eq:visc_energy_bal}
\end{align}
We adapt the dissipation function $\mathcal{D}$ from \citet{Lamb1945}, in which the strain tensor is denoted by $\bm{\varepsilon}_i$. The solution to the balance equation \eqref{eq:visc_energy_bal} provides the damping rates due to the side wall ($\lambda_{\ii{wall}}$), interface ($\lambda_{\ii{int}}$), and irrotational viscous stresses ($\lambda_{\ii{irr}}$). The total damping rate for a wave of mode $(m,n)$ is composed as
\begin{equation}
	\lambda_{\ii{v}} = \lambda_{\ii{wall}}+\lambda_{\ii{int}}+\lambda_{\ii{irr}},
	\label{eq:visc_damp_sum}
\end{equation}
and is assumed to have a much larger time scale than the natural frequency of the rotating wave. All calculations are given in detail in \autoref{appB}. The solutions of $\lambda_{\ii{wall}}$, $\lambda_{\ii{int}}$ and $\lambda_{\ii{irr}}$ are
\begin{align}
	\nn 
	\lambda_{\ii{wall}} (m,n)  
	&=\sum_{i=1,2}
	\frac{ \dfrac{1}{\sqrt{2}}
		\dfrac{\rho_i \sqrt{\omega\nu_i}}{ kL_xL_y }
	}
	{ 
		\left[
		\dfrac{\rho_1}{\tanh[kh_1]} + \dfrac{\rho_2}{\tanh[kh_2]}
		\right]}
	\Biggl( \biggl[
	\dfrac{L_xL_y \; k^2}{2\sinh^2[kh_i]}
	+
	\dfrac{ \delta_{m} h_i(n^2\pi^2  - k^2L_y^2)}{ 2 L_y\:\sinh^2[kh_i]}
	\\[5pt]
	& + \dfrac{ \delta_m (n^2\pi^2 + k^2L_y^2)}{ 2 L_y \: k\:\tanh[kh_i]}
	+
	\dfrac{\delta_n h_i(m^2\pi^2  - k^2L_x^2)}{ 2 L_x\:\sinh^2[kh_i]}
	+
	\dfrac{ \delta_n (m^2\pi^2 + k^2L_x^2)}{ 2 L_x \: k\:\tanh[kh_i]}
	\biggr] \Biggr) , &&
	\label{eq:visc_damp_wall}
	\\
	\lambda_{\ii{int}} (m,n)
	&=\frac{k\sqrt{\omega}}{2\sqrt{2}}
	\:\frac{\bigl( \coth[kh_1] + \coth[kh_2] \bigr)^2}
	{ \biggl( \dfrac{1}{\rho_1\sqrt{\nu_1}} + \dfrac{1}{\rho_2\sqrt{\nu_2}} \biggr)
		\biggl[ \dfrac{\rho_1}{\tanh[kh_1]} + \dfrac{\rho_2}{\tanh[kh_2]} \biggr]
	} , &&
	\label{eq:visc_damp_interface}
	\\
	\nn \lambda_{\ii{irr}} (m,n) &= \frac{2k^2 \left( \dfrac{1}{\rho_1\sqrt{\nu_1}} + \dfrac{1}{\rho_2\sqrt{\nu_2}} \right)^{-1}}{\rho_1\coth{[kh_1]} +
		\rho_2\coth{[kh_2]}} \Biggl(\biggl[ 
	\dfrac{\rho_2\sqrt{\nu_1\nu_2} + \rho_1\nu_1}{\rho_2\sqrt{\nu_2} \tanh{[kh_1]}} +
	\dfrac{\rho_1\sqrt{\nu_1\nu_2} + \rho_2\nu_2}{\rho_1\sqrt{\nu_1} \tanh{[kh_2]}} &&
	\\[2.5pt]
	&  -\bigl(\coth{[kh_1]}+\coth{[kh_2]}\bigr)\bigl(\sqrt{\nu_1} + \sqrt{\nu_2}\bigr) 
	\biggr]\Biggr) . &&
	\label{eq:visc_damp_extra}
\end{align}
 The equation \eqref{eq:visc_damp_wall} is the mode-dependent damping rate owing to the bounding walls. The first term of the five-part sum corresponds to the damping at the top and bottom walls. The remaining terms stem from viscous dissipation in the lateral walls (refer to \eqref{eq:delta_func} for $\delta_m$ and $\delta_n$ functions). In an ARC, with smaller lateral and longitudinal surface areas than the upper and lower ones, dissipation at the top and bottom  surfaces tend to overshadow the contribution from the sidewalls. This is contrary to LMBs which, in general, have a larger sidewall surface area than the upper/lower surface area. The damping at the interface \eqref{eq:visc_damp_interface} is the same as in cylindrical cells \citep{Herreman2019}. Classically, when considering free-surface waves, the irrotational term \eqref{eq:visc_damp_extra} considers the dissipation in the irrotational bulk flow. In our formulation, it also involves some second-order rotational components at the interface. It should be noted that the presented contributions are only applicable when we assume no contact angle hysteresis and no-slip boundary condition at the bounding walls.   

In the single layer limit, interface damping $\lambda_{\ii{int}}$ reduces to zero. Considering unidirectional wave modes $(m,0)$, equations (\ref{eq:visc_damp_wall}) and (\ref{eq:visc_damp_extra}) simplify to 
\begin{align}
	&\lambda_{\ii{v}} = \sqrt{\frac{\omega\nu_2}{2}}
	\frac{1}{L_y} \Biggl[ 1 + \frac{L_y}{L_x} \biggl( 1 + 
	m\pi \frac{L_x-2h_2}{L_x\sinh{[2m\pi h_2/L_x]}}
	\biggr)
	\Biggr]  +2\nu_2k^2 . &&
	\label{eq:damp_1L_wall}
\end{align}   
This single-layer wave limit is fully equivalent to the well-known free-surface damping rate derived by \citet{Keulegan1959} as presented in \cite[chapter~6]{Faltinsen2009}.

\subsection{The magnetic damping rate}
\label{sec:mag_damp}

\subsubsection{Calculation of wave motion-induced currents}
To determine the magnetic damping rate, it is required to calculate the induced electric currents resulting from the term ($\h u \times B_z\h e_z$), that we neglected in \S\,\ref{sec:growth_rate_gen}.  In the leading-order, we can neglect the cell current such as viscous effects $\h J$ and consider free gravity-capillary waves exposed to a vertical magnetic field $B_z$ . Denoting the magnetic damping rate as $\lambda_{\ii m}$, the perturbation ansatz for a single standing wave mode can be rewritten as
\begin{equation}
	[\h u_i, \h j_i, \varphi_i ]= 
	\biggl( 
	C 
	\left[ 
	\widehat{\h{u}}_i, \
	\widehat{\pmb{\mathcal{J}}}_i, 
	\widehat{\Psi}_i \right] 
	+\left[
	\widetilde{\h{u}}_i, 
	\widetilde{\h{j}}_i, 
	\widetilde{\varphi}_i
	\right] 
	\biggr) 
	\;
	\ii e^{ (\ii{i}\overbar{\omega} -\lambda_{\ii m}) t} ,
	\label{eq:mag_ansatz}
\end{equation}
\normalsize
where $\widehat{\Psi}_i$ and $\widehat{\pmb{\mathcal{J}}}_i$ are the quasi-static corrections to the electric potentials and current densities, which are connected via the Ohm's law 
\begin{equation}
	\widehat{\pmb{\mathcal{J}}}_i = \sigma_i(-\bm\nabla \widehat{\Psi}_i + \widehat{\h u}_i \times \h B_z \h e_z) .
	\label{eq:Ohm}
\end{equation}
Same as $\widehat{\h j}$, $\widehat{\pmb{\mathcal{J}}}$ must satisfy the continuity conditions
\begin{equation}
	\nabla \cdot \widehat{\pmb{\mathcal{J}}}_i = 0, \qquad \nabla^2\widehat{\Psi}_i=0,
	\label{eq:laplace_psi}
\end{equation}
and the boundary conditions are adapted from (\ref{eq:bc1} -- \ref{eq:int_bc3}), here translating to 
\begin{subequations}
	\begin{align}
		\partial_x\widehat{{\Psi}}_{i}|_{x=L_x/2} &=
		\partial_x\widehat{{\Psi}}_{i}|_{x=-L_x/2} = \partial_y \widehat{\phi}_i \: B_z ,
		\label{eq:mag_bc1}
		\\
		\partial_y\widehat{{\Psi}}_{i}|_{y=L_y/2} &=
		\partial_y\widehat{{\Psi}}_{i}|_{y=-L_y/2} = -\partial_x \widehat{\phi}_i \: B_z ,
		\label{eq:mag_bc2}
		\\
		\widehat{\Psi}_1 |_{z=h_1}  &= 0  ,
		\label{eq:mag_bc3}
		\\
		\partial_z\widehat{{\Psi}}_2 |_{z=-h_2} &= 0   ,
		\label{eq:mag_bc4}
		\\
		\widehat{\Psi}_1 |_{z=0}  &= \widehat{\Psi}_2 |_{z=0} ,
		\\
		\sigma_1 \partial_z \widehat{\Psi}_1|_{z=0} &= 
		\sigma_2 \partial_z \widehat{\Psi}_2|_{z=0} .
	\end{align}
\end{subequations}
The current no-outflow conditions \eqref{eq:mag_bc1} and \eqref{eq:mag_bc2} are crucial for calculating the induced current as they couple $\widehat{\pmb{\mathcal{J}}}$ with the velocity fields. A pure boundary value problem must therefore be solved again. We insert the ansatz \eqref{eq:mag_ansatz} into the governing equation \eqref{eq:gov_momentum}, where we again ignore the viscous and lower order ($\h J \times \h b_i$) terms, yielding
\begin{equation}
	C \left[ (\ii{i}\omega - \lambda_{\ii m})\rho_i \widehat{\h{u}}_i + \bm{\nabla}\widehat{p}_i \right]
	+
	\left[ \ii{i}\omega \rho_i\widetilde{\h{u}}_i 
	+ \bm{\nabla}\widetilde{p}_i \right]
	= 	C \left( \; \widehat{\pmb{\mathcal{J}}}_i
	\times B_z\bm{e}_z\right)  .
	\label{eq:mag_damp_balance-eq}
\end{equation}
From this equation one can obtain $\lambda_{\ii m}$, but first the electric potential corrections $\widehat{\Psi}_i$ must be determined by solving the Laplace equation \eqref{eq:laplace_psi}, which can be quite tedious. To simplify the problem, we transform this two-layer formulation to a single-layer one and ignore the induced currents in the electrolyte layer. This is justified because the cryolite’s conductivity is several orders of magnitudes smaller than the conductivity of the aluminium and therefore the contribution of magnetic damping stemming from the cryolite to the overall magnetic damping is negligible. We treat the cryolite as a non-conducting liquid, which simplifies the boundary conditions as follows:
\begin{equation}
	\begin{array}{ccc}
		\widehat{\pmb{\mathcal{J}}}_1 =0,
		&
		\widehat{\Psi}_2 |_{z=0}  \approx 0,
		&
		\partial_z \widehat{\Psi}_2|_{z=0} \approx 0 .
	\end{array}
	\label{eq:mag_bc_final}
\end{equation}
The procedure for solving $\nabla^2\widehat{\Psi}_2=0$ is quite laborious, we provide all the details in \autoref{appC}. The solution emerges as an infinite series of harmonic and hyperbolic functions, also comprising linear and quadratic components:
\begin{align}
	\nn
	\widehat\Psi_{2} &= A_0 x - \frac{A_0 \bigl((-1)^n-(-1)^{m+n} \bigr)}{2L_x}\left(x - \frac{L_x}{2}\right)^2  &&
	\\
	\nn
	& + \sum_{i=0}^{\infty} \sum_{j=0}^{\infty} A_{i\!j}
	\biggl( \frac{1 + (-1)^m}{2\csch{[\xi_x x]}}  +  \frac{1 + (-1)^{m+1}}{2 \sech{[\xi_x x]}}  \Biggr)
	\cos\left[\frac{i\pi}{L_y}\left(y+\frac{L_y}{2}\right)\right]
	\cos\left[\frac{j\pi}{h_2}(z+h_2)\right] &&
	\\
	\nn
	& + B_0 y -\frac{B_0 \bigl((-1)^m-(-1)^{m+n} \bigr)}{2L_y}\left(y - \frac{L_y}{2}\right)^2
	\\
	& + \sum_{i=0}^{\infty} \sum_{j=0}^{\infty} B_{i\!j}
	\biggl( \frac{1 + (-1)^n}{2\csch{[\xi_y y]}}  +  \frac{1 + (-1)^{n+1}}{2 \sech{[\xi_y y]}}  \Biggr) 
	\cos\left[\frac{i\pi}{L_x}\left(x+\frac{L_x}{2}\right)\right]
	\cos\left[\frac{j\pi}{h_2}(z+h_2)\right] ,
	\label{eq:mag_potential}
\end{align}
with the coefficients,
\begin{align}
	\xi_x = \sqrt{ \biggl(\frac{i\pi}{L_y}\biggr)^2 + \biggl(\frac{j\pi}{h_2}\biggr)^2},
	\label{eq:xi_termX}
	\\
	\xi_y = \sqrt{ \biggl(\frac{i\pi}{L_x}\biggr)^2 + \biggl(\frac{j\pi}{h_2}\biggr)^2} ,
	\label{eq:xi_termY}
	\\[5pt]
	A_{0} =  \frac{\omega B_z}{k^2} \; \frac{\bigl((-1)^m-(-1)^{m+n} \bigr)}{L_yh_2}, 
	\\
	B_{0} =  \frac{-\omega B_z}{k^2} \; \frac{\bigl((-1)^n-(-1)^{m+n} \bigr)}{L_xh_2}. 
\end{align}
and
\begin{align}
	A_{i\!j} \! &= 
	\left\{
	\begin{array}{l}
		0, \quad \ii{for} \; i=j=0,
		\\
		\frac{2\delta_{i\times j} \; \omega B_z}{k^2 h_2} 
		\frac{n^2 \bigl( (-1)^{i+m+n} - (-1)^m \bigr)}
		{(i^2-n^2)\xi_x L_y}\:
		\frac{k^2h_2^2 (-1)^j}{k^2h_2^2 + j^2\pi^2}
		\biggl( \frac{1 + (-1)^m}{2\cosh{[\xi_x \frac{L_x}{2}]}}  +  \frac{1 + (-1)^{m+1}}{2 \sinh{[\xi_x \frac{L_x}{2}]} }  \biggr),  
		\quad \ii{else},
	\end{array} \right. 
	\label{eq:mag_Aij}
	\\
	B_{i\!j} \! &= 
	\left\{
	\begin{array}{l}
		0, \quad \ii{for} \; i=j=0,
		\\
		\frac{-2\delta_{i\times j} \; \omega B_z}{k^2 h_2} \frac{m^2\bigl( (-1)^{i+m+n} - (-1)^n \bigr)}
		{(i^2-m^2)\xi_y L_x}\:
		\frac{k^2h_2^2 (-1)^j}{k^2h_2^2 + j^2\pi^2}
		\biggl( \frac{1 + (-1)^n}{2\cosh{[\xi_y \frac{L_y}{2}]}}  +  \frac{1 + (-1)^{n+1}}{2 \sinh{[\xi_y \frac{L_y}{2}]} }  \biggr) , 
		\;\: \ii{else}.
	\end{array} \right .
	\label{eq:mag_Bij}
\end{align}
See \eqref{eq:delta_func} for $\delta_{i\times j}$. From the electric potential (\ref{eq:mag_potential}) and the flow potential solution (\ref{eq:vel_potential}) the wave-induced current $\widehat{\pmb{\mathcal{J}}}$ can now be evaluated through Ohm's law (\ref{eq:Ohm}).

\subsubsection{Eddy current distribution}

\begin{figure}
	\centerline{\includegraphics{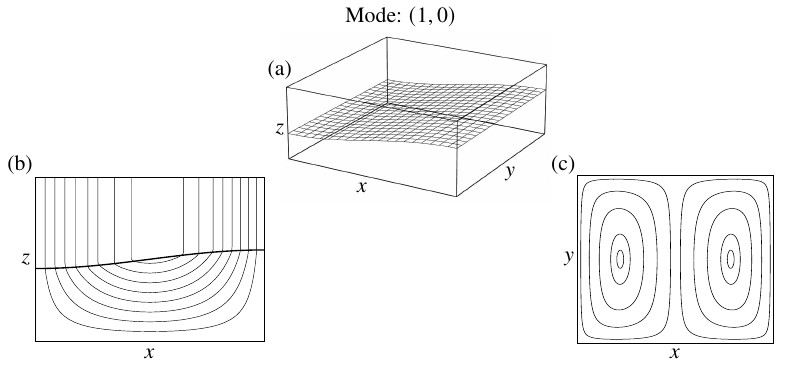}}
	\caption{The wave elevation \eqref{eq:eta_eq}, current excess \eqref{eq:elec_potential} and eddy current \eqref{eq:mag_potential} for wave mode $(1,0)$ in a square cell $L_x/L_y=1$. Figure (a) shows the 3d surface of the wave elevation, (b) shows the streamlines of the closing current in the $x-z$ cross-section  at $y=0$, (c) shows the horizontal eddy current streamlines in the $x-y$ cross-section located at $z=-h_2/2$.}
	\label{fig:mode10_eddy}
\end{figure}
\begin{figure}
	\centerline{\includegraphics{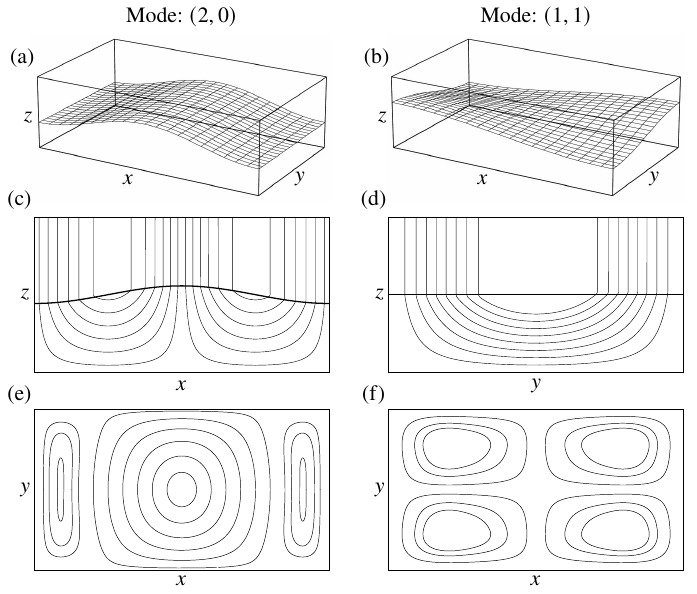}}
	\caption{The wave elevation \eqref{eq:eta_eq}, current excess \eqref{eq:elec_potential} and eddy current \eqref{eq:mag_potential} for wave mode pair $(2,0)$ (left) and $(1,1)$ (right) in a cell of aspect ratio $(L_x/L_y)=\sqrt{3}$. Figure (a)\textendash(b) show the 3d surface of the wave elevations, (c)\textendash(d) show the streamlines of the closing currents in the $x-z$ cross-section  at $y=0$ and $y-z$ cross-section  at $x=0$ respectively, (e)\textendash(f) show the horizontal eddy current streamlines in the $x-y$ cross-section located at $z=-h_2/2$.}
	\label{fig:mode2011_eddy}
\end{figure}
\begin{figure}
	\centerline{\includegraphics{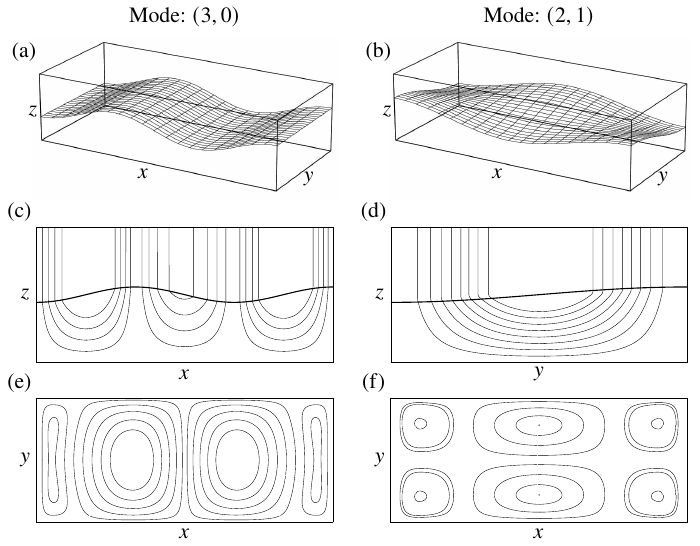}}
	\caption{The wave elevation \eqref{eq:eta_eq}, current excess \eqref{eq:elec_potential} and eddy current \eqref{eq:mag_potential} for wave mode pair $(3,0)$ (left) and $(2,1)$ (right) in a cell of aspect ratio $(L_x/L_y)=\sqrt{5}$. Figure (a)\textendash(b) show the 3d surface of the wave elevations, (c)\textendash(d) show the streamlines of the closing currents in the $x-z$ cross-section  at $y=0$ and $y-z$ cross-section  at $x=0$ respectively, (e)\textendash(f) show the horizontal eddy current streamlines in the $x-y$ cross-section located at $z=-h_2/2$.}
	\label{fig:mode3021_eddy}
\end{figure}

\renewcommand{\figureautorefname}{Figure}

Since the expanded solution (\ref{eq:mag_potential}) is very intricate and the eddy current patterns induced by interfacial wave motions have not yet been analysed, we want to delve deeper into the underlying physics and examine the solutions graphically. For this purpose, we also take a look at the wave elevation $\eta$ (\ref{eq:eta_eq}) and the previously calculated destabilising part of the wave-induced current $\widehat{\bm{j}}$ \eqref{eq:elec_potential}. \autoref{fig:mode10_eddy} shows all three solutions for the wave mode (1,0) that is the most unstable mode in square cells $(L_x/L_y)=1$. Where the excess current closes within one loop in the aluminium layer, the quasi-horizontal flow-induced current $\h u \times B_z\h e_z$ interestingly forms two current eddies aligned to the transverse direction to the wave slope. Precisely these horizontal  current eddies cause magnetic damping in interaction with $B_z$ that we  calculate in the next section.
\renewcommand{\figureautorefname}{Figures}
\autoref{fig:mode2011_eddy} and \ref{fig:mode3021_eddy} show the induced currents for the most unstable wave mode pairs $(2,0)+(1,1)$ and $(3,0)+(2,1)$ in $L_x/L_y=\sqrt{3}$ and $L_x/L_y=\sqrt{5}$ cells. While the destabilising currents $\widehat{\bm{j}}$ follow a simple pattern in which the number of closing loops corresponds directly to the wave numbers $m$ and $n$ (counting independently in the $x$- and $y$-direction), the eddy patterns show a more complex behaviour, both in their spacial arrangement and size distribution.
By analysing the solutions in detail, we were able to deduce a simple formula for the total number of all forming eddies $n_{\ii{eddy}}$ as a function of the wave numbers:
\begin{equation}
	n_{\ii{eddy}} = (m+1)\,(n+1) .
	\label{eq:eddy_num}
\end{equation}
From (\ref{eq:eddy_num}) it is easy to recognise that bidirectional waves $(m>0,n>0)$ induce a much higher number of eddies than unidirectional waves $(m>0,0)$ or $(0,n>0)$, which affects the energy dissipation of the induced current in the conducting medium.
For the case of rotating MPR waves, the total number of current eddies constantly alternates between $(m+1)(n+1)$ and $(m'+1)(n'+1)$.

\subsubsection{Calculation of the magnetic damping rate}
In order to finally obtain the dissipation rate, we have to solve equation \eqref{eq:mag_damp_balance-eq} in the lower layer ($\Omega_2$). For this purpose, we follow the same approach as in \S\,\ref{sec:growth_rate_gen} for deriving the growth rate $\alpha$. The Fredholm alternative applied to a single magnetically damped wave mode can be obtained as follows:  
\begin{equation}
	2K \, \lambda_{\ii m} =
	-\underbrace{\int_{\Omega_2} {\nabla} \widehat{\phi}^*_2\cdot \bigl(-\sigma_2 {\nabla}\widehat{\Psi}_2 \times B_z\bm{e}_z \bigr)}_{\mathcal{Q}_{\ii{wall}}}
	+
	\underbrace{
		\sigma_2\int_{\Omega_2}
		\norm{ \nabla\widehat{\phi}_2\times B_z\bm e_z }^2
	}_{\mathcal{Q}_{\ii{ind}}} ,
	\label{eq:mag_dissipation}
\end{equation}
Therefore, the magnetic damping rate is
\begin{equation}
	\lambda_{\ii m} = \frac{ 
		\mathcal{Q}_{\ii{ind}} - \mathcal{Q}_{\ii{wall}} }
	{2 K}, 
	\qquad \Bigl\{ \ii{refer \eqref{eq:K_11} for $K$ } .
	\label{eq:magnetic_damp}
\end{equation}
with $\mathcal{Q}_{\ii{ind}}$ and $\mathcal{Q}_{\ii{wall}}$ being the solutions of the two integral from equation \eqref{eq:mag_dissipation}. They are
\begin{equation}
	\mathcal{Q}_{\ii{ind}} =
	\frac{\sigma_2\omega^2 B_z^2}{4\delta_{m\times n}\:k}
	\Bigl( \coth{[kh_2]} + kh_2\csch^2{[kh_2]} \Bigr) L_xL_y , \quad \left( \ii{for $\delta_{m\times n}$, see \eqref{eq:delta_func}} \right) 
	\label{eq:Q_ind}
\end{equation}
\begin{align}
	\nn
	&\mathcal{Q}_{\ii{wall}} = \frac{A_0\sigma_2\omega B_z L_x}{k^2
		\biggl(
		f_m[m-0.5] \bigl((-1)^n-(-1)^{2n}\bigr)\bigl(1-(-1)^m\bigr)^2\:
		+ f_m[0.5-m] \bigl(1-(-1)^n\bigr) (m\pi)^2 \biggr)^{-1}} 
	\\
	\nn
	& + \sum_{i=0}^{\infty} \sum_{j=0}^{\infty} 
	\dfrac{2A_{i\!j}\sigma_2 \omega B_z h_2^2
		\Bigl( (im\pi)^2 + (n L_x \xi_x)^2 \Bigr) 
		\bigl((-1)^{1+i+j+m+n} + (-1)^{j+m}\bigr)  }
	{(i^2-n^2)(m^2\pi^2 + L_x^2\xi_x^2) (k^2h_2^2 + j^2\pi^2)
		\left( \dfrac{1 + (-1)^m}{2\csch{\left[\xi_x \frac{L_x}{2}\right]}}  +  \dfrac{1 + (-1)^{m+1}}{2 \sech{\left[\xi_x \frac{L_x}{2}\right]}}  \right)^{-1}
	}
	\\
	\nn
	& - \frac{B_0\sigma_2\omega B_z L_y}{k^2
		\biggl( f_n[n-0.5]
		\bigl((-1)^m- (-1)^{2m}\bigr)\bigl(1-(-1)^n\bigr)^2\:
		+ f_n[0.5-n] \bigl( 1-(-1)^m \bigr)(n\pi)^2 \biggr)^{-1}
	} 
	\\
	& + \sum_{i=0}^{\infty} \sum_{j=0}^{\infty} 
	\frac{2B_{ij}\sigma_2 \omega B_zh_2^2
		\Bigl((i n\pi)^2 + (mL_y\xi_y)^2\Bigr) 
		\bigl((-1)^{i+j+m+n} + (-1)^{1+j+n}\bigr)}
	{(i^2-m^2)(n^2\pi^2 + \xi_y^2L_y^2) (k^2h_2^2 + j^2\pi^2)
		\left( \dfrac{1 + (-1)^n}{2\csch{\left[\xi_y \frac{L_y}{2}\right]}}  +  \dfrac{1 + (-1)^{n+1}}{2 \sech{\left[\xi_y \frac{L_y}{2}\right]}}  \right)^{-1}
	} .
	\label{eq:Q_wall}
\end{align}
For simpler representation, we had to define an auxiliary function
\[
f_q[r] = \left\{\begin{array}{cc}
	\dfrac{1}{(q\pi)^2}, & r>0 \\[10pt]
	0, & r<0
\end{array}
\right. .
\]
Here, $\mathcal{Q}_{\ii{ind}}$ is the pure induction term originating from $\h u \times B_z\h e_z$ that is constantly withdrawing energy from the interfacial waves, whereas $\mathcal{Q}_{\ii{wall}}$ can be considered as a correction to $\mathcal{Q}_{\ii{ind}}$ that accounts for the effects of insulating sidewalls and has an opposite sign to $\mathcal{Q}_{\ii{ind}}$. In cells with perfectly conducting sidewalls, $\mathcal{Q}_{\ii{wall}} = 0, \; \because \widehat{\Psi}_2=0$ and the damping rate will be much larger owing to the recirculation of the induced current in the sidewalls \citep{Sreenivasan2005}.

\section{Comparison and validation}
\label{sec:results}
\begin{table}
	\begin{center}
		\def~{\hphantom{0}}
		\begin{tabular}{lll}
			Parameters from \citep{Politis2021} & & \\[3pt]
			Length of cell ($L_x$) & \SI{6.325}{\meter} & \\
			Width of cell ($L_y$) 	& $6.325,\dots, 2.108$ m & \\[3pt]
			& Cryolite ($\Omega_1$) & Aluminium ($\Omega_2$) \\
			Density ($\rho_1,\, \rho_2$) & \SI{2130}{\kilogram\per\meter\tothe{3}}
			& \SI{2330}{\kilogram\per\meter\tothe{3}} \\
			Kinematic viscosity ($\nu_1, \, \nu_2$) 
			& \SI{4.7e-7}{\meter\tothe{2}\per\second}
			& \SI{8.8e-7}{\meter\tothe{2}\per\second}
			\\
			Electrical conductivity ($\sigma_1, \, \sigma_2$)
			& \SI{210}{\siemens\per\meter}
			& \SI{3.3e6}{\siemens\per\meter}
			\\
			Layer thickness ($h_1,\, h_2$) & \SI{0.05}{\meter} & \SI{0.25}{\meter}
			\\[3pt]
			\hline
			Parameters from \citep{LaRocca2002} & & \\[3pt]
			Dimensions of cell ($L_x = L_y$) & \SI{0.5}{\meter} & \\[3pt]
			
			& $\Omega_1$ & $\Omega_2$ \\
			Density ($\rho_1,\, \rho_2$) & \SI{850}{\kilogram\per\meter\tothe{3}}
			& \SI{1000}{\kilogram\per\meter\tothe{3}} \\
			Kinematic viscosity ($\nu_1, \, \nu_2$) 
			& \SI{3.6e-5}{\meter\tothe{2}\per\second}
			& \SI{1.0e-6}{\meter\tothe{2}\per\second} \\
			Layer thickness ($h_1,\, h_2$) & \SI{0.06}{\meter} & \SI{0.08}{\meter}
			\\[3pt]
			\hline
			Parameters from \citep{Borisov2010} & & \\[3pt]
			Cross-section area of cell ($S$) & \SI{0.002}{\meter\tothe{2}} & \\[3pt]
			& Nitric acid (alky) ($\Omega_1$) & Nitric acid (aq) ($\Omega_2$) \\
			Density ($\rho_1,\, \rho_2$) at $35^{\circ}$C
			& \SI{859}{}
			& \SI{1050}{} \\
			\hspace{2.125cm} at $45^{\circ}$C
			& \SI{847}{}
			& \SI{1046}{} \\
			\hspace{2.125cm} at $55^{\circ}$C
			& \SI{834}{\kilogram\per\meter\tothe{3}}
			& \SI{1041}{\kilogram\per\meter\tothe{3}} \\
			Kinematic viscosity ($\nu_1, \, \nu_2$) at $35^{\circ}$C
			& \SI{2.9e-6}{}
			& \SI{7.4e-7}{} \\
			\hspace{3.61cm} at $45^{\circ}$C
			& \SI{2.2e-6}{}
			& \SI{6.1e-7}{} \\
			\hspace{3.61cm} at $55^{\circ}$C
			& \SI{1.8e-6}{\meter\tothe{2}\per\second}
			& \SI{5.2e-7}{\meter\tothe{2}\per\second} 
			\\
			Electrical conductivity ($\sigma_1, \, \sigma_2$)
			& \SI{0.67}{\siemens\per\meter}
			& \SI{47.5}{\siemens\per\meter}
			\\
			Layer thickness ($h_1 = h_2$) & \SI{0.014}{\meter} &
			\\[3pt]
			\hline
			Parameters from \citep{Sreenivasan2005} & & \\[3pt]
			Length of cell ($L_x$) & \SI{0.15}{\meter} & \\
			Breadth of cell ($L_y$) 	& \SI{0.04}{\meter} & \\[3pt]
			& Mercury ($\Omega_2$) & \\
			Density ($\rho_2$)
			& \SI{13546}{\kilogram\per\meter\tothe{3}} &
			\\
			Kinematic viscosity ($\nu_2$)
			& \SI{1.15e-7}{\meter\tothe{2}\per\second} & 
			\\
			Electrical conductivity ($\sigma_2$)
			& \SI{1e6}{\siemens\per\meter} &
		\end{tabular}
		\captionsetup{justification=justified, margin=0.1cm}
		\caption{Cell parameters and material properties from the different studies, for comparison with the theoretical results.}
		\label{tab:table_properties}
	\end{center}
\end{table}

This chapter is dedicated to the validation of our model, which essentially consists of the two-wave MPR growth rate \eqref{eq:main_gr} and corresponding stability thresholds \eqref{eq:beta_crit-onset} combined with the independently derived viscous \eqref{eq:visc_damp_sum} and magnetic \eqref{eq:magnetic_damp} damping rates. To this end we compare several predictions with different previous studies, whereby we also want to scrutinise some interesting aspects of the MPR physics deducible from our model.

First, in \S\,\ref{sec:visc_LaRocca}, we compare the viscous damping rate with the two-layer sloshing experiments and simulations by \citet{LaRocca2002}. In \S\,\ref{sec:Sreen}, the magnetic damping rate is then independently validated against the liquid metal wave damping experiments and the shallow-water analytical model presented by \citet{Sreenivasan2005}. In \S\,\ref{sec:fractal}, we show that our model is equivalent to the most recent MPR theory by \citet{Politis2021} in the limits of shallow-water gravity waves ($\gamma_{\ii{int}}=0$) with constant damping and $\sigma_2 \gg \sigma_1$. Finally, in \S\,\ref{sec:Borisov}, our predictions for critical Sele parameters are compared with experimental measurements obtained in the two-layer MPR experiment undertaken by \citet{Borisov2010} at several different aspect ratios. \autoref{tab:table_properties} summarises all of the physical parameters and material properties taken from the respective studies that are required to reproduce all the results presented in this section.



\subsection{The viscous damping rate}
\label{sec:visc_LaRocca}
We have already shown that the viscous damping rate $\lambda_{\ii{v}}$ agrees with the established linear damping rate (\ref{eq:damp_1L_wall}) derived by \cite{Keulegan1959} within its single-layer limit describing free-surface waves. However, interfacial waves are physically quite different because leading-order viscous boundary layers form above and below the interface (the viscous boundary layer below a free-surface is of higher order \citep{Joseph2006}), and Lamb's dissipation integral must be treated in a different way. Therefore, we wish to compare our predictions with studies directly referring to interfacial wave damping. The literature in this field is unfortunately rather limited.

\renewcommand{\figureautorefname}{figure}

\begin{figure}
	\centerline{\includegraphics{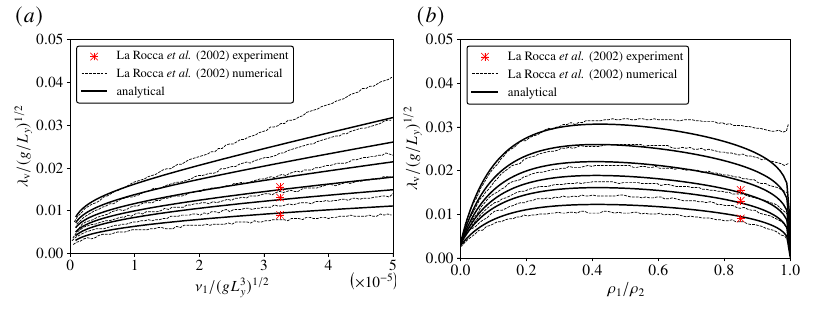}}
	\caption{($a$) non-dimensionalised viscous damping rate as a function of the non-dimensionalised kinematic viscosity for six different unidirectional wave modes ($m,0$), $m=1,2,...,6$ at a constant density ratio of $\rho_1/\rho_2 = 0.85$. ($b$) non-dimensionalised damping rate as a function of the density ratio $\rho_1/\rho_2$ at a constant viscosity ratio ($\nu_1/\nu_2 \approx 36$) for the same wave modes. Additional measurement points are included for the first three modes $m=1,2,3$.}
	\label{fig:rocca_comparison}
\end{figure}

\citet{LaRocca2002} conducted interfacial sloshing experiments in an enclosed square tank filled with two stably stratified liquid layers. These authors put one focus on dissipative effects and carried out direct decay rate measurements of different previously-excited interfacial wave modes. A theoretical model for viscous dissipation was also developed, without, however, providing an analytical expression for the damping rates. In \autoref{fig:rocca_comparison} we compare (\ref{eq:visc_damp_sum}) with La Rocca's results.\renewcommand{\figureautorefname}{Figure} \autoref{fig:rocca_comparison}(a) shows the non-dimensionalised viscous damping rate as a function of the non-dimensionalised kinematic viscosity for six different unidirectional wave modes ($m,0$), $m=1,2,...,6$. For lower wave modes $m=1,2,3$, our linear damping theory slightly overshoots La Rocca's predictions but agrees even better with the three experimental measurements carried out at $\nu_1/(gL_y^3)^{1/2} = \SI{3.25e-5}{}$ and $\rho_1/\rho_2 = 0.85$. For high viscosities and wave modes, our model predicts significantly lower damping rates. This may be due to the fact that higher-order terms become important for larger wave numbers, which we have neglected in our approach.

The applicability of La Rocca's model, however is also limited, as is evident from \ref{fig:rocca_comparison}(b), showing the damping rate as a function of the density ratio $\rho_1/\rho_2$. This figure portrays the transition from free-surface to interfacial wave damping. In the limit of quasi free surfaces $\rho_1/\rho_2 \lesssim 0.1$ we find, as expected, very good agreement with our model. In the regions corresponding to interfacial waves, we observe significant deviations for higher modes just as before, but the limit $\rho_1 \rightarrow \rho_2$ is described incorrectly by \cite{LaRocca2002}. Their damping rate estimates tend to diverge at $\rho_1 = \rho_2$, which is physically incorrect as the wave's natural frequencies become infinitely small in the limit $\rho_1 \rightarrow \rho_2$ and, correspondingly, the waves must then attenuate infinitely slowly. Our solution exhibits the correct behaviour and $\lambda_{\ii{v}}$ vanishes at $\rho_1 = \rho_2$. Since we have  chosen exactly the same ansatz as \cite{Herreman2019}, who derived $\lambda_{\ii{v}}$ for cylindrical geometries, and the decay rates have been successfully validated against measurements by \cite{Horstmann2019}, we are convinced that our treatment of viscous damping is reliable given the applied approximations.

\subsection{The magnetic damping rate}
\label{sec:Sreen}

\begin{figure}
	\centerline{\includegraphics{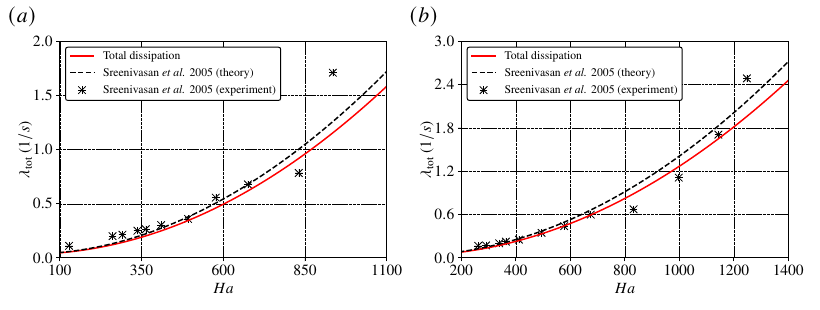}}
	\caption{Comparison of decay rates $\lambda_{\rm tot} = \lambda_{\ii m} + \lambda_{\ii v}$ calculated through  (\ref{eq:visc_damp_sum}) and (\ref{eq:magnetic_damp}) with experimental measurements and theoretical predictions by \cite{Sreenivasan2005} as a function of the Hartmann number $Ha$ for two different pool layer heights $h_2=0.022\,{\rm m}$ ($a$) and $h_2=0.035\,{\rm m}$ ($b$).}
	\label{fig:sreeni_comp}
\end{figure}

The magnetic damping rates $\lambda_{\ii m}$ cannot be compared independently of viscous damping rates $\lambda_{\ii v}$ because waves that move in external magnetic fields are concomitantly subjected to viscous dissipation. Therefore, the two damping contributions must be considered in superposition $\lambda_{\rm tot} = \lambda_{\ii m} + \lambda_{\ii v}$.
\cite{Sreenivasan2005} have dedicated an entire study to the magnetic attenuation of standing gravity waves in rectangular containers and measured the decay rates of surface waves in liquid mercury that were exposed to vertical magnetic fields of up to  $1.3\,{\rm T}$. Moreover, the authors provide a theoretical solution for magnetic damping times that takes into account bulk Ohmic dissipation, viscous side wall dissipation and Hartmann layer dissipation (at the bottom). Their model was the first to properly address the delicate problem of electrically insulating side walls, but was only formulated for the first unidirectional standing wave mode $(m=1,n=0)$ using the shallow-water approximation. For this case, we can compare both theories. \autoref{fig:sreeni_comp} shows the total damping rate $\lambda_{\ii{v}} + \lambda_{\ii m}$ calculated from (\ref{eq:visc_damp_sum}) and (\ref{eq:magnetic_damp}) as a function of the Hartmann number
\begin{equation}
	H\!a = \sqrt{\frac{\sigma_2B_z^2L_y^2}{\rho_2\nu_2}} 
\end{equation}
in comparison with the theoretical estimates and measurements by \cite{Sreenivasan2005} for two different metal layer heights.
To calculate $\lambda_{\ii{v}} = \lambda_{\ii{wall}} + \lambda_{\ii{irr}}$, we have used the simplified free-surface limit (\ref{eq:damp_1L_wall}).\renewcommand{\figureautorefname}{figures} In both \autoref{fig:sreeni_comp}(a)
and \ref{fig:sreeni_comp}(b) we find a close agreement between both theories and experiments. Only for large $H\!a$ numbers our damping rate underestimates the measurement points because we have ignored Hartmann layer dissipation in our model. However, such high $H\!a$ numbers are irrelevant for ARCs, where the magnetic fields hardly ever exceed $B_z = 10\, {\rm mT}$. This comparison therefore validates our approach within the intended range of applicability. Before concluding this section, we want to add two more important remarks. \cite{Sreenivasan2005} show damping times instead of damping rates, whereby deviations between theory and experiment become more visible at small $H\!a$. At these points, our theoretical predictions do not make any improvements either, but in this regime damping is more strongly determined by viscous dissipation effects, which are very difficult to control in liquid metal experiments because the metal surface can easily be contaminated and surface pinning effects may occur along the side walls. Also, \cite{Sreenivasan2005} present some measurements of damping in deep fluid pools from which their theory deviates considerably. The authors attribute the deviation to the shallow water approximation and introduce an empirical non-shallow correction parameter. However, we have noticed that our damping rate, which is also valid for non-shallow geometries, still agrees quite well with \cite{Sreenivasan2005}'s theory even for deep layers. Due to the quasi two-dimensional structure of the induced eddy currents, the shallow water approximation turned out not to be a major limitation for calculating magnetic damping rates. The observed deviations must have another origin.

\subsection{Comparison with \cite{Politis2021} and fractality}
\label{sec:fractal}
We will show that the linear MPR growth rate and stability onset predictions by \cite{Politis2021} are fully contained in our model as a limiting case for $h_1 ,h_2 \ll L_x, L_y$, $\sigma_1 \ll \sigma_2$, $\gamma_{\ii{int}} = 0$ and constant damping coefficients. First, as a special case, we consider the growths of dissipationless MPR waves in inherently unstable cell configurations, for which \cite{Politis2021} derived a compact analytical expression. A cell is called inherently unstable whenever there exist two wave modes fulfilling $\delta \omega = 0$.
In shallow geometries, gravity waves are non-dispersive ($\omega = ck$, $\omega' = ck'$) and the frequency equality $\delta \omega = 0$ reduces to a wave number equality $k=k'$, imposing the following constraint on the lateral aspect ratio: 
\begin{align}
	k = \pi\sqrt{\frac{m^2}{L_x^2} + \frac{n^2}{L_y^2}} &= \pi\sqrt{\frac{m'^2}{L_x^2} + \frac{n'^2}{L_y^2}} = k' , \nonumber \\
	\Rightarrow \frac{L_x^2}{L_y^2} = \frac{m'^2 - m^2}{n'^2 - n^2} &= \frac{(m'-m)(m'+m)}{(n-n')(n+n')}.
\end{align}
This condition allows us to easily understand why the unstable aspect ratios form an absolutely discontinuous dense set of points. 
For a simple explanation, let us first consider the wave mode function $\Theta$ (\ref{eq:form_function}), which can only be non-zero if both wave number pairs $(m,m')$ and $(n,n')$ are of opposite parity. Wave pairs whose wave numbers $(m,m')$ or $(n,n')$  are both even or odd can therefore never be destabilised. From number theory it is known that all sums and differences
of numbers with opposite parity yield odd numbers $m \pm m' = {\rm odd}$ and $n \pm n' = {\rm odd}$. Likewise, products of odd numbers are again always odd, which is why we can assign
\begin{align}
	\frac{(m'-m)(m'+m)}{(n-n')(n+n')} = \frac{\bm{m}}{\bm{n}}, \label{eq:fractal}
\end{align} 
where $\bm{m}$ and $\bm{n}$ are any two odd numbers. This proves the fractality of unstable aspect ratios $L_x/L_y = \sqrt{\bm{m}/\bm{n}}$. From equation (\ref{eq:fractal}) we can deduce the most unstable wave pair for all cells fulfilling $L_x/L_y = \sqrt{\bm{m}/\bm{n}}$ in the inviscid limit: 
\begin{equation}
	m = \frac{\bm m -1}{2}, \quad m' = \frac{\bm m + 1}{2}, \qquad
	n = \frac{\bm n +1}{2}, \quad n' = \frac{\bm n - 1}{2}	.
	\label{eq:mn_odd}
\end{equation}
By inserting the most unstable wave numbers (\ref{eq:mn_odd}) into the growth rate solution \eqref{eq:main_gr} and applying all the mentioned limits we obtain an expression for $\alpha_{\rm MPR}$ as a function of the fractal numbers
\begin{equation}
	\alpha_{\rm MPR} = \pm \, \beta_{\rm S} \: \frac{c}{\sqrt{L_xL_y}} \frac{8}{\pi^3}
	\sqrt{\frac{\delta_{\bm{m}-1}\delta_{\bm{n}-1}}
		{(\bm{m} + \bm{n})(\bm{m}\bm{n}+1)\sqrt{\h m\h n}}},
	\qquad 
	\left\{\ii{ see \eqref{eq:delta_func} for $\delta_{\bm{m}-1}$, $\delta_{\bm{n}-1}$} \right. .
	\label{eq:priede_dimension}
\end{equation}
Equation \eqref{eq:priede_dimension} is the dimensional form of the growth rate solution (3.28) of \citet{Politis2021}, which verifies the equivalence of both models.

\begin{figure}
	\centerline{\includegraphics{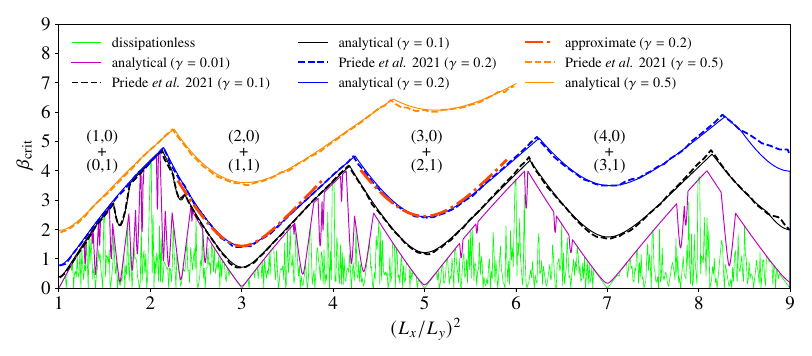}}
	\caption{Critical Sele parameters $\beta_{\ii{crit}}$ from \cite{Politis2021} and our analytical prediction \eqref{eq:beta_crit-onset} as a function of the aspect ratio squared $(L_x/L_y)^2$, with and without considering the effects of damping here modelled by constant damping coefficients $\gamma$. For the comparison, the non-dimensional coefficients $\gamma$ were re-dimensionalised as $\lambda=0.5\gamma \, c/L_x$ and $\lambda'=0.5\gamma\, c/L_y$. The red curves show the approximate solution (\ref{eq:beta_approx}) for $\gamma=0.2$ around $L_x/L_y = \sqrt{3}$ and $L_x/L_y = \sqrt{5}$.}
	\label{fig:priede_fractal}
\end{figure}

It remains to validate our predictions of stability onsets, which \cite{Politis2021} have presented in the form of critical Sele parameters $\beta_{\ii{crit}}$ and studied their dependence on the aspect ratio. For this purpose, the authors also modelled damping by introducing a phenomenological friction coefficient $\gamma$. However, the resulting eigenvalue problem was only solved numerically using the Galerkin method, and for this reason we compare the onset solutions graphically in the following. Although we have derived an explicit analytical solution for $\beta_{\rm crit}$, equation \eqref{eq:beta_crit-onset}, it is not evident \textit{a priori} which mode pair is most unstable and thus defines $\beta_{\rm crit}$. Considering damping and operating beyond the shallow-water approximation, the most unstable mode pair cannot be determined analytically. Our simple strategy for solving this problem is to calculate $\beta_{\rm crit}$ for all possible combinations of wave pairs and then to assign $\beta_{\rm crit}$ to the particular wave pair that yields the lowest onset value. The smallest possible critical Sele parameter defines the global instability onset. However, this iterative method is not a major limitation for our model because large-scale wave modes are always most unstable in practice, such that only wave numbers up to about $m,n \lesssim 10$ must be examined. This is because both viscous and magnetic damping increase markedly with increasing wave numbers and also because small-scale wave motions are gradually suppressed by interfacial tension ($\beta_{\gamma_{\ii{int}}}$ \eqref{eq:beta_crit-onset} diminishes with increasing $k$).

\renewcommand{\figureautorefname}{Figure} \renewcommand{\tableautorefname}{table}

\autoref{fig:priede_fractal} shows iteratively calculated onset values $\beta_{\rm crit}$ as a function of the squared lateral aspect ratio $(L_x/L_y)^2$ compared with numerical solutions by \cite{Politis2021} for different friction coefficients $\gamma$. All geometrical parameters and material properties of the considered cell are given in \autoref{tab:table_properties}. We find a quasi-perfect agreement between both models for any given friction coefficient. The pattern of the green fractal curve ($\gamma = 0$) is determined solely by the chosen resolution. All other curves can be unambiguously determined; the non-differentiable inflection points indicate the transitions between different most unstable wave pairs. Here, a small deviation between both theories emerges at $\gamma = 0.2$ for $(L_x/L_y)^2 > 8.45$. The curve from \cite{Politis2021} shows here $\beta_{\ii{crit}}$ for the mode pair $(5,0)+(4,1)$. However, our solution predicts that the pair $(3,0)+(0,1)$ becomes unstable first, and accordingly our calculated curve is slightly lower. Apart from this point, however, the most unstable wave pair is always correctly predicted by the wave number tuples (\ref{eq:mn_odd}). For these cases, a simple approximate solution for $\beta_{\rm crit}$ can be derived from our theory that is valid in the vicinity of the most unstable aspect ratios ($L_x^2/L_y^2 = 1,3,5,\dots$). Given these aspect ratios, we can set $\bm{n}=1$, because $L_x^2/L_y^2 = \bm{m}/1$ with $\bm{m}=1,3,5,\dots$, and, by introducing (\ref{eq:mn_odd}) into (\ref{eq:beta_crit_shallow}), ignoring the term $i\delta\omega\delta\lambda$ and re-dimensionalising  $\lambda=0.5\gamma \, c/L_x$ and $\lambda'=0.5\gamma\, c/L_y$, $\beta_{\rm crit}$ can be deduced as a function of only $L_x/L_y$ and $\bm{m}$:
\begin{equation}
	\beta_{\ii{crit}}= \frac{\pi^4 \sqrt[4]{\bm{m}}
		(\h m+1) }{16\sqrt{\delta_{\h m-1}}}\sqrt{
		\frac{\gamma^2}{\pi^2} + \frac{L_y}{2L_x} (\h m^2+1) + \frac{L_x}{L_y} - \sqrt{
			\frac{(\h m^2 -1)^2}{4}\,\frac{L_y^2}{L_x^2} + (\h m+1)^2}}. \label{eq:beta_approx}
\end{equation}
\renewcommand{\figureautorefname}{figure} 
In this approximation, the numbers $\bm{m}$ must be taken as constants and the aspect ratio $L_x/L_y$ as a variable. For $\gamma=0.2$, we show (\ref{eq:beta_approx}) by the red curves for $\bm{m}=3$ and $\bm{m}=5$ in \autoref{fig:priede_fractal}. The approximate solution is very accurate, but deviations become evident near the wave mode transition peaks so that the most stable aspect ratios cannot be determined analytically from (\ref{eq:beta_approx}). Onset values for the most unstable aspect ratios $L_x/L_y = 1,3,5,\dots$, however, can readily be derived from (\ref{eq:beta_approx}) by setting $L_x/L_y = \sqrt{\bm{m}}$ yielding the exact solution 
\begin{equation}
	\beta_{\ii{crit}}= \frac{\pi^3 \sqrt[4]{\bm{m}}
		(\h m+1) }{16\sqrt{\delta_{\h m-1}}}\gamma. \label{eq:beta_min}
\end{equation}
The physical relevance of the fractal distribution of critical aspect ratios, which is enveloped by the dissipative curves, has already been discussed in detail by \cite{Politis2021}. We make here a final note that the most unstable wave pairs (smallest $\beta_{\rm crit}$) are not always the wave pairs with the highest growth rate. Therefore, other mode pairs than these predicted at $\beta_{\rm S} = \beta_{\rm crit}$ can develop at higher $\beta_{\rm S} > \beta_{\rm crit}$, which is why the exact pattern of MPR waves is difficult to predict even in the linear regime.
At some certain point $\beta_{\rm S} > \beta_{\ii{crit}}$, it is always the primary rotating mode pair $(1,0)+(0,1)$ that yields the highest growth rate in all possible cell configurations, as already noted by \citet{Munger2008}. But in these high $\beta_{\rm S}$ regimes, linear MPR models are often no longer applicable.

\subsection{Comparison with measured stability onsets by \cite{Borisov2010}}
\label{sec:Borisov}

As \cite{Politis2021} have adopted constant damping parameters, it remains to compare our onset solution (\ref{eq:beta_crit-onset}) in conjunction with wave mode-dependent damping rates (\ref{eq:visc_damp_sum}), which constitute the major amelioration of our model. For this, the overlooked one-of-a-kind experiment by \cite{Borisov2010} is the perfect (and only) option. The authors are the only ones to have ever succeeded in destabilizing MPR waves in two-layer experimental cells that can reproduce the complete MPR magnetohydrodynamics without any noteworthy trade-off. In their set-up, two transparent liquids were stratified (nitric acid ($\rm{HNO}_{\rm3(aq)}$ at the bottom and $\rm{HNO}_{3}$ + ${\rm C}_5\rm{H}_{12}\rm{O}$ at the top), both of which had small but sufficiently contrasting electrical conductivities ($\sigma_1 = 0.67\,{\rm S}\,{\rm m}^{-1}$ and $\sigma_2 = 47.5\,{\rm S}\,{\rm m}^{-1}$). Due to the comparatively small difference in the density of the two liquids, small cell currents  $I \sim 2$ \textendash \ $4\,{\rm A}$, in combination with a strong external magnetic field  $B_z = 0.25\,{\rm T}$, were sufficient to attain large enough Sele parameters $\beta_{\rm S} \sim 2$ to destabilise MPR waves. In order to change the electrical boundary conditions at the bottom electrode, the authors employed a removable insert made of porous material that was impregnated with an aqueous solution of nitric acid. This allowed both sets of idealised boundary conditions to be realised in the experiment, i.e. that the lower liquid be either a much better or much worse electrical conductor than the bottom electrode. Unfortunately, the authors have only published measurements without the insert, for which the ground electrode can be regarded as a perfect electrical conductor. This configuration is not accounted for by our model, but simply by changing the Neumann boundary condition \eqref{eq:bc4} to the Dirichlet boundary condition $\varphi_2 = 0|_{z=-h_2}$, we can easily adapt our solution. The stability onset has the same form as before:
\begin{equation}
	\beta_{\ii{crit}} = \frac{L_x^2 L_y^2}{h_1h_2 \Theta}
	\sqrt{ \frac{\overbar\lambda^2 - \bigl[ \ii i{\delta \omega} -\delta\lambda \bigr]^2} {(\overbar\omega^2 - {\delta\omega}^2) \mathcal{T}_c\mathcal{T}'_c} } ,
	\label{eq:beta_crit_cond}
\end{equation}
where,
\fontsize{8.9pt}{8.9pt}
\begin{align*}
	\mathcal{T}_c &= \dfrac{\Lambda}{k(k^2 - k'^2)}
	\Biggl[ 
	\dfrac{-k'}{\text{tanh}[kh_1]} 
	+ \dfrac{k'\:\text{sech}[k'h_1]}{\text{sinh}[kh_1]}
	+ \dfrac{k}{\text{coth}[k'h_1]}
	+ \dfrac{-k'}{\text{tanh}[kh_2]} 
	+ \dfrac{k'\:\text{sech}[k'h_2]}{\text{sinh}[kh_2]}
	+ \dfrac{k}{\text{coth}[k'h_2]}
	\Biggr] ,
	\\[5pt]
	\mathcal{T}'_c &= \dfrac{\Lambda'}{k'({k'}^2-k^2)}
	\Biggl[ 
	\dfrac{-k}{\text{tanh}[k'h_1]} 
	+ \dfrac{k\: \text{sech}[kh_1]}{\text{sinh}[k'h_1]}
	+ \dfrac{k'}{\text{coth}[kh_1]}
	+\dfrac{-k}{\text{tanh}[k'h_2]} 
	+ \dfrac{k\: \text{sech}[kh_2]}{\text{sinh}[k'h_2]}
	+ \dfrac{k'}{\text{coth}[kh_2]}
	\Biggr] ,
\end{align*}
\normalsize
and the new conductivity jump parameter is
\[
\Lambda = \dfrac{\sigma^{-1}_1-\sigma^{-1}_2}{\sigma^{-1}_1\ii{tanh}[kh_1] + \sigma^{-1}_2\ii{tanh}[k h_2]}	.
\]

\begin{figure}
	\centerline{\includegraphics{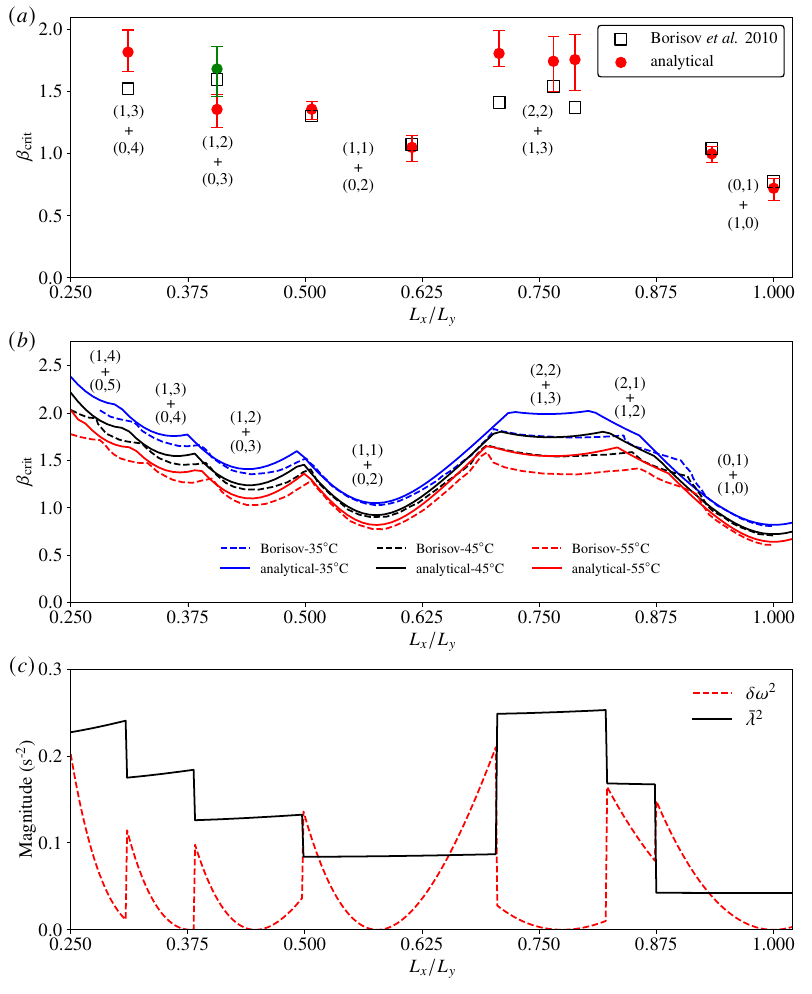}}
	\caption{(a) Comparison with the experimentally observed stability onsets at nine different cell geometries. The error bars cover a temperature range of $35^{\circ}$C to $55^{\circ}$C. (b) shows the analytical stability onset comparisons along with the critical wave modes observed at different cell aspect ratios. (c) distribution of parameters $\delta\omega^2$ and $\overbar\lambda^2$ which play a key role in determining the critical wave mode and hence the critical onset value.}
	\label{fig:borisov_comparison}
\end{figure}

\cite{Borisov2010} measured the critical onsets using a different dimensionless parameter $W = I_0 B_z /[(\rho_2 - \rho_1)gL_x L_y]$ that can easily be rewritten as $\beta_{\rm S}$ through $\beta_{\rm S} = W\:L_xL_yh_1^{-1}h_2^{-1}$. The measured critical Sele parameters are shown together with the most unstable wave pairs $(m,n) + (m',n')$ in 9 different cells of different aspect ratios $L_x/L_y$ in \autoref{fig:borisov_comparison}(a). For a better understanding, \autoref{fig:borisov_comparison}(b) further shows our continuous solution (\ref{eq:beta_crit_cond}) compared with the stability prediction from a linear non-shallow MPR theory developed by \cite{Borisov2008}, who used the Galerkin method to formulate a Cauchy problem that was solved numerically. \renewcommand{\figureautorefname}{Figure} \autoref{fig:borisov_comparison}(c) supplements \renewcommand{\figureautorefname}{figure} \autoref{fig:borisov_comparison}(b) and shows the proportion of the two key parameters $\bar{\lambda}^2$ and $\delta \omega^2$ in equation (\ref{eq:beta_crit-onset}), allowing the reader to determine whether a given onset value $\beta_{\rm crit}$ is influenced more by total damping or frequency mismatching. During their experiments, \cite{Borisov2010} had to contend with the difficulty that the cell was subject to strong temperature fluctuations due to the high Ohmic losses (Joule heating) occurring in both poorly conducting liquids. The authors measured  temperatures between $35^{\circ}-55^{\circ}$C, which we incorporated in \autoref{fig:borisov_comparison}(a) using error bars. At all examined aspect ratios the reported wave pairs coincide with the most unstable pairs predicted by our model. Quantitatively, we also find fair agreement, but there are some considerable deviations, especially for the mode pair $(2,2) + (1,3)$. The measurement point that corresponds to $(1,2) + (0.3)$ lies very close to the jump point between the most unstable pairs $(1,2) + (0.3)$ and $(1,3) + (0.4)$, see \autoref{fig:borisov_comparison}(c). If we pick the neighbouring pair $(1,3) + (0.4)$ (green measuring point) here, although our model still predicts $(1,2) + (0.3)$ to be most unstable, we also find good agreement within the scope of measurement accuracies. In general, it can be observed that those measurements with the highest deviations also differ the most from Borisov's theory. The direct comparison of both theories reveals that our model always yields larger $\beta_{\rm crit}$. Due to difficulties arising from translation, the reason for this discrepancy is quite difficult to identify, but we noticed that \cite{Borisov2010} have entirely neglected sidewall damping. This is a misconception, we have estimated from our damping rates (\ref{eq:visc_damp_wall}), (\ref{eq:visc_damp_interface}) and (\ref{eq:visc_damp_extra}) that sidewall damping in this non-shallow two-layer cell accounts for more than a third of the total damping. Despite this stark difference, both theories concur remarkably well, so there must be other differences in the modelling that we cannot elaborate on further. Due to the completeness of our theory and the compatibility with many previous MPR models, we are convinced that the larger $\beta_{\rm crit}$ parameters predicted by our model are physically reasonable. We hope to dispel any remaining doubts in the future with our own MPR experiment \citep{Hegde2025}, which permits precise and direct measurement of growth rates $\alpha_{\rm MPR}$ and damping rates $\lambda_{\ii v}$ and $\lambda_{\ii m}$ in addition to $\beta_{\rm crit}$.

\vspace*{-0.4cm}
\section{Concluding remarks}
\label{sec:discuss}

Our new explicit perturbation formulation of the linear MPR instability resulting from two superimposed transverse standing wave modes refines existing MPR models in several ways, despite the high degree of maturity that has been attained to date. Our theory overcomes the shallow-water approximation and takes capillary wave regimes into account, allowing it to serve as a benchmark for downscaled MPR model experiments and also for small liquid metal batteries. The most important refinement, however, is that we consider dissipation not only in terms of purely phenomenological damping parameters, but explicitly calculate the wave mode-dependent viscous and magnetic damping rates. Because the resultant dissipation can differ considerably from wave mode to wave mode, see \autoref{fig:borisov_comparison}(c), this makes a sizable difference to the global stability.

We have gone to great lengths to keep all analytical solutions explicit and as tangible as possible so that these can be implemented with minimal efforts and without a wealth of prior knowledge. The main result of our study is provided by equation (\ref{eq:main_gr}), which predicts MPR growth rates $\alpha_{\rm MPR}$ for any given pair of standing wave modes. In a simple way, the global stability behaviour is contained in this single equation: whenever the square root in (\ref{eq:main_gr}) yields a complex number with zero real part, the cell remains stable and only the natural frequencies of the waves are shifted by MHD coupling effects. If the real part of the square root turns nonzero but the real part of $\alpha_{\rm MPR}$ is negative, excited waves still attenuate, but more slowly than they would without MHD coupling, and both waves oscillate at the same frequency. Once the real part of $\alpha_{\rm MPR}$ finally attains a positive value, the cell becomes MPR unstable and an exponentially growing rotating wave motion builds up at the interface. The growth rate $\alpha_{\rm MPR}$ incorporates the individual damping rates $\lambda$ and $\lambda'$ of the two coupled standing waves, which we have calculated separately. To account for viscous damping, we have calculated the leading-order Stokes boundary layers forming at all container walls as well as above and below the interface, and evaluated Lamb's dissipation integral (\ref{eq:dissip}). The total damping rate $\lambda_{\ii{v}}$ (\ref{eq:visc_damp_sum}) is composed of a wall damping term (\ref{eq:visc_damp_wall}), an interfacial damping term (\ref{eq:visc_damp_interface}) and a term resulting from the irrotational flow (\ref{eq:visc_damp_extra}). For calculating the magnetic damping rates, we have derived an exact solution for the electrical potential of the induced eddy currents (\ref{eq:mag_potential}), which close entirely inside the liquid bath as the side walls are assumed to be electrical insulators. We found that an arbitrary bidirectional standing wave $(m,n)$ induces exactly $(m+1)(n+1)$ quasi two-dimensional horizontal current eddies. The resulting magnetic damping rate $\lambda_{\ii m}$ (\ref{eq:magnetic_damp}) is composed of a pure induction term (\ref{eq:Q_ind}) and a very intricate correction term to account for insulating side walls (\ref{eq:Q_wall}). Regardless of the MPR instability, these damping rates can be used for other problems such as for two-liquid sloshing experiments, and can further serve as an easy benchmark for direct numerical solvers, e.g. to verify whether the viscous boundary layers have been resolved sufficiently, which is often a challenge. Due to the complex dependencies of the damping rates on the wave modes, our model cannot in general predict which wave pair will be destabilised first, as is possible in the limiting case of shallow layers and constant damping, see \S\,\ref{sec:fractal}. Therefore, it is necessary to calculate the most unstable wave mode pair, which defines the critical stability parameter $\beta_{\rm crit}$ (\ref{eq:beta_crit-onset}), iteratively. To simplify the calculations for the reader, we have gathered all essential equations in \autoref{appD} and added some explanations for easy implementation.

Several outcomes of the theory were compared with different experimental as well as theoretical studies in \S\,\ref{sec:results} and broad agreement was confirmed. However, a few questions have yet to be resolved. While our viscous damping rate agrees with \cite{Keulegan1959} in the single-layer free-surface limit, the comparison with the two-layer model by \cite{LaRocca2002} revealed some considerable deviations at higher wave modes. If we artificially multiply the irrotational damping contribution (\ref{eq:visc_damp_extra}) by a factor of two, we achieve a very good agreement with \cite{LaRocca2002} for all wave modes. But then our dissipation rates no longer fit Keulegan's well established damping theory, and we can only speculate that a factor two error may have been introduced in one of the studies. Conclusive clarification of this issue requires additional damping experiments to be carried out for higher wave modes. Also, the derived magnetic damping rates could only be validated for one particular unidirectional wave mode ($m=1,n=0$). The damping behaviour of bidirectional wave modes $(m>0,n>0)$, which are important for the MPR instability, is dictated by complex closing current patterns and therefore much more complicated, making it very rewarding to set up dedicated liquid metal wave damping experiments.

We intend to employ the stability theory to calibrate our recent single-layer MPR experiment \citep{Hegde2025} in such a way that it can properly reflect the magnetohydrodynamics of (idealised) aluminium reduction cells. Then, for the first time, we will be able to measure and benchmark MPR growths rates of individual wave pairs. Moreover, our theory is well-positioned for augmenting the cell current (or external magnetic field) with alternating components, with a view to studying active MPR suppression techniques that have recently proved promising \citep{Mohammad2022}. In \cite{Horstmann2025} it was shown that the alternating current MPR suppression rests in a special form of parametric anti-resonance. However, the analysis relied on a highly idealised pendulum model with only two degrees of freedom. The ideal AC frequencies for a minimally-invasive MPR suppression predicted in \cite{Horstmann2025} still need to be confirmed by continuous MPR theories. As a final point, we leave it open for future works to extend our theory to three liquid layers for describing the MPR instability in liquid metal batteries with coupled interfaces; a quite straightforward but very laborious task.

\begin{bmhead}[Acknowledgements.]
	The authors would like to thank William Nash for proofreading the manuscript.
\end{bmhead}
\begin{bmhead}[Funding.]
	This study has received funding from the Deutsche
	Forschungsgemeinschaft (DFG, German Research Foundation) by Award No. 512131026.
\end{bmhead}
\begin{bmhead}[Declaration of interest.]
	The authors report no conflict of interest.
\end{bmhead}

\appendix

\section{Formulation of the Fredholm alternative} \label{appA}
Firstly, we write the average frequency $\overbar\omega$, the frequency difference $\delta\omega$, the average damping rate $\overbar\lambda$ and the difference in damping rate $\delta\lambda$ as the following: 
\begin{equation}
	\overbar\omega = \frac{\omega+\omega'}{2}, \quad
	\delta\omega = \frac{\omega-\omega'}{2}, \quad
	\overbar\lambda = \frac{\lambda+\lambda'}{2}, \quad
	\delta\lambda = \frac{\lambda-\lambda'}{2} ,
\end{equation}
Secondly, the balance equations \eqref{eq:gov_momentum} and \eqref{eq:gov_cont} considering only leading order terms, for a given wave mode $(m,n)$ are
\begin{subequations}
	\begin{equation}
		\rho_i(\ii i\omega - \lambda)\widehat{\h u_i} + \bm\nabla\widehat p_i = 0 ,
		\label{eq:lead_balance_1}
	\end{equation}
	\begin{equation}
		\bm\nabla \cdot \widehat{u_i} = 0 ,
		\label{eq:lead_balance_2}
	\end{equation}
	\begin{equation}
		\widehat{\h j}_i  = -\sigma_i \bm\nabla\widehat\varphi_i , 
		\qquad \bm \nabla \cdot \widehat{\h j}_i=0 .
		\label{eq:static_ohmslaw}
	\end{equation}
\end{subequations}
The complete expression when we input the ansatz \eqref{eq:perturb_method_ansatz} in the momentum equation \eqref{eq:gov_momentum} is given as
\begin{align}
	\begin{bmatrix}
		C\bigl(\ii i\overbar\omega + \alpha - \overbar\lambda \bigr)\rho \widehat{\h{u}} \\
		+ \\
		C'\bigl(\ii i\overbar\omega + \alpha - \overbar\lambda \bigr)\rho \widehat{\h{u}}'
	\end{bmatrix}
	+ 
	\begin{bmatrix}
		(\ii i\overbar\omega + \alpha - \overbar\lambda)\widetilde{\h{u}} + \bm{\nabla}\widetilde{p} \\
		+ \\
		(\ii i\overbar\omega + \alpha - \overbar\lambda)\widetilde{\h{u}}' + \bm{\nabla}\widetilde{p}'
	\end{bmatrix}
	=
	\begin{bmatrix}
		C \Bigl( \:\widehat{\h{j}} \times B_z\bm{e}_z\Bigr) \\
		+ \\
		C' \Bigl( \:\widehat{\h{j}}' \times B_z\bm{e}_z\Bigr)
	\end{bmatrix}. 
	\label{eq:pre_fredholm_step}
	\intertext{On expanding $\overbar\omega$ and $\overbar\lambda$ on the LHS of \eqref{eq:pre_fredholm_step}, we get the following:}
	\begin{bmatrix}
		C\bigl(\ii i\omega - \ii i\delta\omega + \alpha - \lambda + \delta\lambda \bigr)\rho \widehat{\h{u}} \\
		+ \\
		C'\bigl(\ii i\omega - \ii i\delta\omega + \alpha - \lambda + \delta\lambda \bigr)\rho \widehat{\h{u}}'
	\end{bmatrix}
	+ 
	\begin{bmatrix}
		(\ii i\omega - \ii i\delta\omega + \alpha - \lambda + \delta\lambda)\widetilde{\h{u}} + \bm{\nabla}\widetilde{p} \\
		+ \\
		(\ii i\omega - \ii i\delta\omega + \alpha - \lambda + \delta\lambda)\widetilde{\h{u}}' + \bm{\nabla}\widetilde{p}'
	\end{bmatrix}.
	\label{eq:pre_fredholm_step2}
\end{align}
We now make certain leading order approximations  $\omega\widetilde{\h u}_i \gg \delta\omega \widetilde{\h u_i}$ and $\lambda\widetilde{\h u}_i \gg \delta\lambda \widetilde{\h u_i}$, to simplify \eqref{eq:pre_fredholm_step2} as
\begin{align}
	\begin{bmatrix}
		C\bigl(-[\ii{i} \delta\omega - \delta\lambda] + \alpha \bigr)\rho \widehat{\h{u}} \\
		+ \\
		C'\bigl([\ii{i} \delta\omega  - \delta\lambda] + \alpha\bigr)\rho \widehat{\h{u}}'
	\end{bmatrix}
	+ 
	\begin{bmatrix}
		(\ii{i}\omega - \lambda)\widetilde{\h{u}} + \bm{\nabla}\widetilde{p} \\
		+ \\
		(\ii{i}\omega' - \lambda')\widetilde{\h{u}}' + \bm{\nabla}\widetilde{p}'
	\end{bmatrix}
	=
	\begin{bmatrix}
		C \Bigl( \:\widehat{\h{j}} \times B_z\bm{e}_z\Bigr) \\
		+ \\
		C' \Bigl( \:\widehat{\h{j}}' \times B_z\bm{e}_z\Bigr)
	\end{bmatrix} ,
	\\
	\Longrightarrow
	\begin{bmatrix}
		(\ii{i}\omega - \lambda)\rho\widetilde{\h{u}} + 
		\bm{\nabla}\widetilde{p} \\
		+ \\
		(\ii{i}\omega' - \lambda')\rho\widetilde{\h{u}}' + 
		\bm{\nabla}\widetilde{p}'
	\end{bmatrix}
	=
	\begin{bmatrix}
		C \Bigl( \: \widehat{\h{j}} \times B_z\bm{e}_z
		+
		\Bigl( [\ii{i} \delta\omega - \delta\lambda] - \alpha \Bigr)\rho \widehat{\h{u}}\Bigr) \\
		+ \\
		C' \Bigl( \:\widehat{\h{j}}' \times B_z\bm{e}_z
		+
		\Bigl( -[\ii{i} \delta\omega - \delta\lambda] - \alpha \Bigr)\rho \widehat{\h{u}}'
		\Bigr)
	\end{bmatrix} .
	\label{eq:wavepack_first}
\end{align}
Applying the Fredholm alternative to the equations over the two fluid regions as follows:
\begin{equation}
\Sigma_{i=1,2} \int_{V_i} \left[ \widehat{\h u}^*_i \cdot \ii{\eqref{eq:wavepack_first}} + \widehat p^*_i(\bm\nabla \cdot \widetilde{\h u}_i) \right] \ii dV,
\end{equation}
we can write the inner product (for quantities pertaining to wave mode $(m,n)$) as
\begin{equation}
	\begin{array}{rl}
		&
		\sum_{i=1,2} \int_{V_i} \Bigl[ C \widehat{\h u}_i^{*} \cdot 
		\Bigl( 
		[\ii{i} \omega - \lambda]\rho\widetilde{\h u}_i 
		+  \bm{\nabla} \widetilde{p}_i  
		\Bigr) 
		+
		C\widehat{p}_i^{\:*} ( \bm{\nabla} \cdot \widetilde{\h u}_i)
		\Bigr] ,
		\\
		& = \sum_{i=1,2}\int_{V_i} C \widehat{\h u}_i^{*} \cdot 
		\biggl[ C  \Bigl( \: \widehat{\h{j}}_i \times B_z\bm{e}_z +
		\bigl( [\ii{i} \delta\omega - \delta\lambda] - \alpha \bigr)\rho \widehat{\h{u}}_i \Bigr) \biggr] .
	\end{array}  \label{eq:fredholm_step1}
\end{equation}
The left-hand side of the equation \eqref{eq:fredholm_step1} can be expanded to
\begin{align}
	\int_{V_i} C \widehat{\h u}_i^{*} (\bm\nabla \widetilde{p}_i)
	=
	-C \int_{V_i}(\bm{\nabla}\cdot\widehat{\h u}_i^{*})\: \widehat{p}_i + 
	C \oint_{S} \widehat{\h u}_i^{*} \widetilde{p}_i\cdot \h n \:\ii dS,
	\intertext{and}
	\int_{V_i} C  \widehat{p}_i^{\:*}(\bm{\nabla}\cdot \widetilde{\h u}_i )
	=
	C \int_{V_i} -\bm{\nabla}\widehat{p}_i^{\:*}\widetilde{\h u}_i + 
	C \oint_{S} \widehat{p}_i^{\:*}\widetilde{\h u}_i\cdot \h n \:\ii dS,   
	\intertext{where $\widetilde{\h u}\cdot \h n$ can be written as $\widetilde{\h u}_z |_{z=0}$. Also we can write}
	\partial_t \eta = \widetilde{\h u}_z |_{z=0}
	\Longrightarrow \ii i\omega_{11}\widetilde{\eta} + C \bigl( \alpha \widehat{\eta}
	- [\ii i\delta\omega -\delta\lambda] \widehat\eta \:\bigr).
\end{align}
Therefore, from the leading order balance equation \eqref{eq:lead_balance_1} and \eqref{eq:lead_balance_2}, we deduce
\begin{equation}
	\begin{array}{c}
		\int_{V_i} C \Bigl[
		\widetilde{\h u}_i 
		\cancelto{0}{ \bigl( [\ii{i} \omega - \lambda]\rho \widehat{\h u}_i^{*}
			- \bm{\nabla}\widehat{p}_i^{\:*} \bigr)}
		-\cancelto{0}{(\bm{\nabla}\cdot \widehat{\h u}_i^{*})}\quad \widehat{p}_i \Bigr]
		+ C  \oint_{S} 
		(\widehat{p}_i^{\:*}\widetilde{\h u}_i 
		+ \widetilde{p}_i\widehat{\h u}_i^{*} ) 
		\cdot \h n \: \ii dS     ,
		\\
		\Longrightarrow
		\bigl( \alpha - [\ii i\delta\omega - \delta\lambda] \bigr) C^2
		\underbrace{\oint_{S} 
			(\widehat{p}_2^{\:*} - \widehat{p}_1^{\:*})
			|_{z=0}\:\widehat{\eta} \: \ii dS }_{K_{\eta\eta}} .
	\end{array}
	\label{eq:mode1_alpha-C-K}
\end{equation}
From the boundary condition \eqref{eq:int_bc2}, the surface integral of the pressure jump gives
\begin{equation}
	\begin{array}{rcl}
		\oint_{S} 
		(\widehat{p}_2^{\:*} - \widehat{p}_1^{\:*})
		|_{z=0}\widehat{\eta} \: \ii dS 
		& = & 
		\int_{\mathcal{S}} \Bigl( -\gamma_{\ii{int}} |\bm{\nabla}\widehat{\eta}|^2 
		+ (\rho_2 - \rho_1)g |\widehat{\eta}|^2 \Bigr)\Bigl|_{z=0} \ii{d}S
		= K_{\eta\eta} .
	\end{array} 
	\label{eq:mode1_K11-eq}
\end{equation}
By substituting equation \eqref{eq:mode1_K11-eq} in \eqref{eq:mode1_alpha-C-K}, we obtain
\begin{equation}
	\begin{array}{rl}
		\forall \:(m, n), &
		C  \oint_{S} (\widehat{p}_i^{\:*}\widetilde{\h u}_i 
		+ \widetilde{p}_i\widehat{\h u}_i^{*} ) \cdot \h n \: \ii dS
		\rightarrow  \bigl( \alpha - [\ii i\delta\omega - \delta\lambda] \bigr) C^2 K_{\eta\eta} ,\\[5pt]
		\forall \:(m', n'), &
		C'  \oint_{S} (\widehat{p}'^{*}_i\widetilde{\h u}'_i 
		+ \widetilde{p}'_i\widehat{\h u}'^{*}_i ) \cdot \h n \: \ii dS
		\rightarrow  \bigl( \alpha + [\ii i\delta\omega - \delta\lambda] \bigr) {C'}^{2} K_{\eta'\!\eta'} .
	\end{array}
	\label{eq:alpha-C-K_all}
\end{equation}
On the right-hand side of the equation \eqref{eq:fredholm_step1}, we can write the non-Lorentz force terms as
\begin{equation}
	\int_{V_i} C \widehat{\h u}_i^{*} \cdot 
	\biggl[ C \bigl( [\ii{i} \delta\omega - \delta\lambda] - \alpha \bigr)\rho \widehat{\h{u}}_i \biggr]
	=
	C^2 \bigl( [\ii i \delta\omega - \delta\lambda] -\alpha \bigr) 
	\int_{V_i} \rho_i |\widehat{\h u_i}|^2 \ii dV  .
	\label{eq:fredholm_rhs}
\end{equation}
Integration by parts on \eqref{eq:fredholm_rhs} considering the potential wave solution of $\widehat p \;$  \eqref{eq:pot_wave_soln} gives
\begin{align}
	\nn
	\sum_{i=1,2} \int_{V_i} \rho_i |\widehat{\h u_i}|^2 \ii dV
	&= \rho_1\oint_{\delta V_1} 
	\widehat{\phi}^{*}_1 \bm{\nabla}\widehat{\phi}_1 \cdot \h{n}_1 \ii{d}S
	+ \rho_2\oint_{\delta V_2} 
	\widehat{\phi}^{*}_2 \bm{\nabla}\widehat{\phi}_2 \cdot \h{n}_2 \ii{d}S ,
	\\
	\nn
	& = \oint_{\mathcal{S}} \Bigl( -\ii{i}\rho_1\overbar\omega \widehat{\phi}^{*}_1 
	+ \ii{i}\rho_2\overbar\omega \widehat{\phi}^{*}_2 \Bigr) \biggl|_{z=0} \widehat{\eta}
	\ii{d}S ,
	\\
	&=  \oint_{\mathcal{S}} \bigl( \widehat{p}^{*}_2 - \widehat{p}^{*}_1 \bigr)\Bigl|_{z=0} \widehat{\eta} \ii{d}S
	\Longrightarrow K_{\eta\eta} .
	\label{eq:ke-mode1_K11-eq}
\end{align}
The equations \eqref{eq:alpha-C-K_all}, \eqref{eq:fredholm_rhs}, \eqref{eq:A_11} and \eqref{eq:A_22}) gives the matrix equations given below:
\begin{align}
		&\bigl( \alpha - [\ii i\delta\omega - \delta\lambda] \bigr) C^2  K_{\eta\eta} = C^2 I_{\eta\eta} + CC' I_{\eta\eta'} + C^2\bigl ( [\ii{i} \delta\omega - \delta\lambda] - \alpha \bigr) K_{\eta\eta} , 
		\label{eq:balance_eq-post_fred_alt1}
		\\[5pt]
		&\bigl( \alpha + [\ii i\delta\omega - \delta\lambda] \bigr) {C'}^2  K_{\eta'\!\eta'} = CC' I_{\eta'\!\eta} + C'^2 I_{\eta'\!\eta'} + {C'}^2 \bigl( -[\ii{i} \delta\omega - \delta\lambda] - \alpha \bigr) K_{\eta'\!\eta'}  .
	\label{eq:balance_eq-post_fred_alt2}
\end{align}
Solving for $\alpha$ in the above equation results in \eqref{eq:main_gr}.

\subsection{Justification of the ansatz}
\label{appA_1}

The ansatz \eqref{eq:perturb_method_ansatz} when solved using the Fredholm alternative gives us the system of equations \eqref{eq:balance_eq-post_fred_alt1} and \eqref{eq:balance_eq-post_fred_alt2}. In matrix form it can be presented as
\begin{equation}
	\begin{bmatrix}
		\dfrac{I_{\eta\eta}}{2K_{\eta\eta}} + \ii{i}\delta\omega - \delta\lambda -\alpha & \dfrac{I_{\eta\eta'}}{2K_{\eta\eta}} \\
		\\
		\dfrac{I_{\eta'\!\eta}}{2K_{\eta'\!\eta'}} & \dfrac{I_{\eta'\!\eta'}}{2K_{\eta'\!\eta'}} - \ii{i}\delta\omega + \delta\lambda -\alpha
	\end{bmatrix}
	\begin{bmatrix}
		C \\
		\\
		C'
	\end{bmatrix}
	= 0 .
\end{equation}
\label{eq:alpha_matrix_2}
Ignoring the magnetohydrodynamic effects (Lorentz force terms), the determinant of the matrix \eqref{eq:alpha_matrix_2} is
\begin{align}
	&\delta\omega^2 + 2\ii i \delta\omega\delta\lambda - \delta\lambda^2 + \alpha^2 = 0, &&
	\\
	& \alpha^2 = \bigl( \ii i\delta\omega - \delta\lambda \bigr)^2, 
	\quad 
	\therefore \alpha = \pm\bigl(\ii i\delta\omega - \delta\lambda \bigr). && 
	\label{eq:alpha_solution}
\end{align}
Inserting \eqref{eq:alpha_solution} in \eqref{eq:perturb_method_ansatz}, the leading order perturbative terms can be written as
\begin{align}
	C 
	\left[ 
	\widehat{\h{u}}_{i}, \widehat{p}_{i}, \widehat{\eta}, \widehat{\h{j}}_{i} \right] \ii e^{(\ii i \omega -\lambda)t}
	+ 
	C' 
	\left[ 
	\widehat{\h{u}}'_{i}, \widehat{p}'_{i}, \widehat{\eta}', \widehat{\h{j}}'_{i} \right] \ii e^{(\ii i \omega' -\lambda')t}.
	\label{eq:ansatz_proof}
\end{align}
The expression \eqref{eq:ansatz_proof} agrees with the proposition \eqref{eq:wave ansatz}.

\section{Solving the mechanical energy balance equation for viscous decay} \label{appB}
The energy balance equation is as follows: 
\begin{align}
	\dv{}{t} \bigl(E_c + E_p \bigr) &=
	\dv{}{t}
	\Biggl( \sum_{i=1,2} \frac{1}{2} \int_{V_i} \rho_i |\h u_i|^2 \ii{d}V ]\Biggr)
	+
	\dv{}{t}
	\Biggl( \int_S \biggl[ \frac{1}{2} (\rho_2 -\rho_1)g\eta^2 + \gamma\sqrt{1+|\nabla\eta|^2}
	\biggr]
	\ii{d}S  ,
	\label{eq:ke-pe}
	\\
	\mathcal{D} &=
	\sum_{i=1,2} 2\rho_i\nu_i \int_{V_i} \bm{\varepsilon}_i \bm{:} \bm{\varepsilon}_i \; \ii{d}V  .
	\label{eq:strain-doubledot}
\end{align}
 We write the velocity and the elevation profile of a decaying rotating wave given in \eqref{eq:ke-pe}, ignoring the surface tension terms as
\begin{equation}
	\begin{array}{rl}
		\h u_i \hspace{-5pt}&\approx [\widehat{\h u}_i \:\ii e^{\ii i \omega t} + \widehat{\h u}_i \: \ii e^{-\ii i \omega t}]\ii e ^{-\lambda_{\ii v} t}  
		= \widehat{\h u}_i \:\cos[\omega] \: \ii{e}^{-\lambda_{\ii v} t}
		\\[5pt]
		\eta \hspace{-5pt}&\approx [\widehat{\eta} \:\ii e^{\ii i \omega t} - \widehat{\eta} \: \ii e^{-\ii i \omega  t}]\ii e ^{-\lambda_{\ii v} t}  
		= \widehat{\eta} \: \sin[\omega] \: \ii{e}^{-\lambda_{\ii v} t}    
	\end{array} \Biggl\} .
\end{equation}
In a standing wave, the total energy always equals the sum of kinetic and potential energy at a phase difference of $\pi/2$. From \eqref{eq:mode1_K11-eq} and \eqref{eq:ke-mode1_K11-eq}, we write
\begin{align}
	\nn   \dv{}{t} \bigl( E_{\ii{kin}} + E_{\ii{pot}} \bigr) &= -2\lambda_{\ii v} \biggl( \frac{1}{2} K \: \cos^2[\omega] \: \ii{e}^{-2\lambda_{\ii v} t}
	+ \frac{1}{2} K \: \sin^2[\omega] \: \ii{e}^{-2\lambda_{\ii v} t} \biggr) ,
	\\
	&= -\lambda_{\ii v} \: K \: \ii{e}^{-2\lambda_{\ii v} t},
	\qquad \bigl\{ K = K_{\eta\eta} \in (m,n) \lor K_{\eta'\!\eta'} \in (m',n') .
\end{align}
In \eqref{eq:strain-doubledot}, the strain tensor is given by
$\bm{\varepsilon}_i = \dfrac{1}{2} \Bigl( \bm{\nabla} \h u_i + \bm{\nabla} \h u_i^\ii{T} \Bigr) $. Introducing the Lamb's dissipation function as
\begin{align}
	\mathcal{D} = 
	\sum_{i=1,2} 2\rho_i\nu_i \int_{V_i} \bm{\varepsilon}_i \bm{:} \bm{\varepsilon}_i \; \ii{d}V \Longrightarrow
	\sum_{i=1,2}\int_{V_i} \rho_i\Phi_\mathcal{D} \ii{d}V ,
	\\
	\phi_\mathcal{D} = \nu_i 
	\biggl[ \frac{1}{2}\Bigl(\nabla \h u_i^\ii{T} - \nabla \h u_i
	\Bigr)^2\biggr]
	+ 2\nu_i \nabla \h u_i \nabla \h u_i^\ii{T} .
\end{align}
On simplifying $\phi_\mathcal{D}$, we obtain the dissipation integral as follows
\begin{equation}
	\mathcal{D} = \sum_{i=1,2} 
	\rho_i\nu_i \Biggl(
	\underbrace{\int_{V_i} \bigl( \nabla \times \h u_i \bigr)^2\ii{d}V}_{\RN{1}}
	-\underbrace{\int_{\delta V_i} 2\bigl[ \h u_i \times (\nabla \times \h u_i) \bigr].\h n \ii{d}S}_{\RN{2}}
	+  \underbrace{\int_{\delta V_i} \nabla|\h u_i|^2 \cdot \h n \ii{d}S}_{\RN 3}
	\Biggr) .
	\label{eq:dissip}
\end{equation}
Integral $\RN{1}$ emphasises the energy dissipation due to rotational flows and contributes largely to the dissipation in the system. The second integral is of higher order and considers both the bulk and rotational flows. Integral $\RN{3}$ is related to the bulk irrotational stresses (relevant in free surface waves) and takes into account only the flow potential. We solve integral $\RN{1}$ in \eqref{eq:dissip} taking velocity relations \eqref{eq:u_wall}, \eqref{eq:u_int1} and \eqref{eq:u_int2} to obtain $\mathcal{D}_{\ii{wall}}$ and $\mathcal{D}_{\ii{int}}$ and integral $\RN{2}$ and $\RN{3}$ together yields $\mathcal{D}_{\ii{irr}}$. The solutions for the respective dissipation terms are as follows:
\begin{flalign}
	\nn \mathcal{D}_{\ii{wall}} = &\sum_{i=1,2} \Biggl[
	\xi_{m\times n}\Biggl( \frac{\rho_i\omega^2 \sqrt{\omega\nu_i}}{\sqrt{2} \: k^2} \Biggr)
	\Biggl(
	\frac{L_xL_y \; k^2}{2\sinh^2[kh_i]}
	\Biggr) &&  \\
	\nn + 
	& \; \xi_n\Biggl( 
	\frac{\rho_i\omega^2 \sqrt{\omega\nu_i}}{\sqrt{2} \: k^2}
	\Biggr)
	\Biggl(
	\frac{h_i(n^2\pi^2  - k^2L_y^2)}{L_y\:\sinh^2[kh_i]}
	+ \frac{(n^2\pi^2 + k^2L_y^2)}{L_y \: k\:\tanh[kh_i]}
	\Biggr) && \\
	+
	& \; \xi_m\Biggl( 
	\frac{\rho_i\omega^2 \sqrt{\omega\nu_i}}{\sqrt{2} \: k^2}
	\Biggr)
	\Biggl(
	\frac{h_i(m^2\pi^2  - k^2L_x^2)}{L_x\:\sinh^2[kh_i]}
	+ \frac{(m^2\pi^2 + k^2L_x^2)}{L_x \: k\:\tanh[kh_i]}
	\Biggr) \Biggr] ,&&
\end{flalign}
\begin{flalign}
	\mathcal{D}_{\ii{int}} = \xi_{m\times n}
	\dfrac{\omega_{11}^2\sqrt{\omega_{11}}}{2\sqrt{2}}
	\:\frac{\bigl( \coth[K_{\eta\eta}h_1] + \coth[K_{\eta\eta}h_2] \bigr)^2}
	{ \dfrac{1}{L_xL_y}\:\biggl( \dfrac{1}{\rho_1\sqrt{\nu_1}} + \dfrac{1}{\rho_2\sqrt{\nu_2}} \biggr)} ,
	&&
\end{flalign}
\begin{flalign}
	\nn
	&\mathcal{D}_{\ii{ext}} = \frac{2 \xi_{m\times n} \, (\rho_2-\rho_1)gk^2 \left( \dfrac{1}{\rho_1\sqrt{\nu_1}} + \dfrac{1}{\rho_2\sqrt{\nu_2}} \right)^{-1}}{\rho_1\coth{[kh_1]} +
		\rho_2\coth{[kh_2]}} \Biggl(\biggl[ 
	\dfrac{\rho_2\sqrt{\nu_1\nu_2} + \rho_1\nu_1}{\rho_2\sqrt{\nu_2} \tanh{[kh_1]}} +
	\dfrac{\rho_1\sqrt{\nu_1\nu_2} + \rho_2\nu_2}{\rho_1\sqrt{\nu_1} \tanh{[kh_2]}} &&
	\\[2.5pt]
	&  -\bigl(\coth{[kh_1]}+\coth{[kh_2]}\bigr)\bigl(\sqrt{\nu_1} + \sqrt{\nu_2}\bigr) 
	\biggr]\Biggr) ,&&
\end{flalign}
Here, $ \qquad
\xi_r \left\{ \begin{array}{cc}
	1/4, &  r>0\\
	1/2, & r=0
\end{array} \right. .
$
\begin{equation}
	\therefore -\lambda K \ii e^{-2\lambda_{\ii v} t} = -\mathcal{D} \; \ii{e}^{-2\lambda_{\ii v} t} \Longrightarrow \lambda_{\ii v} = \frac{\mathcal{D}}{K} .
\end{equation}

\section{The quasi-static correction of the electric potential}
\label{appC}

We use separation of variables to solve for $\nabla^2\widehat{\Psi}_2=0 \; \in (m,n)$, provided $m+n\neq 0$. Assessing the boundary conditions (\ref{eq:mag_bc1} -- \ref{eq:mag_bc4}) and \eqref{eq:mag_bc_final}, we can deduce that the solution presents itself as hyperbolic ordinary differential equations (ODEs) in the x and y-plane, elliptic ODEs in the z-plane. Proceeding as is will make it difficult to obtain the eigenvalue coefficients needed to solve the quasi-static correction of the electric potential. To simplify this problem, we slice $\widehat\Psi_{mn}$ as
\begin{equation}
	\widehat{\Psi}_{mn} \; \in (m,n)
	= \widehat{\Psi}_{m}  \;\in (m,0) + \widehat{\Psi}_{n} \; \in (0,n) .
	\label{eq:mag_potential_superposed} 
\end{equation}
From principle of superposition, we get 
\begin{align}
	\Bigl[ \underbrace{\nabla^2\widehat{\Psi}_m  = 0 }_{\RN{1}} \Bigr]
	\,+\,
	\Bigl[\underbrace{\nabla^2\widehat{\Psi}_n  = 0}_{\RN{2}} \Bigr]
	=
	\Bigl[
	\nabla^2\widehat{\Psi}_{mn} = 0 \Bigr] .
	\label{eq:mag_laplace_superposed}
\end{align}
The solution to term $\RN{1}$ takes the form of (Refer \ref{eq:xi_termX} for $\xi_x$),
\begin{equation}
	\left.
	\begin{array}{rll}
		\widehat\Psi_m & \hspace{-8pt} = \hspace{-8pt}&
		\sum\limits_{i=0}^{\infty}\sum\limits_{j=0}^{\infty}
		A_{i\!j} \cosh{[\xi_x x]} 
		\cos{\biggl[ \dfrac{i\pi}{L_y} \Bigl(y+\dfrac{L_y}{2}\Bigr) \biggr]}
		\cos{\biggl[ \dfrac{j\pi}{h_2} \bigl(z+h_2\bigr) \biggr]}
		,\quad (m \in \mathbb{Z}^+_{\ii{odd}} )
		\\[10pt]
		& \hspace{-8pt} = \hspace{-8pt}&
		\sum\limits_{i=0}^{\infty}\sum\limits_{j=0}^{\infty}
		A_{i\!j} \sinh{[\xi_x x]} 
		\cos{\biggl[ \dfrac{i\pi}{L_y} \Bigl(y+\dfrac{L_y}{2}\Bigr) \biggr]}
		\cos{\biggl[ \dfrac{j\pi}{h_2} \bigl(z+h_2\bigr) \biggr]}
		,\quad (m \in \mathbb{Z}^+_{\ii{even}} )
	\end{array} 
	\right\} .
\end{equation}
The potential $\widehat\Psi_n$ too takes a similar form with the hyperbolic terms in the y-plane  with coefficient $B_{ij}$. The coefficient $A_{i\!j}$ can\underline{} be obtained from applying the Neumann boundary conditions at the lateral walls (see \ref{eq:mag_bc1}). To remove the infinite sum, we apply Fourier's orthogonality condition on either sides considering the superposition \ref{eq:mag_potential_superposed}. Let
\begin{equation}
	\mathcal{F}(y,z) = \cos{\biggl[ \dfrac{i'\pi}{L_y} \Bigl(y'+\dfrac{L_y}{2}\Bigr) \biggr]}
	\cos{\biggl[ \dfrac{j'\pi}{h_2} \bigl(z'+h_2\bigr) \biggr]} .
\end{equation}
At $x=\pm 0.5 L_x$,
\begin{align}
	\int_{-h_2}^{0}\int_{-0.5L_y}^{0.5L_y} 
	\pdv{\widehat\phi_2 B_z }{y}
	\mathcal{F}(y,z)
	=
	\int_{-h_2}^{0}\int_{-0.5L_y}^{0.5L_y}
	\Biggl(
	\dpdv{\widehat{\Psi}_m}{x}
	\mathcal{F}(y,z)
	+
	\underbrace{
		\pdv{\widehat{\Psi}_n}{x}
		\mathcal{F}(y,z)
	}_{0}
	\Biggr) .
\end{align}  
On the right hand side, the terms are non-zero when $i=i'$ and $j=j'$. Repeating this process taking the condition \eqref{eq:mag_bc2} gives us $B_{i\!j}$, as presented in equation \eqref{eq:mag_Bij}. Equation \eqref{eq:mag_potential_superposed} for odd and even wave modes takes the general form 
\begin{flalign}
	\nn
	\widehat\Psi_{mn} &=  \sum_{i=0}^{\infty} \sum_{j=0}^{\infty} A_{i\!j}
	\biggl( \frac{1 + (-1)^m}{2\csch{[\xi_x x]}}  +  \frac{1 + (-1)^{m+1}}{2 \sech{[\xi_x x]}}  \Biggr)
	\cos\left[\frac{i\pi}{L_y}\left(y+\frac{L_y}{2}\right)\right]
	\cos\left[\frac{j\pi}{h_2}(z+h_2)\right] &&
	\\
	& + \sum_{i=0}^{\infty} \sum_{j=0}^{\infty} B_{i\!j}
	\biggl( \frac{1 + (-1)^n}{2\csch{[\xi_y y]}}  +  \frac{1 + (-1)^{n+1}}{2 \sech{[\xi_y y]}}  \Biggr) 
	\cos\left[\frac{i\pi}{L_x}\left(x+\frac{L_x}{2}\right)\right]
	\cos\left[\frac{j\pi}{h_2}(z+h_2)\right] . &&
	\label{eq:psi_incomplete}
\end{flalign}
The above relation for the correction satisfies the boundary conditions \eqref{eq:mag_bc1} and \eqref{eq:mag_bc2} when the wave is symmetric i.e. $(m \land n) \in \mathbb{Z}^+_{\ii{even}}$. Note that $A_{i\!j}$ fails to provide a real value when ($i=0,\, j=0$). The formula in its current form gives a constant difference in non-symmetric cases when $(m \lor n) \in \mathbb{Z}^+_{\ii{odd}}$. This can be quantified by linear terms which offset the solution by
\begin{equation}
	\pm A_0 = \pdv{\widehat\phi_2 B_z}{y} -  \pdv{\widehat\Psi_2}{x} 
	\biggl|_{n_{\ii{odd}}}, \qquad
	\pm B_0 = -\pdv{\widehat\phi_2 B_z}{x} -  \pdv{\widehat\Psi_2}{y} 
	\biggl|_{m_{\ii{odd}}}, \qquad \left\{ 
	\begin{array}{l}
		i=0 \\
		j=0
	\end{array}
	\right. .
\end{equation}
The terms $A_0$ or $B_0$ only appear in the solution when the wave modes $n$ or $m$ are odd respectively and reduce to 0 otherwise. Therefore we add to general equation \eqref{eq:psi_incomplete}
\begin{equation}
	A_0x - \frac{
		A_0\cos{[n\pi]}(1-\cos{[m\pi]}) \biggl(x-\dfrac{L_x}{2}\biggr)^2}{2L_x} 
	+B_0x - \frac{
		B_0\cos{[m\pi]}(1-\cos{[n\pi]}) \biggl(y-\dfrac{L_y}{2}\biggr)^2 }{2L_y} .
	\label{eq:linear_term_psi}
\end{equation}
The second and fourth terms change the sign convention of the coefficient, compensating for the asymmetrical boundary conditions at the opposing bounding walls in the x- or y-plane. This term solves the Laplace equation as it reduces to 
\begin{equation}
	\frac{A_0 \cos{[n\pi]}(1-\cos(m\pi))}{2L_x} = \frac{B_0 \cos{[m\pi]}(1-\cos(n\pi))}{2L_y} .
\end{equation}
Equation \eqref{eq:linear_term_psi} and \eqref{eq:psi_incomplete} together give \eqref{eq:mag_potential}.

\section{Determining the critical wave mode and the stability onset}
\label{appD}

\subsection{Choosing the lateral wave numbers}
In our formulation it is not directly possible to explicitly obtain the most unstable wave mode which may either refer to the onset of instability ( $\beta_{\rm crit}$) or the largest possible growth rate ($\ii{max}\; \alpha_{\ii{MPR}}$).  We begin with choosing arbitrary, initial wave numbers and calculate the growth rate $\alpha_{\ii{MPR}}$ and the critical instability onset $\beta_{\ii{crit}}$, and iteratively find the wave modes which satisfy the former condition. Though it might seem that there are infinite permutations and combinations of numbers to choose from, there are certain conditions that reduce the sample space. They are:
\begin{enumerate}
	\item  The mode selection function $\Theta > 0$ only if $m+n = \mathbb{N}_{\ii{odd}}$, 
	$m'+n' = \mathbb{N}_{\ii{odd}}$
	\item  Large scale wave modes considerably increase damping and the instability threshold. Therefore wave numbers up to $(m,n), (m',n') \leq 10$ can be taken into consideration
\end{enumerate} 

\subsection{Calculating the stability onset}
\begin{enumerate}
	\item \hspace{1mm} Calculate wave number $k, k'$.
	\[
	k=\pi\sqrt{\frac{m^2}{L_x^2} + \frac{n^2}{L_y^2}}, \qquad
	k'=\pi\sqrt{\frac{m'^2}{L_x^2} + \frac{n'^2}{L_y^2}} .
	\]
	\item \hspace{1mm} Calculate angular frequencies $\omega, \omega'$.
	\[
	\omega = \pm \sqrt{
		\dfrac{(\rho_2-\rho_1)gk + \gamma_{\scaleto{\ii{int}}{5pt}}{k}^3}
		{\rho_1 \coth[k\: h_1]
			+ \rho_2 \coth[k\: h_2]}
	}, \qquad
	\omega' = \pm \sqrt{
		\dfrac{(\rho_2-\rho_1)gk' + \gamma_{\scaleto{\ii{int}}{5pt}}{k'}^3}
		{\rho_1 \coth[k'\: h_1]
			+ \rho_2 \coth[k'\: h_2]} } .
	\]
	\item \hspace{1mm} Note: $\overbar\omega = \dfrac{\omega+\omega'}{2}$, $\delta\omega = \dfrac{\omega-\omega'}{2}$ and $\delta_r = \left\{ \begin{array}{cc}
		1, & r=0\\
		2, & r>0
	\end{array} \right. \!\!. $
	
	\item \hspace{1mm} Determine the average viscous damping, $\overbar\lambda_{\ii v} = \dfrac{ \Bigl(\lambda_{\ii{wall}}+\lambda_{\ii{int}}+\lambda_{\ii{irr}}\Bigr)  
		+ \Bigl(\lambda'_{\ii{wall}}+\lambda'_{\ii{int}}+\lambda'_{\ii{ext}} \Bigr) }
	{2} $,
	\begin{flalign*}
		&\lambda_{\ii{wall}}  
		=\sum_{i=1,2}
		\frac{ \dfrac{1}{\sqrt{2}}
			\dfrac{\rho_i \sqrt{\omega\nu_i}}{ kL_xL_y }
		}
		{ 
			\left[
			\dfrac{\rho_1}{\tanh[kh_1]} + \dfrac{\rho_2}{\tanh[kh_2]}
			\right]}
		\Biggl( \biggl[
		\dfrac{L_xL_y \; k^2}{2\sinh^2[kh_i]}
		+
		\dfrac{ \delta_{m} h_i(n^2\pi^2  - k^2L_y^2)}{ 2 L_y\:\sinh^2[kh_i]}
		\\[5pt]
		& + \dfrac{ \delta_m (n^2\pi^2 + k^2L_y^2)}{ 2 L_y \: k\:\tanh[kh_i]}
		+
		\dfrac{\delta_n h_i(m^2\pi^2  - k^2L_x^2)}{ 2 L_x\:\sinh^2[kh_i]}
		+
		\dfrac{ \delta_n (m^2\pi^2 + k^2L_x^2)}{ 2 L_x \: k\:\tanh[kh_i]}
		\biggr] \Biggr) , &&
	\end{flalign*}
	\begin{flalign*}
		\lambda_{\ii{int}} &= \frac{k\sqrt{\omega}}{2\sqrt{2}}
		\:\frac{\bigl( \coth[kh_1] + \coth[kh_2] \bigr)^2}
		{ \biggl( \dfrac{1}{\rho_1\sqrt{\nu_1}} + \dfrac{1}{\rho_2\sqrt{\nu_2}} \biggr)
			\biggl[ \dfrac{\rho_1}{\tanh[kh_1]} + \dfrac{\rho_2}{\tanh[kh_2]} \biggr]
		} , &&
		\\[10pt]
		\lambda_{\ii{irr}} &= \frac{2k^2 \left( \dfrac{1}{\rho_1\sqrt{\nu_1}} + \dfrac{1}{\rho_2\sqrt{\nu_2}} \right)^{-1}}{\rho_1\coth{[kh_1]} +
			\rho_2\coth{[kh_2]}} \Biggl(\biggl[ 
		\dfrac{\rho_2\sqrt{\nu_1\nu_2} + \rho_1\nu_1}{\rho_2\sqrt{\nu_2} \tanh{[kh_1]}} +
		\dfrac{\rho_1\sqrt{\nu_1\nu_2} + \rho_2\nu_2}{\rho_1\sqrt{\nu_1} \tanh{[kh_2]}} 
		&&
		\\[2.5pt]
		&  -\bigl(\coth{[kh_1]}+\coth{[kh_2]}\bigr)\bigl(\sqrt{\nu_1} + \sqrt{\nu_2}\bigr) 
		\biggr]\Biggr) .&&
	\end{flalign*}
	\vspace*{-0.5cm}
	\item  Determine the average magnetic damping,
	\vspace*{-0.4cm}
	 $\overbar\lambda_{\ii m} = \dfrac{1}{2} \left( \dfrac{ \Bigl(\mathcal{Q}_{\ii{ind}}-\mathcal{Q}_{\ii{wall}}\Bigr)}{2K}
	+ \dfrac{\Bigl(\mathcal{Q}'_{\ii{ind}}-\mathcal{Q}'_{\ii{wall}}\Bigr)}{2K'} \right) $,
	
	\begin{flalign*}
		&\mathcal{Q}_{\ii{ind}} =
		\frac{\sigma_2\omega^2 B_z^2}{4\delta_{m\times n} k}
		\Bigl( \coth{[kh_2]} + kh_2\csch^2{[kh_2]} \Bigr) L_xL_y  ,&&
	\end{flalign*}
	\begin{flalign*}
		\nn
		&\mathcal{Q}_{\ii{wall}} = \frac{A_0\sigma_2\omega B_z L_x}{k^2
			\biggl(
			f_m[m-0.5] \bigl((-1)^n-(-1)^{2n}\bigr)\bigl(1-(-1)^m\bigr)^2\:
			+ f_m[0.5-m] \bigl(1-(-1)^n\bigr) (m\pi)^2 \biggr)^{-1}} 
		&&
		\\
		\nn
		& + \sum_{i=0}^{\infty} \sum_{j=0}^{\infty} 
		\dfrac{2A_{i\!j}\sigma_2 \omega B_z h_2^2
			\Bigl( (im\pi)^2 + (n L_x \xi_x)^2 \Bigr) 
			\bigl((-1)^{1+i+j+m+n} + (-1)^{j+m}\bigr)  }
		{(i^2-n^2)(m^2\pi^2 + L_x^2\xi_x^2) (k^2h_2^2 + j^2\pi^2)
			\left( \dfrac{1 + (-1)^m}{2\csch{\left[\xi_x \frac{L_x}{2}\right]}}  +  \dfrac{1 + (-1)^{m+1}}{2 \sech{\left[\xi_x \frac{L_x}{2}\right]}}  \right)^{-1}
		}
		&&
		\\
		\nn
		& - \frac{B_0\sigma_2\omega B_z L_y}{k^2
			\biggl( f_n[n-0.5]
			\bigl((-1)^m- (-1)^{2m}\bigr)\bigl(1-(-1)^n\bigr)^2\:
			+ f_n[0.5-n] \bigl( 1-(-1)^m \bigr)(n\pi)^2 \biggr)^{-1}
		} 
		\\
		& + \sum_{i=0}^{\infty} \sum_{j=0}^{\infty} 
		\frac{2B_{ij}\sigma_2 \omega B_zh_2^2
			\Bigl((i n\pi)^2 + (mL_y\xi_y)^2\Bigr) 
			\bigl((-1)^{i+j+m+n} + (-1)^{1+j+n}\bigr)}
		{(i^2-m^2)(n^2\pi^2 + \xi_y^2L_y^2) (k^2h_2^2 + j^2\pi^2)
			\left( \dfrac{1 + (-1)^n}{2\csch{\left[\xi_y \frac{L_y}{2}\right]}}  +  \dfrac{1 + (-1)^{n+1}}{2 \sech{\left[\xi_y \frac{L_y}{2}\right]}}  \right)^{-1}
		} .
		&&
	\end{flalign*}
	Where,
	\[
	f_q[r] = \left\{\begin{array}{cc}
		\dfrac{1}{(q\pi)^2}, & r>0 \\[10pt]
		0, & r<0
	\end{array}
	\right. .
	\]
	\begin{align*}
		\xi_x &= \sqrt{ \biggl(\frac{i\pi}{L_y}\biggr)^2 + \biggl(\frac{j\pi}{h_2}\biggr)^2} ,
		\quad
		\xi_y = \sqrt{ \biggl(\frac{i\pi}{L_x}\biggr)^2 + \biggl(\frac{j\pi}{h_2}\biggr)^2} ,
		\\[5pt]
		A_{0} &=  \frac{\omega B_z}{k^2} \; \frac{\bigl((-1)^m-(-1)^{m+n} \bigr)}{L_yh_2} ,
		\quad
		B_{0} =  \frac{-\omega B_z}{k^2} \; \frac{\bigl((-1)^n-(-1)^{m+n} \bigr)}{L_xh_2} ,
	\end{align*}
	\begin{align*}
		A_{i\!j} \! &= 
		\left\{
		\begin{array}{l}
			0, \quad \ii{for} \; i=j=0,
			\\
			\frac{2\delta_{i\times j} \; \omega B_z}{k^2 h_2} 
			\frac{n^2 \bigl( (-1)^{i+m+n} - (-1)^m \bigr)}
			{(i^2-n^2)\xi_x L_y}\:
			\frac{k^2h_2^2 (-1)^j}{k^2h_2^2 + j^2\pi^2}
			\biggl( \frac{1 + (-1)^m}{2\cosh{[\xi_x \frac{L_x}{2}]}}  +  \frac{1 + (-1)^{m+1}}{2 \sinh{[\xi_x \frac{L_x}{2}]} }  \biggr),  
			\quad \ii{else},
		\end{array} \right. 
		\\
		B_{i\!j} \! &= 
		\left\{
		\begin{array}{l}
			0, \quad \ii{for} \; i=j=0,
			\\
			\frac{-2\delta_{i\times j} \; \omega B_z}{k^2 h_2} \frac{m^2\bigl( (-1)^{i+m+n} - (-1)^n \bigr)}
			{(i^2-m^2)\xi_y L_x}\:
			\frac{k^2h_2^2 (-1)^j}{k^2h_2^2 + j^2\pi^2}
			\biggl( \frac{1 + (-1)^n}{2\cosh{[\xi_y \frac{L_y}{2}]}}  +  \frac{1 + (-1)^{n+1}}{2 \sinh{[\xi_y \frac{L_y}{2}]} }  \biggr) , 
			\;\: \ii{else},
		\end{array} \right .
	\end{align*}
	\item \hspace{1mm} The total damping in the system, $\overbar\lambda = \dfrac{
		\lambda_{\ii v} + \lambda'_{\ii v} + \lambda_{\ii m} + \lambda'_{\ii m}}{2}$ and $\delta\lambda = \dfrac{
		\lambda_{\ii v} - \lambda'_{\ii v} + \lambda_{\ii m} - \lambda'_{\ii m}}{2}$.
	\vspace{3pt}
	\item \hspace{1mm} The onset of instability occurs now can be calculated by
	\[
	\beta_{\ii{crit}} = \frac{L_x^2 L_y^2}{h_1h_2 \Theta}
	\sqrt{ \frac{\overbar\lambda^2 - \bigl[ \ii i{\delta \omega} -\delta\lambda \bigr]^2} {(\overbar\omega^2 - {\delta\omega}^2) \mathcal{T}\mathcal{T}'} }.
	\]
	\item \hspace{1mm} The growth rate
	\[
	\alpha_{\ii {MPR}} = \pm \; 
	\sqrt{ J^2{B_z}^2{\Theta}^2
		\; (\overbar\omega^2 - \delta\omega^2)
		\;\frac{\mathcal{T}\mathcal{T}'}{\mathcal{U}\mathcal{U}'}
		+ \bigl[ \ii i{\delta \omega} -\delta\lambda \bigr]^2
	} \;-\; \overbar\lambda ,	
	\]
	where
	\begin{flalign*}
		&\Theta = \pm \; \sqrt{\delta_{m\times m'}\;\delta_{n \times n'}} \; 
		\frac{(-1^{m+m'}-1)(-1^{n+n'}-1)(m^2{n'}^2 - {m'}^2n^2)}
		{(m^2 - {m'}^2)(n^2 - {n'}^2) },&&
		\\
		&\mathcal{T} = \frac{\Lambda}{k(k^2 - k'^2)}
		\Biggl[ 
		\frac{k}{\tanh[k'h_2]} - \frac{k'}{\tanh[kh_2]}
		+ \frac{k'\:\sech[k'h_1]}{\sinh[kh_1]}
		-\frac{k'}{\tanh[kh_1]}
		+ k\tanh[k'h_1] \Biggr] ,&&
		\\
		& \mathcal{T}' = \frac{\Lambda'}{k'({k'}^2-k^2)}
		\Biggl[ 
		\frac{k'}{\tanh[kh_2]} - \frac{k}{\tanh[k'h_2]}
		+ \frac{k\: \sech[kh_1]}{\sinh[k'h_1]}
		-\frac{k}{\tanh[k'h_1]}
		+ k'\tanh[kh_1] \Biggr]  ,&&
	\end{flalign*}
	\[
	\mathcal{U} = L_xL_y\Bigl(\gamma_{\ii{int}} \: k^2 + (\rho_2 - \rho_1)g \Bigr), \qquad
	\mathcal{U}' = L_xL_y\Bigl(\gamma_{\ii{int}}\: {k'}^2 + (\rho_2 - \rho_1)g \Bigr) ,
	\]
	\[
	\Lambda = \frac{(\sigma_1^{-1} - \sigma_2^{-1})}{\sigma_1^{-1} \:\tanh[kh_1] + \sigma_2^{-1} \:\coth[kh_2]}, \qquad 
	\Lambda' = \frac{(\sigma_1^{-1} - \sigma_2^{-1})}{\sigma_1^{-1} \:\tanh[k'h_1] + \sigma_2^{-1} \:\coth[k'h_2]} .
	\]
	\item \hspace{1mm} Repeat step (\Rn{1}) to obtain the lowest possible $\beta_{\ii{crit}}$ or the highest possible $\alpha_{\ii{MPR}}$ for all wave mode combinations. 
\end{enumerate}

\bibliographystyle{jfm}
\bibliography{jfm}

\begin{thebibliography}{45}
\expandafter\ifx\csname natexlab\endcsname\relax\def\natexlab#1{#1}\fi
\def\au#1{#1} \def\ed#1{#1} \def\yr#1{#1}\def\at#1{#1}\def\jt#1{\textit{#1}}
  \def\bt#1{#1}\def\bvol#1{\textbf{#1}} \def\vol#1{#1} \def\pg#1{#1}
  \def\publ#1{#1}\def\arxiv#1{#1}\def\org#1{#1}\def\st#1{\textit{#1}}

\bibitem[Bojarevics \& Evans(2015)]{bojarevics2015}
{\sc \au{Bojarevics, V.} \& \au{Evans, J.~W.}} \yr{2015}  \at{Mathematical
  {{Modelling}} of {{Hall-H{\'e}roult Pot Instability}} and {{Verification}} by
  {{Measurements}} of {{Anode Current Distribution}}}.  \bt{In {\em Light
  {{Metals}} 2015\/} (ed. \ed{Margaret Hyland})},  \pg{pp. 783--788}.
  \publ{Cham: Springer International Publishing}.

\bibitem[Bojarevics \& Pericleous(2006)]{Bojarevics2006}
{\sc \au{Bojarevics, V.} \& \au{Pericleous, K.}} \yr{2006}  \at{Comparison of
  {{MHD}} modles for aluminium reduction cells.}  \jt{Proc. TMS Light Met.}
  \pg{pp. 347--352}.

\bibitem[Bojarevics \& Romerio(1994)]{Bojarevics1994}
{\sc \au{Bojarevics, V.} \& \au{Romerio, M.~V.}} \yr{1994}  \at{Long waves
  instability of liquid metal-electrolyte interface in aluminium electrolysis
  cells: {{A}} generalization of {{Sele}}'s criterion}.  \jt{Eur. J. Mech. -
  B/Fluids}  \bvol{13}.

\bibitem[Borisov {\em et~al.\/}(2008)Borisov, Poslavsky \& Rudnev]{Borisov2008}
{\sc \au{Borisov, I.~D.}, \au{Poslavsky, S.~A.} \& \au{Rudnev, Y.~I.}}
  \yr{2008}  \at{{Wave processes in a two-layer system of immiscible conductive
  liquids}}.  \jt{Adv. Appl. Math. Mech.} ~(826),  \pg{165--184}, Available at:
  \url{ http://vestnik-math.univer.kharkov.ua/Vestnik-Khnu-826-2008-bor.pdf}.

\bibitem[Borisov {\em et~al.\/}(2010)Borisov, Poslavsky \& Rudnev]{Borisov2010}
{\sc \au{Borisov, I.~D.}, \au{Poslavsky, S.~A.} \& \au{Rudnev, Y.~I.}}
  \yr{2010}  \at{{Experimental study of wave processes in a two-layer system of
  immiscible current-carrying fluids.}}  \jt{Appl. Hydromechanics}
  \bvol{12}~(1),  \pg{3--10}, Available at:
  \url{http://dspace.nbuv.gov.ua/handle/123456789/87720}.

\bibitem[Davidson(2001)]{Davidson2001}
{\sc \au{Davidson, P.~A.}} \yr{2001} {\em An {{Introduction}} to
  {{Magnetohydrodynamics}}\/}.  \publ{Cambridge: Cambridge University Press}.

\bibitem[Davidson \& Lindsay(1998)]{Davidson1998}
{\sc \au{Davidson, P.~A.} \& \au{Lindsay, R.~I.}} \yr{1998}  \at{Stability of
  interfacial waves in aluminium reduction cells}.  \jt{J. Fluid. Mech.}
  \bvol{362},  \pg{273--295}.

\bibitem[Duczek {\em et~al.\/}(2024)Duczek, Horstmann, Ding, Einarsrud,
  Gelfgat, {Godinez-Brizuela}, Kjos, Landgraf, Lappan, Monrrabal, Nash,
  Personnettaz, Sarma, Sommerseth, Trtik, Weber \& Weier]{Duczek2024}
{\sc \au{Duczek, C.}, \au{Horstmann, G.~M.}, \au{Ding, W.}, \au{Einarsrud,
  K.~E.}, \au{Gelfgat, A.~Y.}, \au{{Godinez-Brizuela}, O.~E.}, \au{Kjos,
  O.~S.}, \au{Landgraf, S.}, \au{Lappan, T.}, \au{Monrrabal, G.}, \au{Nash,
  W.}, \au{Personnettaz, P.}, \au{Sarma, M.}, \au{Sommerseth, C.}, \au{Trtik,
  P.}, \au{Weber, N.} \& \au{Weier, T.}} \yr{2024}  \at{Fluid mechanics of
  {{Na-Zn}} liquid metal batteries}.  \jt{Appl. Phys. Rev.}  \bvol{11}~(4),
  \pg{041326}.

\bibitem[Eltishchev {\em et~al.\/}(2024)Eltishchev, Losev \&
  Frick]{Eltishchev2024}
{\sc \au{Eltishchev, V.}, \au{Losev, G.} \& \au{Frick, P.}} \yr{2024}
  \at{Maintenance mechanism of a circular surface wave in a magnetohydrodynamic
  cell and limits of its existence}.  \jt{Phys. Rev. Fluids}  \bvol{9}~(8),
  \pg{083702}.

\bibitem[Eltishchev {\em et~al.\/}(2022)Eltishchev, Losev, Kolesnichenko \&
  Frick]{Eltishchev2022}
{\sc \au{Eltishchev, V.}, \au{Losev, G.}, \au{Kolesnichenko, I.} \& \au{Frick,
  P.}} \yr{2022}  \at{Circular surface wave in a cylindrical {{MHD}} cell}.
  \jt{Exp. Fluids}  \bvol{63}~(8),  \pg{127}.

\bibitem[Evans \& Ziegler(2007)]{Evans2007}
{\sc \au{Evans, J.~W.} \& \au{Ziegler, D.~P.}} \yr{2007}  \at{The
  {{Electrolytic Production}} of {{Aluminum}}}.  \bt{In {\em Encyclopedia of
  {{Electrochemistry}}\/}}.  \publ{John Wiley \& Sons, Ltd}.

\bibitem[Faltinsen \& Timokha(2009)]{Faltinsen2009}
{\sc \au{Faltinsen, O.~M.} \& \au{Timokha, A.}} \yr{2009} {\em Sloshing\/}, 1st
  edn.  \publ{Cambridge University Press}.

\bibitem[Gerbeau {\em et~al.\/}(2006)Gerbeau, Bris \&
  Leli{\`e}vre]{gerbeau2006a}
{\sc \au{Gerbeau, J.-F.}, \au{Bris, C.~L.} \& \au{Leli{\`e}vre, T.}} \yr{2006}
  {\em Mathematical {{Methods}} for the {{Magnetohydrodynamics}} of {{Liquid
  Metals}}\/}.  \publ{Clarendon Press}.

\bibitem[Grants \& Baranovskis(2021)]{Grants2021}
{\sc \au{Grants, I.} \& \au{Baranovskis, R.}} \yr{2021}  \at{Experimental
  observation of metal-electrolyte interface stability in a model of liquid
  metal battery}.  \jt{Magnetohydrodynamics}  \bvol{57}~(2),  \pg{171--180}.

\bibitem[Hegde {\em et~al.\/}(2025)Hegde, Gundrum \& Horstmann]{Hegde2025}
{\sc \au{Hegde, P.}, \au{Gundrum, T.} \& \au{Horstmann, G.~M.}} \yr{2025}
  \at{A model experiment to study the metal pad roll instability under ambient
  conditions}.  \jt{Exp. Fluids}  \bvol{66}~(4),  \pg{76}.

\bibitem[Herreman {\em et~al.\/}(2019)Herreman, Nore, Guermond, Cappanera,
  Weber \& Horstmann]{Herreman2019}
{\sc \au{Herreman, W.}, \au{Nore, C.}, \au{Guermond, J.-L.}, \au{Cappanera,
  L.}, \au{Weber, N.} \& \au{Horstmann, G.~M.}} \yr{2019}  \at{Perturbation
  theory for metal pad roll instability in cylindrical reduction cells}.
  \jt{J. Fluid. Mech.}  \bvol{878},  \pg{598--646}.

\bibitem[Herreman {\em et~al.\/}(2023)Herreman, Wierzchalek, Horstmann,
  Cappanera \& Nore]{Herreman2023}
{\sc \au{Herreman, W.}, \au{Wierzchalek, L.}, \au{Horstmann, G.~M.},
  \au{Cappanera, L.} \& \au{Nore, C.}} \yr{2023}  \at{Stability theory for
  metal pad roll in cylindrical liquid metal batteries}.  \jt{J. Fluid. Mech.}
  \bvol{962},  \pg{A6}.

\bibitem[Horstmann {\em et~al.\/}(2025)Horstmann, Kuhn \&
  Dohnal]{Horstmann2025}
{\sc \au{Horstmann, G.~M.}, \au{Kuhn, J.} \& \au{Dohnal, F.}} \yr{2025}
  \at{Suppression of magnetohydrodynamic interfacial wave instabilities by
  means of parametric anti-resonance}.  \jt{Nonlinear Dyn.}  \bvol{113},
  \pg{14449–14469}.

\bibitem[Horstmann {\em et~al.\/}(2018)Horstmann, Weber \&
  Weier]{Horstmann2018}
{\sc \au{Horstmann, G.~M.}, \au{Weber, N.} \& \au{Weier, T.}} \yr{2018}
  \at{Coupling and stability of interfacial waves in liquid metal batteries}.
  \jt{J. Fluid. Mech.}  \bvol{845},  \pg{1--35}.

\bibitem[Horstmann {\em et~al.\/}(2019)Horstmann, Wylega \&
  Weier]{Horstmann2019}
{\sc \au{Horstmann, G.~M.}, \au{Wylega, M.} \& \au{Weier, T.}} \yr{2019}
  \at{Measurement of interfacial wave dynamics in orbitally shaken cylindrical
  containers using ultrasound pulse-echo techniques}.  \jt{Exp. Fluids}
  \bvol{60}~(4),  \pg{56}.

\bibitem[Joseph(2006)]{Joseph2006}
{\sc \au{Joseph, D.~D.}} \yr{2006}  \at{Potential flow of viscous fluids:
  {{Historical}} notes}.  \jt{Int. J. Multiph. Flow}  \bvol{32}~(3),
  \pg{285--310}.

\bibitem[Kelley \& Weier(2018)]{Kelley2018}
{\sc \au{Kelley, D.~H.} \& \au{Weier, T.}} \yr{2018}  \at{Fluid {{Mechanics}}
  of {{Liquid Metal Batteries}}}.  \jt{Appl. Mech. Rev.}  \bvol{70}~(020801).

\bibitem[Kermeli {\em et~al.\/}(2015)Kermeli, {ter Weer}, {Crijns-Graus} \&
  Worrell]{Kermeli2015}
{\sc \au{Kermeli, K.}, \au{{ter Weer}, P.-H.}, \au{{Crijns-Graus}, W.} \&
  \au{Worrell, E.}} \yr{2015}  \at{Energy efficiency improvement and {{GHG}}
  abatement in the global production of primary aluminium}.  \jt{Energ Effic.}
  \bvol{8}~(4),  \pg{629--666}.

\bibitem[Keulegan(1959)]{Keulegan1959}
{\sc \au{Keulegan, G.~H.}} \yr{1959}  \at{Energy dissipation in standing waves
  in rectangular basins}.  \jt{J. Fluid. Mech.}  \bvol{6}~(1),  \pg{33--50}.

\bibitem[La~Rocca {\em et~al.\/}(2002)La~Rocca, Sciortino \&
  Boniforti]{LaRocca2002}
{\sc \au{La~Rocca, M.}, \au{Sciortino, G.} \& \au{Boniforti, M.~A.}} \yr{2002}
  \at{Interfacial gravity waves in a two-fluid system}.  \jt{Fluid Dyn. Res.}
  \bvol{30}~(1),  \pg{31}.

\bibitem[Lamb(1945)]{Lamb1945}
{\sc \au{Lamb, H.}} \yr{1945} {\em Hydrodynamics\/}.  \publ{New York,: Dover
  publications}.

\bibitem[Lukyanov {\em et~al.\/}(2001)Lukyanov, El \& Molokov]{Lukyanov2001}
{\sc \au{Lukyanov, A.}, \au{El, G.} \& \au{Molokov, S.}} \yr{2001}
  \at{Instability of {{MHD-modified}} interfacial gravity waves revisited}.
  \jt{Phys. Lett. A}  \bvol{290}~(3),  \pg{165--172}.

\bibitem[Mohammad {\em et~al.\/}(2022)Mohammad, Dupuis, Funkenbusch \&
  Kelley]{Mohammad2022}
{\sc \au{Mohammad, Ibrahim}, \au{Dupuis, Marc}, \au{Funkenbusch, Paul~D.} \&
  \au{Kelley, Douglas~H.}} \yr{2022}  \at{Oscillating {{Currents Stabilize
  Aluminum Cells}} for {{Efficient}}, {{Low Carbon Production}}}.  \jt{JOM}
  \bvol{74}~(5),  \pg{1908--1915}.

\bibitem[Molokov(2018)]{Molokov2018}
{\sc \au{Molokov, S.}} \yr{2018}  \at{The nature of interfacial instabilities
  in liquid metal batteries in a vertical magnetic field}.  \jt{EPL}
  \bvol{121}~(4),  \pg{44001}.

\bibitem[Molokov {\em et~al.\/}(2011)Molokov, El \& Lukyanov]{Molokov2011}
{\sc \au{Molokov, S.}, \au{El, G.} \& \au{Lukyanov, A.}} \yr{2011}
  \at{Classification of instability modes in a model of aluminium reduction
  cells with a uniform magnetic field}.  \jt{Theor. Comput. Fluid Dyn.}
  \bvol{25}~(5),  \pg{261--279}.

\bibitem[Munger \& Vincent(2006)]{Munger2006}
{\sc \au{Munger, D.} \& \au{Vincent, A.}} \yr{2006}  \at{Direct simulations of
  {{MHD}} instabilities in aluminum reduction cells}.  \jt{MHD}  \bvol{42}~(4),
   \pg{417--425}.

\bibitem[Munger \& Vincent(2008)]{Munger2008}
{\sc \au{Munger, D.} \& \au{Vincent, A.}} \yr{2008}  \at{A cylindrical model
  for rotational {{MHD}} instabilities in aluminum reduction cells}.
  \jt{Theor. Comput. Fluid Dyn.}  \bvol{22}~(5),  \pg{363--382}.

\bibitem[Nore {\em et~al.\/}(2021)Nore, Cappanera, Guermond, Weier \&
  Herreman]{Nore2021}
{\sc \au{Nore, C.}, \au{Cappanera, L.}, \au{Guermond, J.-L.}, \au{Weier, T.} \&
  \au{Herreman, W.}} \yr{2021}  \at{Feasibility of {{Metal Pad Roll Instability
  Experiments}} at {{Room Temperature}}}.  \jt{Phys. Rev. Lett.}
  \bvol{126}~(18),  \pg{184501}.

\bibitem[Pedcenko {\em et~al.\/}(2017)Pedcenko, Molokov \&
  Bardet]{Pedcenko2017}
{\sc \au{Pedcenko, A.}, \au{Molokov, S.} \& \au{Bardet, B.}} \yr{2017}  \at{The
  {{Effect}} of ``{{Wave Breakers}}'' on the
  {{Magnetohydrodynamic~Instability}} in {{Aluminum Reduction Cells}}}.
  \jt{Metall. Mater. Trans. B}  \bvol{48}~(1),  \pg{6--10}.

\bibitem[Pedchenko {\em et~al.\/}(2009)Pedchenko, Molokov, Priede, Lukyanov \&
  Thomas]{Pedchenko2009}
{\sc \au{Pedchenko, A.}, \au{Molokov, S.}, \au{Priede, J.}, \au{Lukyanov, A.}
  \& \au{Thomas, P.~J.}} \yr{2009}  \at{Experimental model of the interfacial
  instability in aluminium reduction cells}.  \jt{EPL}  \bvol{88}~(2),
  \pg{24001}.

\bibitem[Politis \& Priede(2021)]{Politis2021}
{\sc \au{Politis, G.} \& \au{Priede, J.}} \yr{2021}  \at{Fractality of metal
  pad instability threshold in rectangular cells}.  \jt{J. Fluid. Mech.}
  \bvol{915},  \pg{A101}.

\bibitem[Sele(1977)]{Sele1977}
{\sc \au{Sele, T.}} \yr{1977}  \at{Instabilities of the metal surface in
  electrolytic alumina reduction cells}.  \jt{Metall. Trans. B}  \bvol{8}~(4),
  \pg{613--618}.

\bibitem[Sneyd \& Wang(1994)]{Sneyd1994}
{\sc \au{Sneyd, A.~D.} \& \au{Wang, A.}} \yr{1994}  \at{Interfacial instability
  due to {{MHD}} mode coupling in aluminium reduction cells}.  \jt{J. Fluid.
  Mech.}  \bvol{263},  \pg{343--360}.

\bibitem[Sreenivasan {\em et~al.\/}(2005)Sreenivasan, Davidson \&
  Etay]{Sreenivasan2005}
{\sc \au{Sreenivasan, B.}, \au{Davidson, P.~A.} \& \au{Etay, J.}} \yr{2005}
  \at{On the control of surface waves by a vertical magnetic field}.  \jt{Phys.
  Fluids}  \bvol{17}~(11),  \pg{117101}.

\bibitem[Tucs {\em et~al.\/}(2018)Tucs, Bojarevics \& Pericleous]{Tucs2018}
{\sc \au{Tucs, A.}, \au{Bojarevics, V.} \& \au{Pericleous, K.}} \yr{2018}
  \at{Magnetohydrodynamic stability of large scale liquid metal batteries}.
  \jt{J. Fluid. Mech.}  \bvol{852},  \pg{453--483}.

\bibitem[Urata(1985)]{Urata1985}
{\sc \au{Urata, N.}} \yr{1985}  \at{Magnetics and {{Metal Pad Instability}}}.
  \bt{In {\em Essential {{Readings}} in {{Light Metals}}: {{Volume}} 2
  {{Aluminum Reduction Technology}}\/} (ed. \ed{G.~Bearne, M.~Dupuis \&
  G.~Tarcy})},  \pg{pp. 330--335}.  \publ{Cham: Springer International
  Publishing}.

\bibitem[Xiang \& Zikanov(2019)]{Xiang2019}
{\sc \au{Xiang, L.} \& \au{Zikanov, O.}} \yr{2019}  \at{Numerical simulation of
  rolling pad instability in cuboid liquid metal batteries}.  \jt{Phys. Fluids}
   \bvol{31}~(12),  \pg{124104}.

\bibitem[Zikanov(2018)]{Zikanov2018}
{\sc \au{Zikanov, O.}} \yr{2018}  \at{Shallow water modeling of rolling pad
  instability in liquid metal batteries}.  \jt{Theor. Comput. Fluid Dyn.}
  \bvol{32}~(3),  \pg{325--347}.

\bibitem[Zikanov {\em et~al.\/}(2004)Zikanov, Sun \& Ziegler]{Zikanov2004}
{\sc \au{Zikanov, O.}, \au{Sun, H.} \& \au{Ziegler, D.~P.}} \yr{2004} Shallow
  {{Water Model}} of {{Flows}} in {{Hall-Heroult Cells}}.  \bt{In {\em Light
  {{Met}}.\/}},  \pg{pp. 445--451}.

\bibitem[Zikanov {\em et~al.\/}(2000)Zikanov, Thess, Davidson \&
  Ziegler]{Zikanov2000}
{\sc \au{Zikanov, O.}, \au{Thess, A.}, \au{Davidson, P.~A.} \& \au{Ziegler,
  D.~P.}} \yr{2000}  \at{A new approach to numerical simulation of melt flows
  and interface instability in hall-h{\'e}roult cells}.  \jt{Metall. Mater.
  Trans. B}  \bvol{31}~(6),  \pg{1541--1550}.

\end{thebibliography}



\end{document}